\documentclass[12pt]{article}
\usepackage{a4,epsfig}

\newcommand{\gev}{\ensuremath{\mathrm{\, GeV}}}
\newcommand{\gevc}{\ensuremath{\mathrm{\, GeV}/{\rm c}}}
\newcommand{\gevcc}{\ensuremath{\mathrm{\, GeV}/{\rm c}^2}}
\newcommand{\tevcc}{\ensuremath{\mathrm{\, TeV}/{\rm c}^2}}
\newcommand{\cm}{\ensuremath{\mathrm{\, cm}}}

\newcommand{\invpb}{\ensuremath{\mathrm{\, pb^{-1}}}}

\newcommand{\nbar}{\ensuremath{\mathrm{\bar{N}_{95}}}}
\newcommand{\alephcoll}{{\tt ALEPH}}
\newcommand{\sq}{\ensuremath{\tilde{\mathrm q}}}
\newcommand{\q}{\ensuremath{{\mathrm q}}}
\newcommand{\J}{\ensuremath{{\mathrm J}}}
\newcommand{\lepton}{\ensuremath{{\mathrm l}}}
\newcommand{\LQ}{\ensuremath{{\mathrm {LQ}}}}
\newcommand{\e}{\ensuremath{{\mathrm e}}}
\newcommand{\snu}{\ensuremath{\tilde{\nu}}}
\newcommand{\slep}{\ensuremath{\tilde{\mathrm{\ell}}}}
\newcommand{\Sf}{\ensuremath{\tilde{\mathrm{f}}}}
\newcommand{\se}{\ensuremath{\tilde{\mathrm{e}}}}

\newcommand{\sbq}{\ensuremath{\mathrm{\tilde{b}}}}
\newcommand{\stq}{\ensuremath{\mathrm{\tilde{t}}}}
\newcommand{\sfrm}{\ensuremath{\mathrm{\tilde{f}}}}
\newcommand{\ee}{\ensuremath{\mathrm{e^+e^-}}}

\newcommand{\w}{\ensuremath{\mathrm{W}}}
\newcommand{\bottom}{\ensuremath{\mathrm{b}}}
\newcommand{\Z}{\ensuremath{\mathrm{Z}}}
\newcommand{\Higgs}{\ensuremath{\mathrm{H}}}

\newcommand{\qq}{\ensuremath{\mathrm{q\bar{q}}}}

\newcommand{\Pf}{\ensuremath{\mathrm{f}}}
\newcommand{\PW}{\ensuremath{\mathrm{W}}}
\newcommand{\PZ}{\ensuremath{\mathrm{Z}}}
\newcommand{\Pe}{\ensuremath{\mathrm{e}}}

\newcommand{\lqq}{\ensuremath{\mathrm{lqq}}}
\newcommand{\nqq}{\ensuremath{\mathrm{\nu qq}}}
\newcommand{\nch}{\ensuremath{N_{\mathrm{ch}}}}
\newcommand{\mvis}{\ensuremath{M_{\mathrm{vis}}}}

\newcommand{\elep}{\ensuremath{E_{\mathrm{lep}}}}

\newcommand{\ehad}{\ensuremath{E_{\mathrm{had}}}}

\newcommand{\newc}{\newcommand}
\newc{\emiss}{{\not \!\! E}}
\newc{\R}{$R$}
\newc{\charginom}{M_{\chi^{+}}}
\newc{\mue}{\mu_{\tilde{e}_{iL}}}
\newc{\mud}{\mu_{\tilde{d}_{jL}}}
\newc{\beq}{\begin{equation}}
\newc{\eeq}{\end{equation}}
\newc{\barr}{\begin{eqnarray}}
\newc{\earr}{\end{eqnarray}}
\newc{\ra}{\rightarrow}
\newc{\Da}{\Downarrow}
\newc{\lam}{\lambda}
\newc{\eps}{\epsilon}
\def\tevc{{\rm \, Te\kern-0.125em V/{\rm c}^2}}
\newc{\eq}[1]{(\ref{eq:#1})}
\newc{\eqs}[2]{(\ref{eq:#1},\ref{eq:#2})}
\newc{\etal}{{\it et al.}\ }
\newc{\Hbar}{{\bar H}}
\newc{\Ubar}{{\bar U}}
\newc{\Dbar}{{\bar D}}
\newc{\Ebar}{{\bar E}}
\newc{\eg}{{\it e.g.}\ }
\newc{\ie}{{\it i.e.}\ }
\newc{\nonum}{\nonumber}
\newc{\lab}[1]{\label{eq:#1}}
\newc{\lle}[3]{L_{#1}L_{#2}\Ebar_{#3}}
\newc{\lqd}[3]{L_{#1}Q_{#2}\Dbar_{#3}}
\newc{\udd}[3]{\Ubar_{#1}\Dbar_{#2}\Dbar_{#3}}
\newc{\dpr}[2]{({#1}\cdot{#2})}
\newc{\rpv}{{\not R_p}}
\newc{\rpvm}{{\not \! R_p}}
\newc{\rp}{$R_p$}
\def\gsim{\hbox{\lower -.06cm \hbox{
{\hbox{$\scriptstyle >$}}{\hbox{\kern -.21cm\lower .15cm \hbox{$\scriptstyle
\sim$}}}}}}
\def\lsim{\hbox{\lower -.06cm \hbox{
{\hbox{$\scriptstyle <$}}{\hbox{\kern -.21cm\lower .15cm \hbox{$\scriptstyle 
\sim$}}}}}}

\font\ninerm=cmr9

\begin{document}
\begin{titlepage}
\title{
Search for Supersymmetry with a dominant R-Parity\\
violating $LQ{\bar D}$ Coupling in ${\mathrm e}^+{\mathrm e}^-$ Collisions\\
at centre-of-mass energies of 130~GeV to 172~GeV
\vspace{1cm}}
\author{The ALEPH Collaboration$^\ast)$} 
\date{ }

\maketitle

\begin{picture}(160,1)
\put(10,105){\rm EUROPEAN LABORATORY FOR PARTICLE PHYSICS (CERN)}
\put(115,84){\parbox[t]{45mm}{CERN-EP/98-147}}
\put(115,78){\parbox[t]{45mm}{21 September 1998}}
\end{picture}

\vspace{0cm}
\begin{abstract}
\vspace{.5cm}
A search for pair-production of supersymmetric particles under the 
assumption that R-parity is violated via a dominant $LQ{\bar D}$ coupling 
has been performed using the
data collected by \alephcoll{} at centre-of-mass energies of 130--172$\gev$. 
The observed candidate events in the data are in agreement with the
Standard Model expectation. This result is translated into lower limits on the masses 
of charginos, neutralinos, sleptons, sneutrinos and squarks.  For instance, for 
$m_0=500\gevcc$ and $\tan{\beta}=\sqrt{2}$ charginos with masses smaller than
$81 \gevcc$ and neutralinos with masses smaller 
than $29 \gevcc$  are excluded at the $95\%$ confidence level
 for any generation structure of the $LQ\bar{D}$ coupling.

\end{abstract}
\vfill
\centerline{\em (Submitted to European Physical Journal C)}
\vskip .5cm
\noindent
--------------------------------------------\hfil\break
{\ninerm $^\ast)$ See next pages for the list of authors}

\end{titlepage}

\newpage
\pagestyle{empty}
\newpage
\small
%
%
\newlength{\saveparskip}
\newlength{\savetextheight}
\newlength{\savetopmargin}
\newlength{\savetextwidth}
\newlength{\saveoddsidemargin}
\newlength{\savetopsep}
\setlength{\saveparskip}{\parskip}
\setlength{\savetextheight}{\textheight}
\setlength{\savetopmargin}{\topmargin}
\setlength{\savetextwidth}{\textwidth}
\setlength{\saveoddsidemargin}{\oddsidemargin}
\setlength{\savetopsep}{\topsep}
%
%
\setlength{\parskip}{0.0cm}
\setlength{\textheight}{25.0cm}
\setlength{\topmargin}{-1.5cm}
\setlength{\textwidth}{16 cm}
\setlength{\oddsidemargin}{-0.0cm}
\setlength{\topsep}{1mm}
\pretolerance=10000
\centerline{\large\bf The ALEPH Collaboration}
\footnotesize
\vspace{0.5cm}
{\raggedbottom
\begin{sloppypar}
\samepage\noindent
R.~Barate,
D.~Buskulic,
D.~Decamp,
P.~Ghez,
C.~Goy,
S.~Jezequel,
J.-P.~Lees,
A.~Lucotte,
F.~Martin,
E.~Merle,
\mbox{M.-N.~Minard},
\mbox{J.-Y.~Nief},
P.~Perrodo,
B.~Pietrzyk
\nopagebreak
\begin{center}
\parbox{15.5cm}{\sl\samepage
Laboratoire de Physique des Particules (LAPP), IN$^{2}$P$^{3}$-CNRS,
F-74019 Annecy-le-Vieux Cedex, France}
\end{center}\end{sloppypar}
\vspace{2mm}
\begin{sloppypar}
\noindent
R.~Alemany,
M.P.~Casado,
M.~Chmeissani,
J.M.~Crespo,
M.~Delfino,
E.~Fernandez,
M.~Fernandez-Bosman,
Ll.~Garrido,$^{15}$
E.~Graug\`{e}s,
A.~Juste,
M.~Martinez,
G.~Merino,
R.~Miquel,
Ll.M.~Mir,
P.~Morawitz,
A.~Pacheco,
I.C.~Park,
A.~Pascual,
I.~Riu,
F.~Sanchez
\nopagebreak
\begin{center}
\parbox{15.5cm}{\sl\samepage
Institut de F\'{i}sica d'Altes Energies, Universitat Aut\`{o}noma
de Barcelona, 08193 Bellaterra (Barcelona), E-Spain$^{7}$}
\end{center}\end{sloppypar}
\vspace{2mm}
\begin{sloppypar}
\noindent
A.~Colaleo,
D.~Creanza,
M.~de~Palma,
G.~Gelao,
G.~Iaselli,
G.~Maggi,
M.~Maggi,
S.~Nuzzo,
A.~Ranieri,
G.~Raso,
F.~Ruggieri,
G.~Selvaggi,
L.~Silvestris,
P.~Tempesta,
A.~Tricomi,$^{3}$
G.~Zito
\nopagebreak
\begin{center}
\parbox{15.5cm}{\sl\samepage
Dipartimento di Fisica, INFN Sezione di Bari, I-70126 Bari, Italy}
\end{center}\end{sloppypar}
\vspace{2mm}
\begin{sloppypar}
\noindent
X.~Huang,
J.~Lin,
Q. Ouyang,
T.~Wang,
Y.~Xie,
R.~Xu,
S.~Xue,
J.~Zhang,
L.~Zhang,
W.~Zhao
\nopagebreak
\begin{center}
\parbox{15.5cm}{\sl\samepage
Institute of High-Energy Physics, Academia Sinica, Beijing, The People's
Republic of China$^{8}$}
\end{center}\end{sloppypar}
\vspace{2mm}
\begin{sloppypar}
\noindent
D.~Abbaneo,
U.~Becker,$^{22}$
G.~Boix,$^{24}$
M.~Cattaneo,
F.~Cerutti,
V.~Ciulli,
G.~Dissertori,
H.~Drevermann,
R.W.~Forty,
M.~Frank,
F.~Gianotti,
R.~Hagelberg,
A.W.~Halley,
J.B.~Hansen,
J.~Harvey,
P.~Janot,
B.~Jost,
I.~Lehraus,
O.~Leroy,
P.~Maley,
P.~Mato,
A.~Minten,
L.~Moneta,$^{20}$
A.~Moutoussi,
F.~Ranjard,
L.~Rolandi,
D.~Rousseau,
D.~Schlatter,
M.~Schmitt,$^{1}$
O.~Schneider,
W.~Tejessy,
F.~Teubert,
I.R.~Tomalin,
E.~Tournefier,
M.~Vreeswijk,
H.~Wachsmuth
\nopagebreak
\begin{center}
\parbox{15.5cm}{\sl\samepage
European Laboratory for Particle Physics (CERN), CH-1211 Geneva 23,
Switzerland}
\end{center}\end{sloppypar}
\vspace{2mm}
\begin{sloppypar}
\noindent
Z.~Ajaltouni,
F.~Badaud
G.~Chazelle,
O.~Deschamps,
S.~Dessagne,
A.~Falvard,
C.~Ferdi,
P.~Gay,
C.~Guicheney,
P.~Henrard,
J.~Jousset,
B.~Michel,
S.~Monteil,
\mbox{J-C.~Montret},
D.~Pallin,
P.~Perret,
F.~Podlyski
\nopagebreak
\begin{center}
\parbox{15.5cm}{\sl\samepage
Laboratoire de Physique Corpusculaire, Universit\'e Blaise Pascal,
IN$^{2}$P$^{3}$-CNRS, Clermont-Ferrand, F-63177 Aubi\`{e}re, France}
\end{center}\end{sloppypar}
\vspace{2mm}
\begin{sloppypar}
\noindent
J.D.~Hansen,
J.R.~Hansen,
P.H.~Hansen,
B.S.~Nilsson,
B.~Rensch,
A.~W\"a\"an\"anen
\begin{center}
\parbox{15.5cm}{\sl\samepage
Niels Bohr Institute, 2100 Copenhagen, DK-Denmark$^{9}$}
\end{center}\end{sloppypar}
\vspace{2mm}
\begin{sloppypar}
\noindent
G.~Daskalakis,
A.~Kyriakis,
C.~Markou,
E.~Simopoulou,
A.~Vayaki
\nopagebreak
\begin{center}
\parbox{15.5cm}{\sl\samepage
Nuclear Research Center Demokritos (NRCD), GR-15310 Attiki, Greece}
\end{center}\end{sloppypar}
\vspace{2mm}
\begin{sloppypar}
\noindent
A.~Blondel,
\mbox{J.-C.~Brient},
F.~Machefert,
A.~Roug\'{e},
M.~Rumpf,
R.~Tanaka,
A.~Valassi,$^{6}$
H.~Videau
\nopagebreak
\begin{center}
\parbox{15.5cm}{\sl\samepage
Laboratoire de Physique Nucl\'eaire et des Hautes Energies, Ecole
Polytechnique, IN$^{2}$P$^{3}$-CNRS, \mbox{F-91128} Palaiseau Cedex, France}
\end{center}\end{sloppypar}
\vspace{2mm}
\begin{sloppypar}
\noindent
E.~Focardi,
G.~Parrini,
K.~Zachariadou
\nopagebreak
\begin{center}
\parbox{15.5cm}{\sl\samepage
Dipartimento di Fisica, Universit\`a di Firenze, INFN Sezione di Firenze,
I-50125 Firenze, Italy}
\end{center}\end{sloppypar}
\vspace{2mm}
\begin{sloppypar}
\noindent
R.~Cavanaugh,
M.~Corden,
C.~Georgiopoulos,
T.~Huehn,
D.E.~Jaffe
\nopagebreak
\begin{center}
\parbox{15.5cm}{\sl\samepage
Supercomputer Computations Research Institute,
Florida State University,
Tallahassee, FL 32306-4052, USA $^{13,14}$}
\end{center}\end{sloppypar}
\vspace{2mm}
\begin{sloppypar}
\noindent
A.~Antonelli,
G.~Bencivenni,
G.~Bologna,$^{4}$
F.~Bossi,
P.~Campana,
G.~Capon,
V.~Chiarella,
P.~Laurelli,
G.~Mannocchi,$^{5}$
F.~Murtas,
G.P.~Murtas,
L.~Passalacqua,
M.~Pepe-Altarelli$^{12}$
\nopagebreak
\begin{center}
\parbox{15.5cm}{\sl\samepage
Laboratori Nazionali dell'INFN (LNF-INFN), I-00044 Frascati, Italy}
\end{center}\end{sloppypar}
\vspace{2mm}
\begin{sloppypar}
\noindent
M.~Chalmers,
L.~Curtis,
J.G.~Lynch,
P.~Negus,
V.~O'Shea,
C.~Raine,
J.M.~Scarr,
P.~Teixeira-Dias,
A.S.~Thompson,
E.~Thomson,
J.J.~Ward
\nopagebreak
\begin{center}
\parbox{15.5cm}{\sl\samepage
Department of Physics and Astronomy, University of Glasgow, Glasgow G12
8QQ,United Kingdom$^{10}$}
\end{center}\end{sloppypar}
\pagebreak
\begin{sloppypar}
\noindent
O.~Buchm\"uller,
S.~Dhamotharan,
C.~Geweniger,
P.~Hanke,
G.~Hansper,
V.~Hepp,
E.E.~Kluge,
A.~Putzer,
J.~Sommer,
K.~Tittel,
S.~Werner,
M.~Wunsch
\nopagebreak
\begin{center}
\parbox{15.5cm}{\sl\samepage
Institut f\"ur Hochenergiephysik, Universit\"at Heidelberg, D-69120
Heidelberg, Germany$^{16}$}
\end{center}\end{sloppypar}
\vspace{2mm}
\begin{sloppypar}
\noindent
R.~Beuselinck,
D.M.~Binnie,
W.~Cameron,
P.J.~Dornan,$^{12}$
M.~Girone,
S.~Goodsir,
N.~Marinelli,
E.B.~Martin,
J.~Nash,
J.K.~Sedgbeer,
P.~Spagnolo,
M.D.~Williams
\nopagebreak
\begin{center}
\parbox{15.5cm}{\sl\samepage
Department of Physics, Imperial College, London SW7 2BZ,
United Kingdom$^{10}$}
\end{center}\end{sloppypar}
\vspace{2mm}
\begin{sloppypar}
\noindent
V.M.~Ghete,
P.~Girtler,
E.~Kneringer,
D.~Kuhn,
G.~Rudolph
\nopagebreak
\begin{center}
\parbox{15.5cm}{\sl\samepage
Institut f\"ur Experimentalphysik, Universit\"at Innsbruck, A-6020
Innsbruck, Austria$^{18}$}
\end{center}\end{sloppypar}
\vspace{2mm}
\begin{sloppypar}
\noindent
A.P.~Betteridge,
C.K.~Bowdery,
P.G.~Buck,
P.~Colrain,
G.~Crawford,
G.~Ellis,
A.J.~Finch,
F.~Foster,
G.~Hughes,
R.W.L.~Jones,
A.N.~Robertson,
M.I.~Williams
\nopagebreak
\begin{center}
\parbox{15.5cm}{\sl\samepage
Department of Physics, University of Lancaster, Lancaster LA1 4YB,
United Kingdom$^{10}$}
\end{center}\end{sloppypar}
\vspace{2mm}
\begin{sloppypar}
\noindent
P.~van~Gemmeren,
I.~Giehl,
C.~Hoffmann,
K.~Jakobs,
K.~Kleinknecht,
M.~Kr\"ocker,
H.-A.~N\"urnberger,
G.~Quast,
B.~Renk,
E.~Rohne,
H.-G.~Sander,
S.~Schmeling,
C.~Zeitnitz,
T.~Ziegler
\nopagebreak
\begin{center}
\parbox{15.5cm}{\sl\samepage
Institut f\"ur Physik, Universit\"at Mainz, D-55099 Mainz, Germany$^{16}$}
\end{center}\end{sloppypar}
\vspace{2mm}
\begin{sloppypar}
\noindent
J.J.~Aubert,
C.~Benchouk,
A.~Bonissent,
J.~Carr,$^{12}$
P.~Coyle,
A.~Ealet,
D.~Fouchez,
F.~Motsch,
P.~Payre,
M.~Talby,
M.~Thulasidas,
A.~Tilquin
\nopagebreak
\begin{center}
\parbox{15.5cm}{\sl\samepage
Centre de Physique des Particules, Facult\'e des Sciences de Luminy,
IN$^{2}$P$^{3}$-CNRS, F-13288 Marseille, France}
\end{center}\end{sloppypar}
\vspace{2mm}
\begin{sloppypar}
\noindent
M.~Aleppo,
M.~Antonelli,
F.~Ragusa
\nopagebreak
\begin{center}
\parbox{15.5cm}{\sl\samepage
Dipartimento di Fisica, Universit\`a di Milano e INFN Sezione di
Milano, I-20133 Milano, Italy.}
\end{center}\end{sloppypar}
\vspace{2mm}
\begin{sloppypar}
\noindent
R.~Berlich,
V.~B\"uscher,
H.~Dietl,
G.~Ganis,
K.~H\"uttmann,
G.~L\"utjens,
C.~Mannert,
W.~M\"anner,
\mbox{H.-G.~Moser},
S.~Schael,
R.~Settles,
H.~Seywerd,
H.~Stenzel,
W.~Wiedenmann,
G.~Wolf
\nopagebreak
\begin{center}
\parbox{15.5cm}{\sl\samepage
Max-Planck-Institut f\"ur Physik, Werner-Heisenberg-Institut,
D-80805 M\"unchen, Germany\footnotemark[16]}
\end{center}\end{sloppypar}
\vspace{2mm}
\begin{sloppypar}
\noindent
J.~Boucrot,
O.~Callot,
S.~Chen,
M.~Davier,
L.~Duflot,
\mbox{J.-F.~Grivaz},
Ph.~Heusse,
A.~H\"ocker,
A.~Jacholkowska,
M.~Kado,
D.W.~Kim,$^{2}$
F.~Le~Diberder,
J.~Lefran\c{c}ois,
L.~Serin,
\mbox{J.-J.~Veillet},
I.~Videau,$^{12}$
J.-B.~de~Vivie~de~R\'egie,
D.~Zerwas
\nopagebreak
\begin{center}
\parbox{15.5cm}{\sl\samepage
Laboratoire de l'Acc\'el\'erateur Lin\'eaire, Universit\'e de Paris-Sud,
IN$^{2}$P$^{3}$-CNRS, F-91898 Orsay Cedex, France}
\end{center}\end{sloppypar}
\vspace{2mm}
\begin{sloppypar}
\noindent
\samepage
P.~Azzurri,
G.~Bagliesi,$^{12}$
S.~Bettarini,
T.~Boccali,
C.~Bozzi,
G.~Calderini,
R.~Dell'Orso,
R.~Fantechi,
I.~Ferrante,
A.~Giassi,
A.~Gregorio,
F.~Ligabue,
A.~Lusiani,
P.S.~Marrocchesi,
A.~Messineo,
F.~Palla,
G.~Rizzo,
G.~Sanguinetti,
A.~Sciab\`a,
G.~Sguazzoni,
R.~Tenchini,
C.~Vannini,
A.~Venturi,
P.G.~Verdini
\samepage
\begin{center}
\parbox{15.5cm}{\sl\samepage
Dipartimento di Fisica dell'Universit\`a, INFN Sezione di Pisa,
e Scuola Normale Superiore, I-56010 Pisa, Italy}
\end{center}\end{sloppypar}
\vspace{2mm}
\begin{sloppypar}
\noindent
G.A.~Blair,
J.T.~Chambers,
J.~Coles,
G.~Cowan,
M.G.~Green,
T.~Medcalf,
J.A.~Strong,
J.H.~von~Wimmersperg-Toeller
\nopagebreak
\begin{center}
\parbox{15.5cm}{\sl\samepage
Department of Physics, Royal Holloway \& Bedford New College,
University of London, Surrey TW20 OEX, United Kingdom$^{10}$}
\end{center}\end{sloppypar}
\vspace{2mm}
\begin{sloppypar}
\noindent
D.R.~Botterill,
R.W.~Clifft,
T.R.~Edgecock,
P.R.~Norton,
J.C.~Thompson,
A.E.~Wright
\nopagebreak
\begin{center}
\parbox{15.5cm}{\sl\samepage
Particle Physics Dept., Rutherford Appleton Laboratory,
Chilton, Didcot, Oxon OX11 OQX, United Kingdom$^{10}$}
\end{center}\end{sloppypar}
\vspace{2mm}
\begin{sloppypar}
\noindent
\mbox{B.~Bloch-Devaux},
P.~Colas,
B.~Fabbro,
G.~Fa\"\i f,
E.~Lan\c{c}on,$^{12}$
\mbox{M.-C.~Lemaire},
E.~Locci,
P.~Perez,
H.~Przysiezniak,
J.~Rander,
\mbox{J.-F.~Renardy},
A.~Rosowsky,
A.~Trabelsi,$^{23}$
B.~Tuchming,
B.~Vallage
\nopagebreak
\begin{center}
\parbox{15.5cm}{\sl\samepage
CEA, DAPNIA/Service de Physique des Particules,
CE-Saclay, F-91191 Gif-sur-Yvette Cedex, France$^{17}$}
\end{center}\end{sloppypar}
\vspace{2mm}
\begin{sloppypar}
\noindent
S.N.~Black,
J.H.~Dann,
H.Y.~Kim,
N.~Konstantinidis,
A.M.~Litke,
M.A. McNeil,
G.~Taylor
\nopagebreak
\begin{center}
\parbox{15.5cm}{\sl\samepage
Institute for Particle Physics, University of California at
Santa Cruz, Santa Cruz, CA 95064, USA$^{19}$}
\end{center}\end{sloppypar}
\pagebreak
\vspace{2mm}
\begin{sloppypar}
\noindent
C.N.~Booth,
S.~Cartwright,
F.~Combley,
M.S.~Kelly,
M.~Lehto,
L.F.~Thompson
\nopagebreak
\begin{center}
\parbox{15.5cm}{\sl\samepage
Department of Physics, University of Sheffield, Sheffield S3 7RH,
United Kingdom$^{10}$}
\end{center}\end{sloppypar}
\vspace{2mm}
\begin{sloppypar}
\noindent
K.~Affholderbach,
A.~B\"ohrer,
S.~Brandt,
J.~Foss,
C.~Grupen,
G.~Prange,
L.~Smolik,
F.~Stephan
\nopagebreak
\begin{center}
\parbox{15.5cm}{\sl\samepage
Fachbereich Physik, Universit\"at Siegen, D-57068 Siegen, Germany$^{16}$}
\end{center}\end{sloppypar}
\vspace{2mm}
\begin{sloppypar}
\noindent
G.~Giannini,
B.~Gobbo
\nopagebreak
\begin{center}
\parbox{15.5cm}{\sl\samepage
Dipartimento di Fisica, Universit\`a di Trieste e INFN Sezione di Trieste,
I-34127 Trieste, Italy}
\end{center}\end{sloppypar}
\vspace{2mm}
\begin{sloppypar}
\noindent
J.~Putz,
J.~Rothberg,
S.~Wasserbaech,
R.W.~Williams
\nopagebreak
\begin{center}
\parbox{15.5cm}{\sl\samepage
Experimental Elementary Particle Physics, University of Washington, WA 98195
Seattle, U.S.A.}
\end{center}\end{sloppypar}
\vspace{2mm}
\begin{sloppypar}
\noindent
S.R.~Armstrong,
E.~Charles,
P.~Elmer,
D.P.S.~Ferguson,
Y.~Gao,
S.~Gonz\'{a}lez,
T.C.~Greening,
O.J.~Hayes,
H.~Hu,
S.~Jin,
P.A.~McNamara III,
J.M.~Nachtman,$^{21}$
J.~Nielsen,
W.~Orejudos,
Y.B.~Pan,
Y.~Saadi,
I.J.~Scott,
J.~Walsh,
Sau~Lan~Wu,
X.~Wu,
G.~Zobernig
\nopagebreak
\begin{center}
\parbox{15.5cm}{\sl\samepage
Department of Physics, University of Wisconsin, Madison, WI 53706,
USA$^{11}$}
\end{center}\end{sloppypar}
}
\footnotetext[1]{Now at Harvard University, Cambridge, MA 02138, U.S.A.}
\footnotetext[2]{Permanent address: Kangnung National University, Kangnung,
Korea.}
\footnotetext[3]{Also at Dipartimento di Fisica, INFN Sezione di Catania,
Catania, Italy.}
\footnotetext[4]{Also Istituto di Fisica Generale, Universit\`{a} di
Torino, Torino, Italy.}
\footnotetext[5]{Also Istituto di Cosmo-Geofisica del C.N.R., Torino,
Italy.}
\footnotetext[6]{Now at LAL, Orsay}
\footnotetext[7]{Supported by CICYT, Spain.}
\footnotetext[8]{Supported by the National Science Foundation of China.}
\footnotetext[9]{Supported by the Danish Natural Science Research Council.}
\footnotetext[10]{Supported by the UK Particle Physics and Astronomy Research
Council.}
\footnotetext[11]{Supported by the US Department of Energy, grant
DE-FG0295-ER40896.}
\footnotetext[12]{Also at CERN, 1211 Geneva 23,Switzerland.}
\footnotetext[13]{Supported by the US Department of Energy, contract
DE-FG05-92ER40742.}
\footnotetext[14]{Supported by the US Department of Energy, contract
DE-FC05-85ER250000.}
\footnotetext[15]{Permanent address: Universitat de Barcelona, 08208 Barcelona,
Spain.}
\footnotetext[16]{Supported by the Bundesministerium f\"ur Bildung,
Wissenschaft, Forschung und Technologie, Germany.}
\footnotetext[17]{Supported by the Direction des Sciences de la
Mati\`ere, C.E.A.}
\footnotetext[18]{Supported by Fonds zur F\"orderung der wissenschaftlichen
Forschung, Austria.}
\footnotetext[19]{Supported by the US Department of Energy,
grant DE-FG03-92ER40689.}
\footnotetext[20]{Now at University of Geneva, 1211 Geneva 4, Switzerland.}
\footnotetext[21]{Now at University of California at Los Angeles (UCLA),
Los Angeles, CA 90024, U.S.A.}
\footnotetext[22]{Now at SAP AG, D-69185 Walldorf, Germany}
\footnotetext[23]{Now at D\'epartement de Physique, Facult\'e des Sciences de Tunis, 1060 Le Belv\'ed\`ere, Tunisia.}
\footnotetext[24]{Supported by the Commission of the European Communities,
contract ERBFMBICT982894.}
%
%
\setlength{\parskip}{\saveparskip}
\setlength{\textheight}{\savetextheight}
\setlength{\topmargin}{\savetopmargin}
\setlength{\textwidth}{\savetextwidth}
\setlength{\oddsidemargin}{\saveoddsidemargin}
\setlength{\topsep}{\savetopsep}
\normalsize
\newpage
\pagestyle{plain}
\setcounter{page}{1}

\section{Introduction}
In minimal supersymmetric extensions of the Standard Model (MSSM) \cite{MSSM} 
it is usually assumed that R-parity is conserved. 
R-parity, a discrete multiplicative  quantum number \cite{fayet} 
defined by\footnote{Here $B$, $L$ and $S$ denote baryon number,
lepton number and the spin of a field.} $R_{p}=-1^{3B+L+2S}$, distinguishes
Standard Model (SM) 
particles with $R_{p}=+1$  from supersymmetric (SUSY) particles with $R_{p}=-1$.
 R-parity conservation has two important consequences for SUSY 
phenomenology. Firstly, SUSY particles must be produced in pairs and, secondly,
the Lightest SUSY Particle (the LSP) must be stable. All SUSY particles decay to
the LSP, and since the LSP is weakly 
interacting it will escape detection and the characteristic signature for 
R-parity conserving SUSY is therefore missing energy.

If R-parity is violated the following additional terms -- which are invariant
under the  $SU(3)_c\times SU(2)_L\times U(1)_Y$ gauge symmetry -- are allowed in the
superpotential \cite{rpsuper}
\begin{equation}
W_{\rpv} = \lam_{ijk}\lle{i}{j}{k}+\lam'_{ijk}\lqd{i}{j}{k}+\lam''_{ijk}\udd{i}{j}{k}.
\label{eqrpv}
\end{equation}
Here $L$ ($Q$) are the lepton (quark) doublet superfields, and
$\Dbar,\Ubar$ ($\Ebar$) are the 
down-like and up-like quark (lepton) singlet superfields, respectively; 
$\lambda, \lambda', \lambda''$ are Yukawa couplings, and $i,j,k=1,2,3$ 
are generation indices. The simultaneous presence of the last two terms 
leads to rapid proton decay, a problem which may be overcome by imposing R-parity
conservation, or alternatively by allowing only a subset of the terms in
(\ref{eqrpv}), as is done in ``R-parity violating'' models \cite{rpv.models}.
The introduction of these terms has two 
major consequences for collider searches:
the LSP is not stable 
and supersymmetric particles (sparticles) can be produced singly. 
The latter possibility is not addressed here and this paper focuses on the 
pair-production of sparticles, which subsequently decay violating R-parity. 
Two simplifying assumptions are made throughout the analysis:
\begin{itemize}
\item{Only one term in  Eq.(\ref{eqrpv}) is non-zero. 
 The analysis presented here is restricted to signals from the $LQ{\bar
  D}$ couplings. Signals from the $LL\bar{E}$ couplings were considered 
in \cite{LLEpaper}.
When the results are translated into limits, 
  it is also assumed that only one of the possible twenty seven
 $\lambda'_{ijk}$ couplings is non-zero. 
 The derived limits correspond to the most conservative choice of the coupling.}
\item{The lifetime of the LSP is negligible, i.e.\ the mean path of flight is less than $1$cm.}
\end{itemize} 
The second assumption restricts this analysis to models satisfying lower 
bounds on $\lambda'$, but these lower bounds are well below upper limits
from low energy constraints. 

The reported search results use data collected by the \alephcoll{} detector in
1995-1996 at centre-of-mass energies from 130 to 172$\gev${}. 
The total data sample used in the analysis corresponds to
 an integrated recorded luminosity of 27.5\invpb.

The outline of this paper is as follows: after reviewing the phenomenology of R-parity
violating SUSY models and existing limits in Sections~\ref{pheno} and
\ref{ex.lims}, a brief description of the \alephcoll{}
detector is given in Section~\ref{aleph.detector}. The data and Monte Carlo (MC)
samples and the search analyses are
described in Sections~\ref{dataandmc} and \ref{searches}, and the results and their interpretation
within the MSSM  are discussed in
Section~\ref{results}. Finally conclusions are drawn in Section~\ref{conclusions}. 

\section{Phenomenology}\label{pheno}
Within minimal Supersymmetry all SM fermions have scalar SUSY partners: the
sleptons, sneutrinos and squarks. The SUSY equivalent of the gauge and Higgs
bosons are the charginos and neutralinos, which are the mass eigenstates of the
(${\tilde \w^+}, {\tilde \Higgs^+}$) and (${\tilde \gamma},{\tilde \Z},{\tilde
  \Higgs^0_1},{\tilde \Higgs^0_2}$) fields, respectively, with obvious notation.
If R-parity is conserved the LSP is stable and 
cosmological arguments \cite{cosmo} consequently require it to be neutral,
i.e.\ the lightest neutralino, the sneutrino or the gravitino.

If R-parity is violated, the LSP can decay to SM particles,  and the
above cosmological arguments do not apply. This analysis considers all
possible LSP candidates with the exception of the gravitino, which is assumed
to be heavy enough to effectively decouple, and the gluino, which cannot be the 
LSP if the gaugino masses are universal at the GUT scale \cite{MSSM}.

\begin{figure}[t]
\begin{center}
\makebox[\textwidth]{
\epsfig{figure=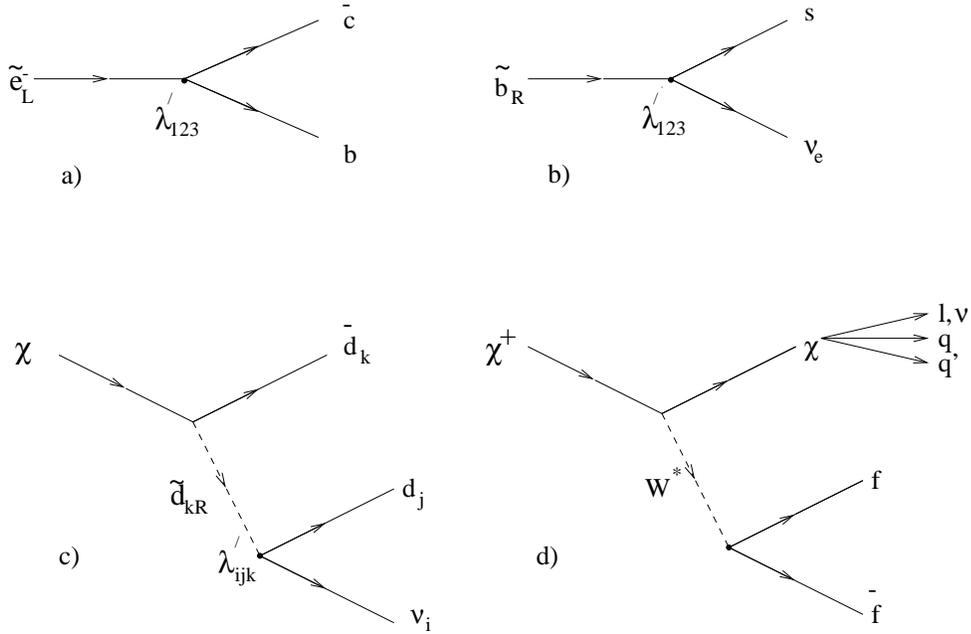,width=.8\textwidth}}
\caption[.]{\em\label{rpv.decays.feyn}{Examples of  decays of supersymmetric particles:
a) {\em direct} decay of a left-handed selectron, b) {\em direct} sbottom decay,
c) {\em direct} neutralino decay via sfermion exchange and 
d) {\em indirect} chargino decay via exchange of a $W^*$.}}
\end{center}
\end{figure}

The production cross sections do not depend on the size of the 
R-parity violating Yukawa coupling $\lambda'$, since the pair-production
of sparticles only involves gauge couplings\footnote{Ignoring 
 t-channel processes in which the R-parity violating coupling appears twice.}. 
 The sparticle decay modes 
are classified according to their topologies:
all decays proceeding via the lightest neutralino are throughout
referred to as the ``indirect'' decay modes. The final states produced
by the other decays, the ``direct'' decay modes, consist of two quarks or
one or two quarks and a lepton\footnote{In the following the 
term ``lepton'' shall denote ``charged lepton''.} or neutrino as summarised in 
Table~\ref{rpv.decays}. 
 Fig.~\ref{rpv.decays.feyn}a and \ref{rpv.decays.feyn}b show examples of {\em direct} selectron and
 sbottom decays;
 Fig.~\ref{rpv.decays.feyn}c and \ref{rpv.decays.feyn}d show 
 examples of a ({\em direct}) neutralino decay and an {\em indirect} chargino decay. 
  The classification into {\em direct} decay modes is made on the basis of the
  topology of the decay, and it is therefore immaterial whether the exchanged
  sfermion in the chargino or neutralino decays  
  is real or virtual. In order to be as model independent as possible,   
all topologies arising from both classes of decays  
are considered in the subsequent analyses.

\begin{table}
\centering
\begin{tabular}{|c|c|}
\hline 
Sparticle  &  Decay Mode ($\lambda'_{ijk}$)\\
\hline 
$\chi^+$ & $\nu_i u_j \bar{d}_k$, $l^+_i \bar{d}_j d_k$, $l^+_i \bar{u}_j u_k$, $\bar{\nu}_i \bar{d}_j u_k$ \\
$\chi$ &  $l^-_i u_j \bar{d}_k$ , $l^+_i \bar{u}_j d_k$ , $\nu_i d_j \bar{d}_k$, $\bar{\nu}_i \bar{d}_j d_k$ \\
$\tilde{d}_{kR}$ & $\bar{\nu}_i d_j$, $l^-_i u_j$ \\
$\tilde{d}_{jL}$ & $\bar{\nu}_i d_k$ \\
$\tilde{u}_{jL}$ & $l^+_i d_k$ \\
 ${\tilde l}^-_{iL}$ & $\bar{u}_j d_k$  \\
 ${\tilde \nu}_{i}$ & $d_j \bar{d}_k$ \\
\hline
\end{tabular}
\caption[.]{{\em Direct} R-parity violating decay modes for a non-zero coupling
  $\lambda'_{ijk}$. 
  Here
  $i,j,k$ are generation indices. For example, the selectron $\tilde{e}^-_L$ can decay
  to   $\bar{c}b$ via the coupling  $\lambda'_{123}$.}
\label{rpv.decays}
\end{table}

Following the above terminology, the lightest neutralino can decay 
{\em directly} to two quarks and a lepton or neutrino
either via 2-body decays to lighter sfermions, or via a 3-body
decay. The flavours of the decay products of the neutralino depend on the flavour 
structure of the Yukawa coupling $\lambda'_{ijk}$.
 Heavier neutralinos can also decay {\em indirectly} to the lightest neutralino:
 $\chi' \ra \Z^* \chi \ra \Pf\bar{\Pf} \chi$  .  

The chargino can decay {\em indirectly} to the neutralino:
$\chi^+ \ra \w^* \chi \ra \Pf\bar{\Pf'} \chi$. The chargino can also decay {\em directly} 
to SM particles: $\chi^+ \ra {\mathrm {u \bar{u} l^+}}$ or $\chi^+ \ra {\mathrm {u
 \bar{d}  \nu}}$. This typically happens when sfermions are
 lighter than the chargino, or when the chargino is the LSP.
 Throughout this paper the gauge unification condition \cite{MSSM}
\beq
{M_1 = \frac{5}{3} \tan^2\theta_\w M_2} \label{gauge.uni}
\eeq
is assumed. Under this assumption the chargino cannot be the LSP if $M_{\chi^+}>45.6\gevcc$ --
 the LEP~1 chargino mass limit \cite{lep1.jfg}--, but it is noted that the
 search analyses cover chargino LSP topologies.

Sfermions can decay {\em indirectly} to the lightest neutralino:
$\slep \ra \lepton \chi$, ${\tilde \nu} \ra \nu \chi$ and 
$\sq \ra \q \chi$. If the chargino is
lighter than the sfermions, the decays $\slep \ra \nu \chi^+$,
${\tilde \nu} \ra \lepton^- \chi^+$ and $\sq \ra \q' \chi^+$ are viable decay modes, 
but are not considered in the following. Sfermions
may also decay {\em directly} to two quarks, in the case of sleptons and sneutrinos,
or a quark and a lepton or neutrino, in the case of squarks.

\section{Existing Limits and the LSP Decay Length}\label{ex.lims}

No direct searches were undertaken at LEP~1 under the 
assumption of a non-zero $LQ\bar{D}$ operator. However {\em direct} decays of sfermions are 
constrained by searches for other particles. Searches for charged Higgs bosons at 
LEP~1 \cite{lep1.ch.higgs} constrain slepton or sneutrino pairs decaying {\em directly}
to four-jet final states leading to a mass limit of $M_{\slep},M_{\tilde{\nu}}>45\gevcc$.

When the {\em direct} decays of squarks are dominant the signature is identical to
leptoquark production. The limits from the Tevatron \cite{TeVLQ} on 
 scalar leptoquarks are $M_{LQ} > 213 \gevcc$ and $184 \gevcc$  
 for $BR(\LQ \ra \e \q)=1$ and $BR(\LQ \ra \mu \q)=1$, respectively, and
 exclude the possibility of seeing $\sq \ra \e\q$ or $\sq \ra \mu \q$ at LEP. 

For charginos and neutralinos and the
{\em indirect} decay modes of the sfermions 
the only existing limits on sparticle masses are those that derive from 
the precision measurements of the Z-width: $M_{\chi^+}>45.6\gevcc$, $M_{\slep}>38\gevcc$,
$M_{\tilde \nu}>41\gevcc$ and \mbox{ $M_{\sq_L}>44\gevcc$}. Allowing for a general mixing
in the squark sector there is no absolute lower bound on squark masses.

In addition to these mass limits,
 upper-bounds on the size of the coupling $\lambda'$  from low energy constraints
 exist \cite{resonant.1}. The most stringent limit requires \cite{neutrino.majorana}:
\beq
\lambda'_{133} < 0.002 \sqrt{\frac{M_{\sbq}} {100 \gevcc}} \label{lam133limit}
\label{low.bound}
\eeq
As discussed in \cite{LLEpaper} this bound and the assumption of negligible
lifetime restrict the sensitivity of this analysis to neutralino masses 
exceeding  $M_{\chi}\gsim$\ 10\gevcc, since pair-produced neutralinos with
smaller masses and couplings satisfying Eq.(\ref{low.bound}) would decay with a
mean path of flight  exceeding $1$ cm.
  Close to the kinematic limit,  gauginos can be
probed down to $\lambda' \gsim\ 10^{-5}$ for $M_{\Sf} = 100 \gevcc$,
and {\em direct} sfermion decays down to $\lambda' \gsim\ 10^{-7}$.

\section{\label{aleph.detector}The ALEPH Detector}
The \alephcoll{} detector is described in detail in
Ref.~\cite{bib:detectorpaper}. An account of the performance of the
detector and a description of the standard analysis algorithms can be
found in Ref.~\cite{bib:performancepaper}. Here, only a brief
description of the detector components and the algorithms relevant for
this analysis is given.

The trajectories of charged particles are measured with
a silicon
vertex detector, a cylindrical drift chamber, and a large time
projection chamber (TPC). The detectors are immersed in a 1.5~T axial field provided
by a superconducting solenoidal coil.
The electromagnetic calorimeter (ECAL), placed between the TPC and the coil,
is a highly segmented sampling calorimeter which is used to identify electrons
and photons and to measure their energy. 
The luminosity monitors extend the calorimetric coverage
down to 34~mrad from the beam axis. 
The hadron calorimeter (HCAL) consists of the iron return yoke of the magnet
instrumented with streamer tubes. It provides a measurement of hadronic energy
and, together with the external muon chambers, muon identification.

The calorimetry and tracking information are combined
in an energy flow algorithm, classifying a set of
energy flow ``particles'' as photons, neutral
hadrons and charged particles. Hereafter, charged particle tracks
reconstructed with at least four hits in the TPC,
and originating from within a cylinder of length~20~cm and radius~2~cm
coaxial with the beam and centred at the nominal collision point,
will be referred to as {\it good tracks}.

Lepton identification is described in~\cite{bib:performancepaper}.
Electrons are identified
using the transverse and longitudinal shower shapes in ECAL.
Muons are separated from hadrons by their characteristic 
pattern in HCAL and the presence of hits in the muon chambers.

\section{Data and Monte Carlo Samples}\label{dataandmc}
This analysis uses data collected by \alephcoll{} in 1996 at centre-of-mass
energies of 161.3~GeV (11.1\invpb), 170.3~GeV (1.1\invpb) and
172.3~GeV (9.6\invpb). In the search for sfermions the sensitivity is 
increased  by including also the LEP~1.5
data recorded in 1995 at \mbox{$\sqrt{s}=130$--136~GeV} (5.7\invpb).

For the purpose of designing selections and evaluating efficiencies,
samples of signal events for all accessible final states have been
generated using {\tt SUSYGEN}~\cite{susygen} for a wide range of signal
masses. A subset of these has been processed through the
full \alephcoll{} detector
simulation and reconstruction programs, whereas efficiencies for
intermediate points have been interpolated using a fast, simplified simulation.

For the stop, the decays via loop diagrams to a charm quark and the lightest
neutralino result in a lifetime larger than the typical
hadronisation time scale. The scalar bottom can also develop a 
substantial lifetime in certain regions of parameter space. It is also possible that
the lifetime of squarks decaying {\em directly} is sufficiently long for hadronisation
effects to become important.
This has been 
taken into account by modifying the {\tt SUSYGEN} MC program 
to allow stops and sbottoms to hadronise prior to their decays according 
to the spectator model \cite{laurent}.

Samples of all major backgrounds have been generated and passed through the
full simulation, corresponding to at least 20 times the collected
luminosity in the data. Events from $\gamma\gamma\to$ hadrons, $\ee\to \qq{}$
and four-fermion events
from $\PW\Pe\nu$, $\PZ\gamma^*$ and $\PZ\Pe\Pe$
were produced with {\tt PYTHIA}~\cite{pythia}, with a 
 vector-boson invariant mass cut of $0.2\gevcc$ for $\PZ\gamma^*$ and $\PW\Pe\nu$,
and $2\gevcc$ for $\PZ\Pe\Pe$.
Pairs of W bosons were generated with {\tt KORALW}~\cite{koralw}.
 Pair production of leptons
was simulated with {\tt UNIBAB}~\cite{unibab} (electrons) and
{\tt KORALZ}~\cite{koralz}
(muons and taus), and the process $\gamma\gamma\to$ leptons with
{\tt PHOT02}~\cite{phot02}.

\section{\label{searches}Selection Criteria}
For a dominant $LQ{\bar D}$ operator the event topologies are mainly
characterised by  large hadronic activity, possibly with some leptons and/or
missing energy. In the simplest case the topology consists  of four jet final
states, and in the more complicated scenario of multi-jet and 
multi-lepton and/or multi-neutrino states. 

In the following sections the selections of the various topologies are 
described in turn.
A brief summary of all selections, the
expected number of background events from SM processes, and the number of
candidates selected in the data is shown in Table~\ref{tops}.
\begin{table}
\centering
\begin{tabular}{|l|l|c|c|}
\hline
Selection & Signal Process & Background   & Data \\
\hline
Multi-jets plus Leptons & 
  $\chi^+\chi^-\to \mathrm{qqqq}\chi\chi$ & 2.1 & 3 \\
&  $\chi^+\chi^-\to \mathrm{l\nu qq}\chi\chi$ & &  \\
&  $\chi^+\chi^-\to \mathrm{l\nu l\nu}\chi\chi$ & &  \\ \hline
Four Jets  & $\snu\snu \to \mathrm{qqqq}$   & 44.0 & 42 \\
           & $\slep\slep \to \mathrm{qqqq}$ &      &    \\ \hline
\multicolumn{4}{|l|}{Two Jets (plus Leptons) Selections} \\ \hline
2J+2$\tau$   & $\sq\sq \to \tau \mathrm{q} \tau \mathrm{q}$ & 1.1 & 2 \\
2J+$\tau \nu$& $\sq\sq \to \nu \mathrm{q} \tau \mathrm{q}$  & 2.0 & 0 \\ 
AJ-H          & $\sq\sq \to \nu \mathrm{q} \nu \mathrm{q}$   & 1.7 & 0 \\ \hline
\multicolumn{4}{|l|}{Direct Chargino / Neutralino Decay Selections} \\ \hline
4J(2L)       & $\chi\chi \to \lqq\lqq$        & 1.5 & 1 \\
4J(2$\tau$)  & $\chi\chi \to \tau \mathrm{qq} \tau \mathrm{qq}$ & 1.3 & 1 \\
2L2J(2J)     & $\chi\chi \to \lqq\lqq$        & 1.0 & 0 \\
2$\tau$2J(2J)& $\chi\chi \to \tau \mathrm{qq} \tau \mathrm{qq}$ & 1.6 & 2 \\
4J(2$\nu$)   & $\chi\chi \to \nqq\nqq$        & 2.1 & 3 \\
4J(L$\nu$)     & $\chi\chi \to \lqq\nqq$        & 1.6 & 1 \\
4J2L-low& $\chi\chi \to \lqq \lqq$ & 1.6 & 1 \\ 
4JL$\nu$-low& $\chi\chi \to \lqq \nqq$ & 2.4 & 2 \\ 
4J2$\tau$-low& $\chi\chi \to \tau \mathrm{qq} \tau \mathrm{qq}$ & 2.0 & 2 \\ 
4J2$\nu$-low & $\chi\chi \to \nqq\nqq$        & 2.4 & 2 \\
\hline
\end{tabular}
\caption[.]{\em The selections, the signal processes giving rise to the above
topologies, the number of background events expected, and the number of
candidate events selected in the data ($\sqrt{s}=161-172\gev$).}
\label{tops}
\end{table}

The positions of the most important cuts of all selections have been
chosen such that the expected cross section upper limit 
 (~\nbar)  without the presence of a signal is minimised \cite{n95}.  
This minimum was determined using the Monte Carlo
for background and signal, focussing on signal masses close to the
high end of the sensitivity region.

In some cases high signal efficiencies
are achieved using some of the selections designed to search for supersymmetry
when R-parity is conserved \cite{rpc}. The selections used for this purpose
were  4J-VH, 4J-H and  4J-L to select four-jet final states with small, moderate
and large amounts of missing energy, respectively,  4J-$\gamma$ to select four-jet final states with an
isolated photon and missing energy,  and  
AJ-H  to select acoplanar jet events with a moderate amount of missing energy. 
The reader is referred to \cite{rpc} for further details.

\subsection{Multi-jets plus Leptons}

This topology is expected from the  {\em indirect} decays of charginos to neutralinos,
e.g.  \mbox{$\chi^+ \ra \w^* \chi \ra \w^* \mathrm{l q q}$} or $\chi^+ \ra \w^* \nu
\mathrm{qq}$, and the {\em indirect} decays of squarks, e.g. \mbox{${\sq} \ra \q \chi
\ra \mathrm{qlqq}$}. Depending on
the $\w^*$ phase space and decay mode, the topology may resemble a purely 
hadronic final state, a leptonic final state with
some hadronic activity acompanied by possibly some missing energy, or a mix thereof. Therefore, three 
subselections have been designed to select events with differing amounts
of leptonic and hadronic activity (Table \ref{hlcuts}).
Subselection I is designed to select final states based on the hadronic activity,
eg. $\chi^+\chi^- \ra \mathrm{qqqq} + \chi\chi$. Since large hadronic activity is a feature
of most of the signals of interest this selection is reasonably efficient in
most cases. Subselection II is designed for decays such as 
$\chi^+\chi^- \ra \lepton\nu \mathrm{qq} + \chi\chi$
where the leptonic energy is more important and subselection III is designed 
to select the decays $\chi^+\chi^- \ra \lepton\nu \lepton\nu + \chi\chi$.  
 
For all three
subselections there is a common preselection, requiring a number of charged tracks
$N_{\rm ch}\ge 10$, a visible mass
$M_{\rm vis}>45\gevcc$, and the polar angle of the missing momentum vector
$\theta_{\rm miss}>30^\circ$. To ensure equal treatment of charged leptons 
and neutrinos a number of physical quantities  are calculated
excluding identified electrons or muons. In Table \ref{hlcuts} such quantities are denoted
by primed event variables. The $\qq{}$ background is reduced by selecting spherical 
events using the event thrust, $T$, and the minimum Durham scale $y_{\rm i}$ between
all jets when the event is clustered to $i$ jets. 

\begin{figure}
\begin{center}
\makebox[\textwidth]{
\epsfig{figure=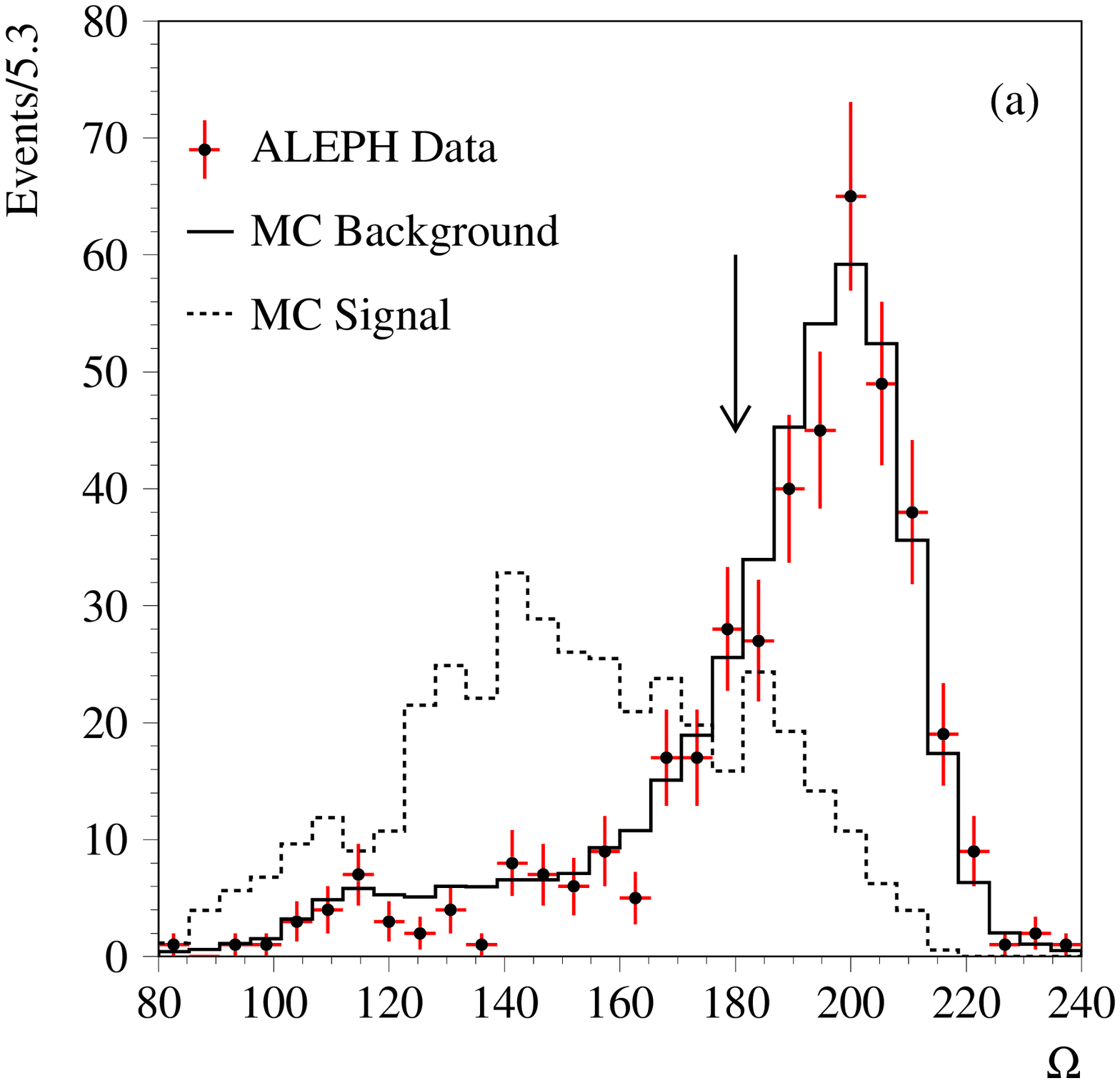,width=0.5\textwidth}\hfill
\epsfig{figure=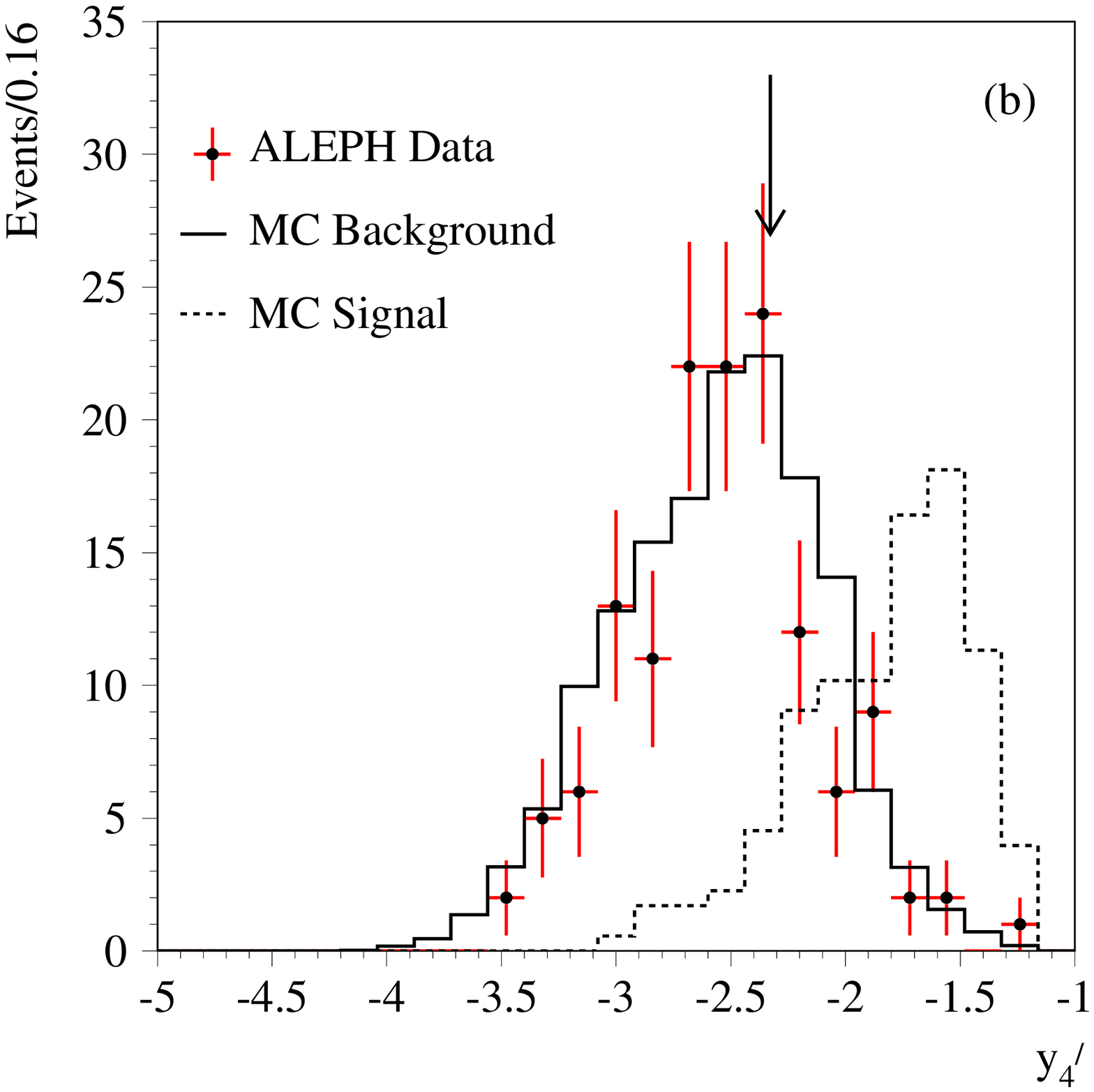,width=0.5\textwidth}}
\makebox[\textwidth]{
\epsfig{figure=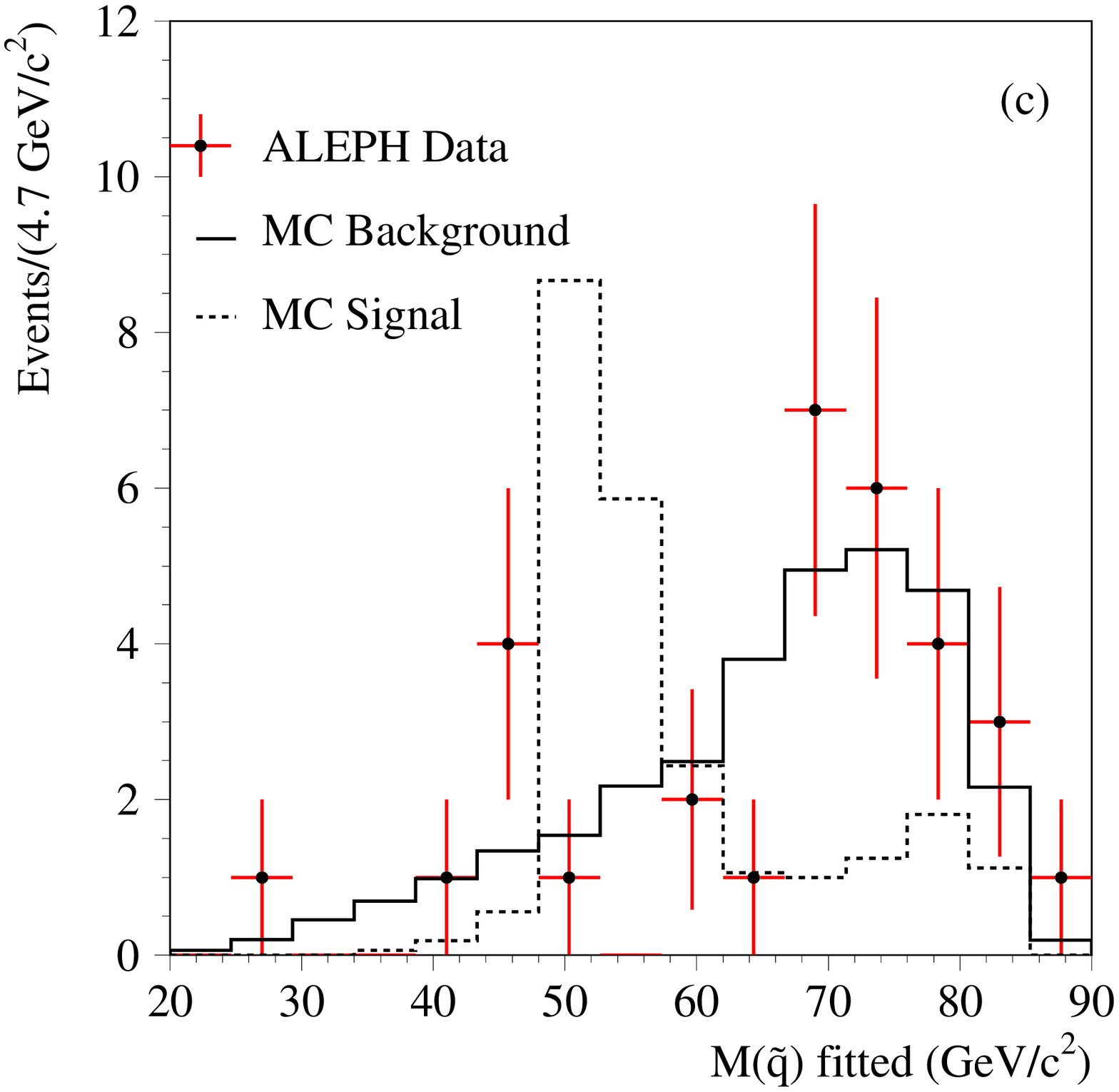,width=0.5\textwidth}\hfill
\epsfig{figure=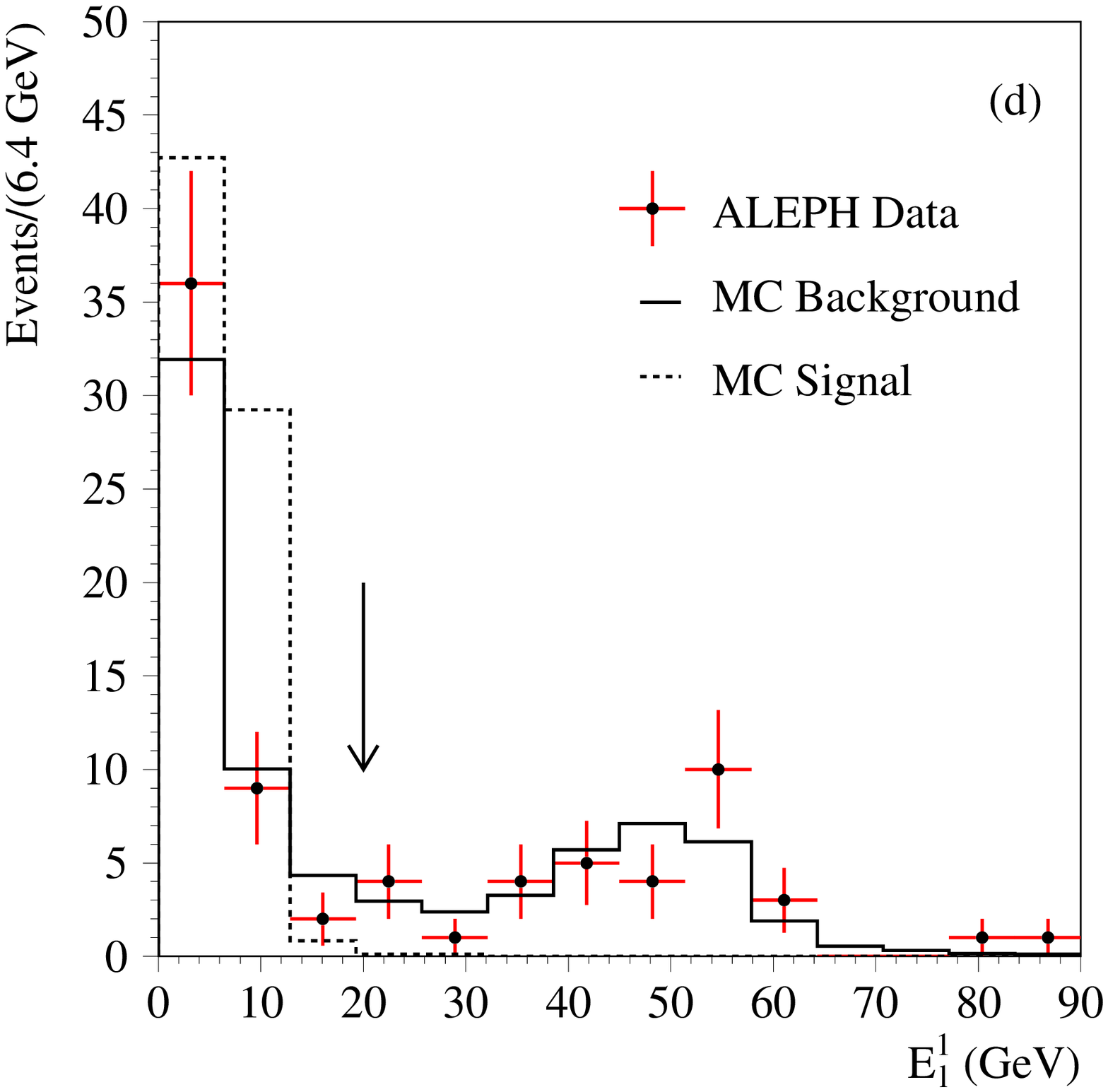,width=0.5\textwidth}}
\caption{\label{datamc}\em The distributions of a)  
$\Omega=\Phi_{\rm{aco}}'+0.7(\mvis'-120)$ as used in subselection I of the 
``Multi-jets plus Leptons''
selection, b) the $y'_4$ variable as used
in subselection II and III of the ``Multi-jets plus Leptons'' selection, 
 c) the reconstructed invariant mass $M_{\sq}$ as used in the
``$2\J+2\tau$'' selection, and d) the  energy of the leading lepton $E^1_{\rm l}$ as
used in the {\em direct} Chargino/Neutralino selections. 
 The data (dots) at $\sqrt{s}=$161--172~GeV are compared to the
 background Monte Carlo (full histograms).
 The dashed histograms show typical signal distributions for $\lambda'_{3jk}$
 in arbitrary normalisation: a) and b) $\chi^+\chi^-\to \w^* \w^* \chi\chi$, 
 c) ${\stq} {\stq} \to \tau \q \tau \q$ with $M_{\stq}=50\gevcc$ and
d) $\chi^+\chi^-\to \mathrm{qq} \tau \mathrm{qq}
 \tau $. In all cases only a subset of the cuts was applied to preserve 
sufficient statistics. Arrows indicate the cut positions.}
\end{center}
\end{figure}

Subselection I reduces the background from hadronic events with initial state
photons seen in the detector by requiring that the electromagnetic energy in
any jet, $E^{\rm{em}}_{{\rm jet}}$, be less than $90\%$ of the jet energy $E_{\rm{jet}}$. 
High transverse energy, $E_{\rm T}$, is required and the isolation of the missing
momentum vector is ensured by removing events with large deposits of energy,
$E^{\rm{iso}}_{10}$, within a $10^\circ$ cone. Finally a two-dimensional 
cut is applied in the $(M'_{\rm vis}, \Phi'_{\rm{aco}})$ plane, where
$\Phi'_{\rm{aco}}$ is the acoplanarity angle of the hadronic
system. Fig.~\ref{datamc}a shows the distribution of a one-dimensional
projection of this variable for data, background Monte Carlo and signal events
 at an intermediate stage of the selection.

Subselections II and III are designed for cases where the neutral hadronic 
energy, $E_{\rm{had}}$, is not too large compared to the leptonic energy, $E_{\rm{lep}}$.
Subselection II further requires a high $y_4'$ value, an
 acoplanar hadronic system or a high value
of $y_6'$ (to reduce the $\qq{}$ background) and $E_{\rm{lep}}<40\gev$ (to reduce the
WW background).  The discriminating power of the Durham scale  $y_4'$ is illustrated in Fig.~\ref{datamc}b.
As in \cite{LLEpaper} the suppression of the $\w^+\w^-\to \lepton\nu \q\bar{\q}$ background
is aided by the definition of the following quantity
\begin{equation}
\chi^2_{\w\w} = \left( \frac{M_{\q\q}-M_\w}{10\gevcc} \right)^2 + 
\left( \frac{M_{\lepton\nu}-M_\w}{10\gevcc} \right)^2 +
\left( \frac{p_\lepton-43\gevc}{\Delta p_\lepton} \right)^2 \label{eqn:chiww}
\label{antiww}
\end{equation}
Here $M_{\q\q}$ is the hadronic mass, i.e.\ the mass of the event after
removing the leading
lepton, $M_{\lepton\nu}$ is the mass of the leading lepton and the missing momentum,
and $p_\lepton$ is the momentum of the leading lepton. The spread~$\Delta p_\lepton$ of
lepton momenta from WW is approximated by 5\gevc{} at $\sqrt{s}=161$~GeV
and 5.8\gevc{} at $\sqrt{s}=172$~GeV.

The total expected  background for the inclusive combination of all three 
subselections is 2.1 events, dominated by WW and $\q{\bar \q}(\gamma)$ processes. 

\begin{table}[t]
\begin{center}
\begin{tabular}{|c|c|c|} \hline
 subselection~I & subselection~II & subselection~III\\
\hline \hline
\multicolumn{3}{|c|}{$\nch\ \geq 10$}\\
\multicolumn{3}{|c|}{$\mvis\ > 45\gevcc$}\\
\multicolumn{3}{|c|}{$\Theta_{\rm{miss}} > 30^\circ$}\\
\hline
$\mvis'>43\%\sqrt{s}$ & $\mvis'<50\%\sqrt{s}$ & $\mvis'<65\gevcc$\\
$T < 0.9$ & $T <0.74$ & $T<0.8$\\
 & $y_4' > 0.0047$ & $y_4'>0.001$\\
$y_5>0.003$ & & \\
$y_6>0.002$ &  & $y_6>0.00035$\\
$E_{\rm T}>60\gev$ & \raisebox{2.5ex}[-2.5ex]
{$\left (\begin{array}{c}
\Phi_{\rm{aco}}'<145^\circ\\
\mathrm{or} \\
y_6 > 0.002
\end{array} \right )$}&\\
\hline
$E^{\rm{em}}_{\rm{jet}} < 90\%E_{\rm{jet}}$ & $\elep\ < 40\gev$ & \\
$E^{\rm iso}_{10} < 5\gev$  & $\ehad\ < 2.5 \elep$ & $\ehad\ < 47\%\elep$\\
\hline
 &\multicolumn{2}{|c|}
{$\chi_{\mathrm{WW}}>~$3.3 (for$\sqrt{s}=161\gev)$}\\
\raisebox{1.5ex}[-1.5ex]{$\Phi_{\rm{aco}}'+0.7(\mvis'-120)<180$} &
\multicolumn{2}{|c|}{$\chi_{\mathrm{WW}}>$~3.5 (for$\sqrt{s}=172\gev)$}\\
\hline
\end{tabular}
\caption{\label{hlcuts}\em The  list of cuts as defined for
the ``Multi-jets plus Leptons'' selection.}
\end{center}
\end{table}

To efficiently select {\em indirect} squark topologies subselection I was 
reoptimised for the data from 130 to $172\gev$. 
The cuts on $y_5$ and $y_6$ were tightened to 0.0044 and 0.0025 
respectively. At centre-of-mass energies from 130 to 136$\gev$ the two dimensional 
cut in the plane $(\Phi'_{\rm{aco}},M'_{\rm{vis}})$ was altered to 
$\Phi'_{\rm{aco}} + 0.8(M'_{\rm{vis}}-105)<180$. The expected background 
in the data from 130 to 172$\gev$ is 1.1 events.

\subsection{Four Jets}\label{fourj.selection}

Pairs of left-handed sleptons and sneutrinos can decay {\em directly} into four-jet final 
states with the property that the invariant di-jet masses are equal: $M_{\rm inv}(q_1,q_2) =
M_{\rm inv}(q_3,q_4)$. 
To select this final state the analysis which was originally developed
for the search for pair production of 
charged Higgs bosons decaying into four jets in \cite{charged.higgs} is used.

After requiring at least 8 good  tracks and a total charged energy
of more than $10\%\sqrt{s}$, events from $\q\bar{\q}(\gamma)$ are rejected
by a two-dimensional cut in the ($p^{\rm z}_{\rm{miss}}$, $M_{\rm{vis}}$) plane,
where $p^{\rm z}_{\rm{miss}}$
is the missing momentum along the beam pipe. Spherical events
with thrust less than 0.9 are then clustered into four jets and kept if
$y_{4}>0.006$.
After vetoing events with photon-like jets,
events that match the equal di-jet mass hypothesis are selected by
cutting on the mass difference of the di-jet systems, and by performing
a 5C-fit (energy-momentum conservation and equal mass constraint) that
is required to lead to a small $\chi^2$.
A total background of 46.8 events is expected at
$\sqrt{s}=130-172\gev$. 

\subsection{Two Jets (plus Leptons)}\label{twoj.selection}

Squarks can decay {\em directly} into a quark plus a lepton or neutrino. The 
resulting topologies are acoplanar jets, two jets and a lepton or 
two jets and two leptons. Because limits on leptoquarks from the Tevatron \cite{TeVLQ}
effectively exclude the possibility of seeing decays to $\e\q$ or $\mu \q$
at LEP only the decays $\sq \ra \tau \q$ and $\sq \ra \nu \q$ are considered.

When both squarks decay to $\nu \q$ the topology is that of acoplanar jets
and the selection AJ-H described in \cite{rpc} is used to select this final state.  
For final states involving taus a tau identification procedure similar
to the one described in 
\cite{mssm.higgs} is applied.  
Only one tau is allowed to be a three prong decay; the other
must be a one prong. The four vectors of the jets and the two taus (or one tau
and the missing momentum vector) are used to perform a constrained fit under
the assumption of equal masses, and a cut is applied on the quality of the fit.
The obtained mass distributions 
 for data, background Monte Carlo and signal events (${\stq} {\stq} \to
\tau \q \tau \q$)  are shown in Fig.~\ref{datamc}c at an intermediate stage of the
 selection.

The selection for the $\tau\tau \mathrm{qq}$ final state includes cuts on 
$p^\tau_{\rm T}$, the transverse momenta of the taus, $\phi^\tau_{\rm isol}$,
the isolation angles from the nearest charged track, and $M_\tau$, the
tau mass. Additional quality requirements are placed on the tau 
candidates using the ratio of the particle momenta parallel to the tau
direction to the total momentum of all energy flow objects in the tau,
 $\tau_{\parallel}$, and the energy in a cone at angles between $18^\circ$
and $32^\circ$ around the tau direction, $E_{\rm iso}$. 
The selection for the $\tau\nu \mathrm{qq}$ final state also includes cuts on the
isolation angles and on the transverse acoplanarity of the jets
$\Phi_{\rm acopT}^{\rm jets}$. The WW background is reduced by vetoing events
using $\chi_{\mathrm{WW}}$ as defined in Eq.(\ref{eqn:chiww}). To remove 
background from $W^+W^- \to \tau\nu \q\bar{\q}$ the quantity
\begin{equation}
\chi'_{\w\w} = \sqrt{\left( \frac{M_{\q\q}-M_\w}{9\gevcc} \right)^2 + 
\left( \frac{M_{\tau\nu}-M_\w}{7\gevcc} \right)^2 },
\end{equation}
where $M_{\q\q}$ is the di-jet mass and $M_{\tau\nu}$ is the recoil mass of
the di-jet system, is used to construct a second WW veto for the 2J+$\tau \nu$ selection.
Table
\ref{2jplusleptons} lists the complete set of cuts for the 2J+2$\tau$ and the 
2J+$\tau \nu$ selections. 

\begin{table}
\centering
\begin{tabular}{|c|c|} \hline
      2J+2$\tau$          & 2J+$\tau\nu$              \\ \hline
\multicolumn{2}{|c|}{$N_{\rm ch} \geq 9$}                 \\
\multicolumn{2}{|c|}{$E_{\rm ch} > 20\% \sqrt{s}$}        \\
\multicolumn{2}{|c|}{$E^{\rm em}_{\rm jet} < 95\% E_{\rm jet}$}   \\ \hline
$\Phi_{\rm miss} > 30^\circ$  & $\Phi_{\rm miss} > 24^\circ$  \\
$M_{\rm vis} < 90\% \sqrt{s}$ &                           \\ 
$p_{\rm T} > 3.5\%\sqrt{s}$     & $p_{\rm T} > 16 \gevc$           \\
$T<0.94$                  &                           \\
$y_4 < 0.002$             &                           \\ \hline
$\Sigma p^\tau_{\rm T}>37\gev$           & \\
$\Sigma \phi^\tau_{\rm isol}>45^\circ$ & $\phi^\tau_{\rm isol}+\phi^\nu_{\rm isol}>45^\circ$ \\
$\phi^\tau_{\rm isol} >15^\circ$       & $\phi^\tau_{\rm isol} >28^\circ$ \\
                                   & $\phi^\nu_{\rm isol} >12^\circ$ \\
$M_\tau < 2.4\gevcc$               & \\
$\tau_{\parallel}>0.99$            & \\
$\tau=e,\mu$ OR $E_{\rm iso}<2\gev$    & \\
                                   & $\Phi_{\rm acopT}^{\rm jets}<167^\circ$ \\ \hline
$\chi_{\mathrm{WW}}>$~3.0 ($\sqrt{s}=161\gev)$ & 
$\chi_{\mathrm{WW}}>$~4.8 ($\sqrt{s}=161\gev)$ \\  
$\chi_{\mathrm{WW}}>$~3.3 ($\sqrt{s}=172\gev)$ & 
$\chi_{\mathrm{WW}}>$~5.8 ($\sqrt{s}=172\gev)$ \\ \hline
& $\chi'_{\mathrm{WW}}>$~1.7 ($\sqrt{s}=161\gev)$ \\
& $\chi'_{\mathrm{WW}}>$~4.6 ($\sqrt{s}=172\gev)$ \\ \hline
\end{tabular}
\caption {\em The list of cuts for the 2\J+2$\tau$ and 2\J+$\tau\nu$ selections.}
\label{2jplusleptons}
\end{table}

\subsection{Chargino/Neutralino Direct Decay Selections}

The {\em direct} decay modes of charginos and neutralinos are listed in
Table \ref{rpv.decays}. The kinematics of these decays strongly depends 
on the sfermion mass spectrum. If a sfermion
is nearly degenerate in mass with the chargino or neutralino some of the 
decay products may be very soft. For this reason a large number of
different selections are required to cover all possible cases in terms
of final state particles and event distributions. In the following the
selections are described in turn. Brackets around the final states of a
selection denote soft particles: e.g. the 4J(2L) selection is designed for four
jet final states with two soft leptons. Table \ref{direct.decay.selections}
lists the complete set of cuts for all the selections. Topologies with
electrons or muons  are selected by the 4J(2L), 2L2J(2J),
4J(L$\nu$) selections with typical efficiencies of 40-60$\%$. Topologies 
with moderate or large missing energy are selected with a similar performance by the 
4J-VH selection of \cite{rpc} or the 4J(2$\nu$) selection. Tau final states are
selected by the 4J(2$\tau$), 2$\tau$2J(2J) and the 4J(2$\nu$) selections with 
typical efficiencies of 15-30$\%$.

\begin{table}
\centering
\begin{tabular}{|c|c|} \hline
      4J(2L)          & 4J(2$\tau$)              \\ \hline
$N_{\rm ch} > 8$, $E_{\rm ch} > 20\% \sqrt{s}$          &  $N_{\rm ch} > 8$, $E_{\rm ch} > 15(39)\% \sqrt{s} $ \\
$M_{\rm vis} > 85\% \sqrt{s}$         &   $68(73)\% \sqrt{s}  < M_{\rm vis} < 98(97)\% \sqrt{s}$\\
$\ge 2$ identified leptons with $\phi_{\rm l}>7^\circ$ & $\ge 1$ identified lepton $\phi_{\rm l}>12(15)^\circ$ \\
$(E_{\rm l}^1+E_{\rm l}^2) < 5 E_{\rm had}$  & $E_{\rm l}^1<20\gev$ \\
$y_4 > 0.01 , y_6 > 0.0008 $&     $y_4>0.012 (0.0051), y_6>0.0014(0.00085) $ \\
&                       $|cos \theta_{\rm miss}| < 0.93$  \\ \hline
2L2J(2J) & 2$\tau$2J(2J) \\ \hline
$N_{\rm ch} > 8$, $E_{\rm ch} > 20\%\sqrt{s}$          & $N_{\rm ch} > 8$, $E_{\rm ch} >
30\%\sqrt{s}$, $M_{\rm vis}<95\%\sqrt{s}$ \\
$\ge 2$ identified leptons with $\phi_{\rm l}>7^\circ$ &  
$\ge 1$ identified lepton with $\phi_{\rm l}>15^\circ$ \\
$E_{\rm l}^1 > 10\gev$ , $E_{\rm l}^2 > 40\% E_{\rm l}^1$  & $E_{\rm l}^1<32 \gev$  \\
$M_{\rm ll} > 20\gevcc$ &  $M_{\rm{l\nu}} <76(73) \gevcc $  \\
$20\gev < E_{\rm miss} + E_{\rm l}^1 < 70\gev$ & $25(35) < E_{\rm miss} +
E_{\rm lep}$ \\
$y_4 > 0.003$ & $y_4>0.012(0.0039) , y_5 > 0.0005(0.0003) $  \\
& $|p_{\rm Z}|<32\gevc, |cos \theta_{\rm miss}| < 0.93$ \\ \hline
\multicolumn{2}{|c|}{4J(2$\nu$) }\\ \hline
\multicolumn{2}{|c|}{$N_{\rm ch} > 23(25) ,   55\% \sqrt{s} < M_{\rm vis} < 93(94)\% \sqrt{s} , M_{\rm miss}<60(70)\gevcc $ }\\
\multicolumn{2}{|c|}{$\Phi_{\rm acop}<175(177)^\circ$ , $p_{\rm T}>12(7)\gevc$ }\\
\multicolumn{2}{|c|}{$M_{\rm W}<90\gevcc$ , $E_{\rm W}^{30}<7(8)\%\sqrt{s}$ }\\
\multicolumn{2}{|c|}{$0.55 (0.56) < InvB$ OR $(InvB \times p_{\rm T})> 6\gevc$}\\

\multicolumn{2}{|c|}{$E_{\rm l}^1<10(15)\gev$ }\\ 
\hline
{4J(L$\nu$) } & {4JL$\nu$-low }\\ \hline
{$N_{\rm ch} > 8$, $E_{\rm ch} > 20\% \sqrt{s} $, $M_{\rm vis}>85\%\sqrt{s}$} & 
 $N_{\rm ch} > 8$,  $55\%<\sqrt{s}<M_{\rm vis}<90\%\sqrt{s}$ \\
{$\ge 1$ identified lepton, $\phi_{\rm l}>20^\circ$ } & 
{$\ge 1$ identified lepton, $\phi_{\rm l}>5^\circ$ } \\
{$(E_{\rm l}^1+E_{\rm l}^2) < 5 E_{\rm had} $} & 
$E_{\rm l}^1>20\gev, E_{\rm miss} > 20\% E_{\rm l}^1$\\
{$y_4>0.01, y_6>0.0008$ }& 
$T > -187 y_5 + 0.93 $\\
& $E^{\rm iso}_{10}< 5\gev$, $p_{\rm T}>8\gevc$, $\Phi_{\rm acop}>155^\circ$\\ 
\hline
4J2L-low & 4J2$\tau$-low  \\ \hline
$N_{\rm ch} >8$, $75\% \sqrt{s} < M_{\rm vis}$ &  {$59\% \sqrt{s} < M_{\rm vis} < 87\% \sqrt{s}$ }\\
& {$|p_{\rm Z}|<16(17) \gevc$}\\
& {$5.0 (5.1) < p_{\rm T} <28(26)\gevc$ }\\
$\ge 1$ identified lepton, $\phi_{\rm l}>5^\circ$  & {$\ge 1$ identified lepton, $\phi_{\rm l}>14^\circ$ }\\
$M_{\rm qq}>110\gevcc, M_{\rm l\nu}<65\gevcc$ &  {$N_{\rm ch}^{\rm 6jet}\ge 1$ }\\
{$E_{\rm l}^1>20 \gev$, $M_{\rm ll}>30\gevcc$}  &  {$E_{\rm l}^1<38(37) \gev$}\\
&  { $T > -34.3(-35.8) y_4 + 1.06 (1.01)$ }\\
\hline
\multicolumn{2}{|c|}{4J2$\nu$-low  }\\
\hline
\multicolumn{2}{|c|}{$p_{\rm T}>11(12)\gevc , M_{\rm miss}>39(43)\gevcc$  }\\
\multicolumn{2}{|c|}{$M_{\rm vis}>35(41)\gevcc$ , $|p_{\rm Z}| < 23(24) \gevc $ }\\ 
\multicolumn{2}{|c|}{$InvB>0.1$, $T<0.71$, $\alpha_{23}>146(149)^\circ$ }\\
\multicolumn{2}{|c|}{$N_{\rm ch}^{\rm 4jet}>1 $}\\
\multicolumn{2}{|c|}{$T > -32.3(-33.3) y_4 + 1.02 (1.01)$ }\\
\hline
\end{tabular}
\caption{\label{direct.decay.selections}\em The list of cuts for the {\em direct} chargino/neutralino
selections at $\sqrt{s}=161\gev$. Numbers in brackets indicate the cut values of the selections 
at $\sqrt{s}=172\gev$ if different.}
\end{table}

\begin{itemize}
\item{\bf 4J(2L)}: This selection is designed for events with at least two 
(soft)  electrons or muons and four jets. For  the 
 preselection a minimum of nine charged tracks are required with a total charged 
 energy of more than $20\%\sqrt{s}$ and a high visible mass $M_{\rm vis}$.
 There must be two identified isolated leptons with a minimum separation 
 angle to the closest charged track ($\phi_{\rm l}$) of $7^\circ$. To reject the
four fermion background the sum of the energy of the two highest energetic leptons
($E_{\rm l}^1+E_{\rm l}^2$) is required to be less than five times the neutral hadronic
energy. Finally the $\qq{}$ background is reduced by requiring large  $y_4, y_6$ values.
  
\item{\bf 4J(2$\tau$)}: The selection for four jets plus two soft taus consists
of a preselection on $N_{\rm ch}, E_{\rm ch}$ and $M_{\rm vis}$. Taus are tagged through
their leptonic decays by demanding at least one well isolated identified
lepton. Background from WW is reduced by requiring that the leading
lepton has an energy less than $20\gev$ (see Fig.~\ref{datamc}d).
 The WW plus the $\qq{}$ 
backgrounds are further reduced by requiring large $y_4, y_6$ values 
and that the missing momentum vector does not point along 
the beam axis.

\item{\bf 2L2J(2J)}: After preselection requirements on $N_{\rm ch}$ and $E_{\rm ch}$, 
 two electron or muon final states with at least two jets are selected by
 requiring two or more identified energetic isolated leptons. $\PZ\gamma^*$
 and $\PZ\Pe\Pe$ backgrounds are suppressed by demanding that the invariant mass
 of the two most energetic leptons $M_{\lepton\lepton}$ be greater than $20\gevcc$. 
 The $\qq{}$ and WW backgrounds are reduced by requiring that the missing
 energy ($E_{\rm miss}$) plus the energy of the leading lepton be in the
 range $20-70\gev$, and that $y_4 > 0.003$.

\item{\bf 2$\tau$2J(2J)}: After a preselection, at least one isolated lepton
with $E_{\rm l}^1<32\gev$ is required. To reject the $WW \ra \mathrm{qql}\nu$ background the
invariant mass of the leading lepton and the missing momentum, $M_{\lepton\nu}$, is
required to be below the W-boson mass. To reject background from hadronic WW
decays the sum ($E_{\rm miss}+E_{\rm lep}$) must be large, and the $\qq{}$ background is
rejected by cuts on the jet-finding variables $y_4,y_5$, and demanding that the
missing momentum vector does not point along the beam axis. 

\item{\bf 4J(L$\nu$)}: The $4q\tau\nu$ topologies are efficiently 
selected by an inclusive combination of the 4J(2$\tau$) and the 4J(2$\nu$)
selections.  For the $4 \q \lepton \nu$ (l=$\e, \mu$) topologies the inclusive
combination of the 4J(2L) and the 4J(2$\nu$) selections gives poor performance,
and therefore a separate selection for this final state was designed. The  4J(L$\nu$)
selection is identical to the 4J(2L), except that only one identified lepton is
required with an isolation angle $\phi_{\rm l} > 20^\circ$.

\item{\bf 4J(2$\nu$)}: The selection is based on the 4J-VH selection of \cite{rpc},
but is optimised to select four jet final states with a {\it moderate} amount of
missing energy. After a preselection, acoplanar events with a momentum
imbalance in the transverse plane are selected. Events with a high energetic
lepton are vetoed. To reject the $WW \ra \mathrm{qq} \tau \nu$ background, the invariant
hadronic mass $M_{\rm W}$ excluding the tau jet is calculated and required to be
$M_{\mathrm{had}}<90\gevcc$. The energy in a $30^\circ$ azimuthal wedge around the direction
of the missing momentum ($E_{\rm W}^{30}$) must be small. And finally the $\qq{}$ background 
is vetoed by requiring either a large inverse boost $InvB$ (where $InvB = (\sqrt{\frac{1}{2}
(\gamma_1^{-2} + \gamma_2^{-2})})$ and  $\gamma_i = E_i/m_i$ for each hemisphere of
the event), or by requiring that the product ($InvB \times p_{\rm T}$) exceeds $6
\gevc$.

\item{\bf 4J2L-low}: The selection is designed for small gaugino masses ($M_\chi
\lsim\ 50\gevcc$), where the gaugino decay products may be heavily boosted.
  Events with a large visible mass $M_{\rm vis}$ are required to have at least 
 one high energetic lepton ($E_{\rm l}^1>20\gev$). The four-fermion  backgrounds 
 are reduced using cuts on the WW-rejection variables $M_{\q\q}, M_{\lepton\nu}$ as
 used in Eq.(\ref{antiww}), and on $M_{\lepton\lepton}$. 
 
\item{\bf 4J2$\tau$-low}:  After a preselection on $M_{\rm vis}$, $|p_{\rm Z}|$ and $p_{\rm T}$, at least
 one well isolated lepton (electron or muon) must be identified, with an energy
 below approximately half the W mass. The event is then clustered  into six jets
 using the Durham algorithm, and the charged multiplicity of any of the six jets  
 ($N_{\rm ch}^{\rm 6jet}$) is required to be $N_{\rm ch}^{\rm 6jet}\ge 1$.  Finally, a
 two dimensional cut in the plane of thrust and $y_4$ is applied.

\item{\bf 4J2$\nu$-low}: The selection employs cuts on the event  shape
variables $p_{\rm T}$, $M_{\rm miss}$, $M_{\rm vis}$, $|p_{\rm Z}|$, $InvB$ and $T$. The $WW \ra \mathrm{qq} \tau\nu$
background is rejected by requiring that the smallest angle between the tau jet
and the other jets ($\alpha_{23}$) is large. After clustering the event into
four jets using the Durham algorithm, the charged multiplicity of any of the
four jets   ($N_{\rm ch}^{\rm 4jet}$) is required to be $N_{\rm ch}^{\rm 4jet}> 1$. Finally, a
 two dimensional cut in the plane of thrust and $y_4$ is applied.

\item{\bf 4JL$\nu$-low}: Events with an energetic lepton are required to have a
missing energy of at  least $E_{\rm miss}>20\%E_{\rm l}^1$. The $\qq{}$ and WW backgrounds are
reduced by requirements on $E^{\rm iso}_{10}, p_{\rm T}$ and $\Phi_{\rm acop}$, and a two
dimensional cut on the thrust $T$ and $y_5$. Finally the WW-veto of Eq.(\ref{antiww})
is applied.   

\end{itemize}

\section{\label{results}Results}
As can be seen from Table \ref{tops} no excess of events was observed
in the data recorded at $\sqrt{s}=$161--172~GeV, corresponding to an
integrated luminosity of 21.7~\invpb. 
Of the events selected by the ``Multi-jets
plus Leptons'' selection, one is consistent with being a $\q\bar{\q}\gamma$, one with 
WW and one with ZZ. Both of the 2J+2$\tau$ candidates are consistent with $\q\bar{\q}$.
The thirteen candidates selected by the direct chargino/neutralino selections 
are all consistent with either $\q\bar{\q}$ or WW backgrounds. 

Of those selections that are employed at
$\sqrt{s}=$130--136~GeV candidates are only found by the ``Four Jet'' selection and the
reoptimised subselection I from the Multi-jets plus Leptons selection. The latter
selects two events
at LEP~1.5 energies and one at $\sqrt{s}=161\gev$. The two candidates at lower energies are
selected by the analysis published in \cite{4jet}; the other is consistent with
$q\bar{q}\gamma$.

In the following sections, the absence of any significant excess of events in the data
with respect to the Standard Model expectation is used to set limits
on the production of charginos and neutralinos, sleptons, sneutrinos and
squarks. The systematic uncertainty on the efficiencies is of the order of 4--5\%,
dominated by the statistical uncertainty due to limited Monte Carlo statistics,
with small additional contributions from lepton identification and
energy flow reconstruction. It is taken into account by conservatively
reducing the selection efficiency by one standard deviation.  
Background subtraction is only used in the Four Jet selection. In this  
case the expected background is conservatively reduced by $20\%$.

\subsection{\label{chargino.limit}Charginos and Neutralinos}

Charginos and heavier neutralinos can decay either {\em indirectly} via the lightest
neutralino, or {\em directly} via (possibly virtual) sleptons or sneutrinos.
The corresponding branching fractions of the {\em direct} and {\em indirect} decays, as well as
the branching fractions of the {\em direct} decays into different final
states (c.f. Table~\ref{rpv.decays}),  in general depend
on the field content and masses of the charginos and neutralinos, the
sfermion mass spectrum and the Yukawa coupling~$\lambda'$. Furthermore,
because of possible mixing in the third generation sfermion sector, staus,
stops and sbottoms can be substantially lighter than their first or second
generation partners. The effect of light staus 
 is to increase the tau branching ratio in the {\em indirect}  decays (e.g. $\chi^+ \ra
 \tau \nu \chi$) with respect to the other {\em indirect} decay modes,
 whereas light stops and sbottoms increase the hadronic
 branching ratios of the {\em indirect} decays. Light sfermions can
 also affect the BRs of the {\em direct} decay modes
 depending on the generation structure of the R-parity
 violating couplings. 

 To constrain a model with such a large number of unknown parameters,
 limits were set that are independent of the various branching ratios.
 For this purpose, the signal topologies are classified into the two extreme 
 cases of {\em direct} topologies (when both charginos decay {\em directly}) and
 {\em indirect} topologies (when both charginos decay {\em indirectly}). 
 {\em Mixed topologies} are not considered in detail here, but example efficiencies
are listed in Table \ref{effics}.
 Additionally, the branching ratios of the various decays involved in both
 {\em indirect} and {\em direct} decays are varied freely, and the limit is set
 using the most conservative choice.

\begin{table}
\centering
\begin{tabular}{|l|c|c|c|}
\hline
Signal Process & Topology & Masses (\gevcc) & Efficiency (\%) \\
\hline
$\chi^+\chi^-\to\PW^*\PW^*\tau \mathrm{qq} \tau \mathrm{qq}$ & indirect &
  $M_{\chi^+}=80$, $M_{\chi}=30$ &$ 48.4 \pm 1.5$ \\
 & & $M_{\chi^+}=80$, $M_{\chi}=70$  & $23.9 \pm 0.7$ \\

$\chi^+\chi^-\to\PW^*\PW^*\nu \mathrm{qq} \nu \mathrm{qq}$ & indirect &
  $M_{\chi^+}=80$, $M_{\chi}=30$ &  $56.4 \pm 1.7$\\
 & & $M_{\chi^+}=80$, $M_{\chi}=70$  & $52.8 \pm 1.6$ \\

$\chi^+\chi^-\to \tau \mathrm{qq} \PW^* \tau \mathrm{qq} $ & mixed &
  $M_{\chi^+}=80$, $M_{\chi}=30$ & $55.8 \pm 1.7$ \\
$\chi^+\chi^-\to \nu \mathrm{qq} \PW^* \nu \mathrm{qq}  $ & mixed &
  $M_{\chi^+}=80$, $M_{\chi}=30$ & $62.9 \pm 1.9$  \\

$\chi^+\chi^-\to \mathrm{qqqq} (+\tau\tau) $& direct & $M_{\chi^+}=80$, 
$\Delta M=(0,10,20)$ &$ (18.6,20.5,29.0)$ \\
$\chi^+\chi^-\to \mathrm{qqqq} (+\nu \nu) $& direct & $M_{\chi^+}=80$, 
$\Delta M=(0,10,20)$ & $(17.5,24.0,33.6)$ \\
$\chi^+\chi^-\to \mathrm{qq}\tau\tau (+\mathrm{qq}) $& direct & $M_{\chi^+}=80$, 
$\Delta M=(0,10,20)$ & $(36.1,25.2,30.3)$ \\

\hline
$\chi\chi\to \tau \mathrm{qq} \tau \mathrm{qq}  $& direct & $M_{\chi^+}=40$
 & $18.8 \pm 0.5$ \\
$\chi\chi\to \nu \mathrm{qq} \nu \mathrm{qq}  $& direct & $M_{\chi^+}=40$
 & $18.7 \pm 0.5$ \\
\hline

$\se\se\to \e \nu  \mathrm{qq} \e \nu  \mathrm{qq}  $ &  indirect & $M_{\slep}=50$, $M_{\chi}=30$
& $35.0\pm 1.1$ \\
$\se\se\to \e \nu  \mathrm{qq} \mathrm{qq} $  &  mixed           & $M_{\slep}=50$, $M_{\chi}=30$
& $41.8 \pm 1.3$ \\
$\se\se\to \mathrm{qqqq}$&  direct & $M_{\slep}=50$ & $37.0 \pm 1.1$ \\
\hline

$\snu\snu\to \nu \tau \mathrm{qq} \nu \tau \mathrm{qq} $&  indirect & $M_{\snu}=50$, $M_{\chi}=30$ &
$23.6 \pm 0.7$  \\
$\snu\snu\to \nu \tau \mathrm{qq} \mathrm{qq} $&  mixed & $M_{\snu}=50$, $M_{\chi}=30$ &
$11.9 \pm 0.4$  \\
$\snu\snu\to \mathrm{qqqq}$&  direct & $M_{\snu}=50$ & $37.0 \pm 1.1$ \\
\hline

$\sq\sq\to \q \tau \mathrm{qq}  \q \tau \mathrm{qq}  $& indirect & $M_{\sq} = 50$, $M_{\chi}=30$
& $20.2\pm 0.6$ \\
$\sq\sq\to \tau \mathrm{qq} \nu \mathrm{qq}$& mixed & $M_{\sq} = 50$, $M_{\chi}=30$  & $16.4 \pm 0.5$ \\
$\sq\sq\to \tau \q \tau \q $& direct & $M_{\sq} = 50$ & $21.5 \pm 0.6$\\
$\sq\sq\to \nu  \q \tau \q $& direct & $M_{\sq} = 50$ & $19.4 \pm 0.6$\\
$\sq\sq\to \nu  \q \nu  \q $& direct & $M_{\sq} = 50$ & $29.9 \pm 0.9$ \\
\hline

\end{tabular}
\caption[.]{\em Selection efficiencies at $\sqrt{s}=172\gev$ for a
representative set of signal processes,
with a lepton flavour composition in the final state leading
to the smallest efficiencies.}
\label{effics}
\end{table}

 Limits have been evaluated in the framework of the MSSM, where the
 masses of the gauginos can be calculated from the three parameters
 $M_2,\mu$ and $\tan{\beta}$. The cross sections of 
 neutralinos (charginos) receive a positive (negative) contribution due to
 t-channel selectron  (electron-sneutrino) exchange, respectively, and thus 
 depend also on $M_{\slep}$ and $M_{\tilde \nu}$. A common 
 sfermion mass~$m_0$ at the GUT scale was assumed, and the 
 renormalisation group equations  were used to calculate the sfermion
 mass spectrum at the electroweak scale. Substantially lighter mass eigenstates
 of the stau, the sbottom and the stop are obtained by varying the mixing
 between the left-handed and right-handed states. The limit is set for the most 
 conservative mixing. 
  
 In summary, the limits derived in this approach are 
 independent of the various branching ratios of the gauginos, with the exception
 of the branching ratio of the {\em direct} and {\em indirect} decays, where they apply to
 either  $100\%$ {\em direct} or $100\%$ {\em indirect} topologies. The
 limit only depends on the  four parameters $M_2, \mu, \tan{\beta}, m_0$, and is
 independent of  mixing between the third generation sfermions, and the
 generation structure of the R-parity violating coupling $\lambda'_{ijk}$.
 The branching ratios which set the limit may not
 correspond to a physically viable model in certain cases (i.e.\ in specific
 points in parameter space $M_2, \mu, \tan{\beta}, m_0$), and hence the real
 limit within a specific model may be stronger than the conservative
 and more general limit presented in this section.

As discussed in Section~\ref{ex.lims}, the lightest neutralino can have
a decay length of more than 1\cm{} when $M_{\chi}\lsim\ 10\gevcc$ for couplings 
which are not already excluded by low energy constraints.
Since long-lived sparticles are not considered
in this analysis, regions in parameter space with $M_{\chi}< 10\gevcc$
are ignored in the following.
 Limits on the charginos and neutralinos are derived in
 Sections~\ref{dom.ind} and \ref{dom.dir} for the two extreme
 cases of $100\%$ {\em indirect} and $100\%$  {\em direct} topologies, respectively.
Due to the large cross section for pair production of charginos, the data
recorded at $\sqrt{s}=$130--136~GeV do not improve the sensitivity
of the analysis, and therefore have not been included here. 

\subsubsection{Dominance of indirect decays \label{dom.ind}}
In this scenario all charginos and neutralinos are assumed to decay
to the lightest neutralino, which then decays violating R-parity
into two quarks and a lepton or neutrino.
The {\em indirect} topologies generally correspond to the cases where the
sfermions are heavier than the charginos and
the neutralinos. When the sfermions are lighter than the
charginos (or the heavier neutralinos) and heavier than the
lightest neutralino, the {\em indirect} decays will also  
dominate provided that the neutralino couples gaugino-like and/or the
coupling $\lambda'$ is small.

To select {\em indirect} decays of charginos proceeding through exchange of a
virtual W or sfermion, the ``Multi-jets plus Leptons'' selection is used.
 A set of efficiencies for choices of the lepton flavour
 corresponding to the smallest efficiencies is shown in Table~\ref{effics}.
Since the kinematic configuration for two-body decays of charginos into
real sfermions resembles (especially in the limit of small mass difference
between gaugino and sfermion) those expected from the {\em indirect} decays
of the sfermions themselves, the corresponding sfermion selections
are employed in these cases. The combination of selections used for
given mass hierarchies and branching fractions is chosen
according to the \nbar\ prescription.

The most important {\em indirect} decay modes of the second lightest neutralino
are decays via (virtual) Z-exchange as well as radiative decays into
 a photon and the lightest neutralino. For $\chi'\chi$ production, the topologies
arising from the former decay channel are selected by the inclusive
combination of the ``Multi-jets plus Leptons'', 4J-L and 4J-H selections,
whereas for the latter 4J-L is replaced by the dedicated 4J-$\gamma$
selection. Which of these two options is used for a given branching
fraction BR($\chi'\to\chi\gamma$) is decided using the \nbar\ prescription.

\begin{figure}
\begin{center}
{\makebox[0.5\textwidth][c]{
{\epsfig{figure=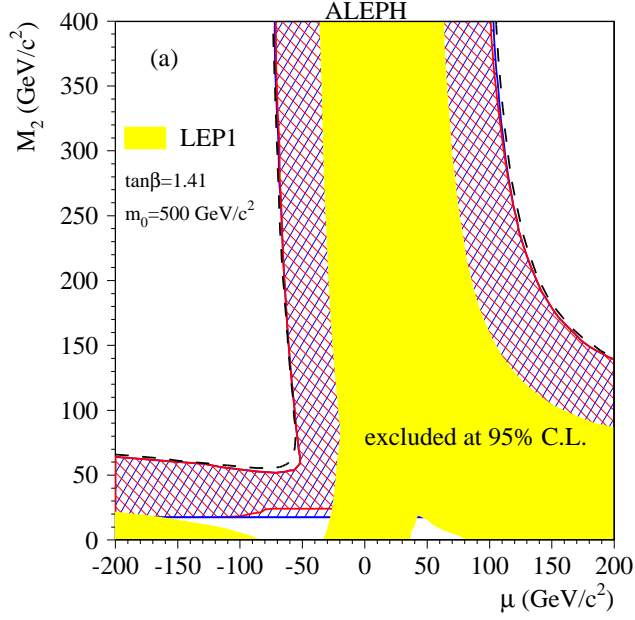,width=0.55\textwidth}\hfill}}}
{\makebox[0.5\textwidth][c]{
{\epsfig{figure=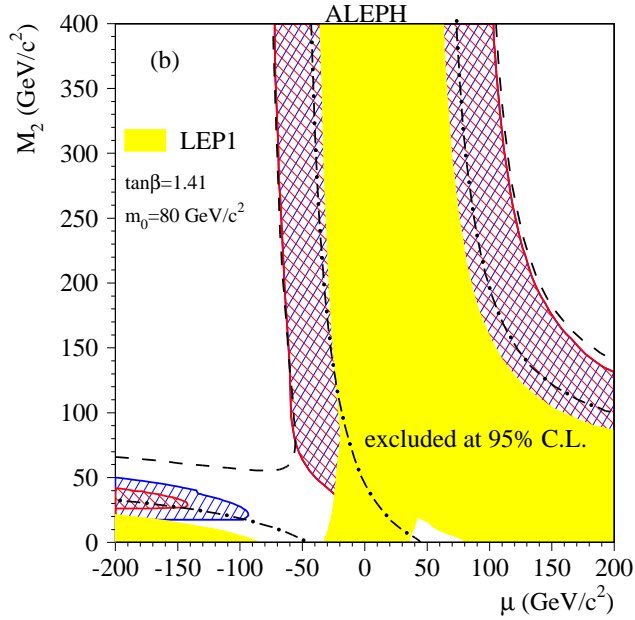,width=0.55\textwidth}\hfill}}}
\caption[.]{\em\label{lqd.indirect}{Regions in the $(\mu,M_2)$ plane excluded
 at $95\%$ C.L. at $\tan\beta=\sqrt{2}$ and a) $m_0 = 500\gevcc$ or
b) $m_0 = 80\gevcc$, assuming that the {\em indirect} decays  dominate (hatched regions).
The white region below $M_2 \lsim\ 15\gevcc$ corresponds to neutralino masses less
than $10\gevcc$, the light shaded region to the LEP~1 limit. 
The superimposed dashed lines show the kinematic limit
$M_{\chi^+}=86$\gevcc. The dash-dotted line shows the 
$M_{\chi^+}=56\gevcc$ contour.}}
\end{center}
\end{figure}

For a given value of $m_0$ and $\tan\beta$, limits are derived in the
$(\mu,M_2)$ plane for the worst case in terms of the branching ratios
of the decays. This corresponds to allowing for all choices of coupling
$\lambda'_{ijk}$, BR($\chi\to \mathrm{lqq}$) and for all third generation
mixing angles. The 
kinematics of the decays and the signal efficiency depend strongly on 
the mass hierarchies of the charginos, neutralinos and sfermions.
The sfermions may be heavier than the charginos and neutralinos yet they
may still be sufficiently light to affect the branching ratios of the chargino
and neutralino decays. In this case the decays remain three body decays
and the kinematics of the events are not affected. The limit corresponding 
to this case is shown as the hatched outer area in Fig.~\ref{lqd.indirect} for
$\tan\beta=\sqrt{2}$ and two values of $m_0$. Allowing for arbitrary mixing
in the third generation, charginos may decay {\em indirectly} via real stops
or staus. For small mass differences between the gaugino
and the sfermion, efficiencies are smaller compared to the 3-body decay
efficiencies and the corresponding worst case limit, shown as the inner hatched 
area in Fig.~\ref{lqd.indirect}, is weaker than in the previous case.

\begin{figure}[t]
\begin{center}
\makebox[\textwidth][l]{
\epsfig{figure=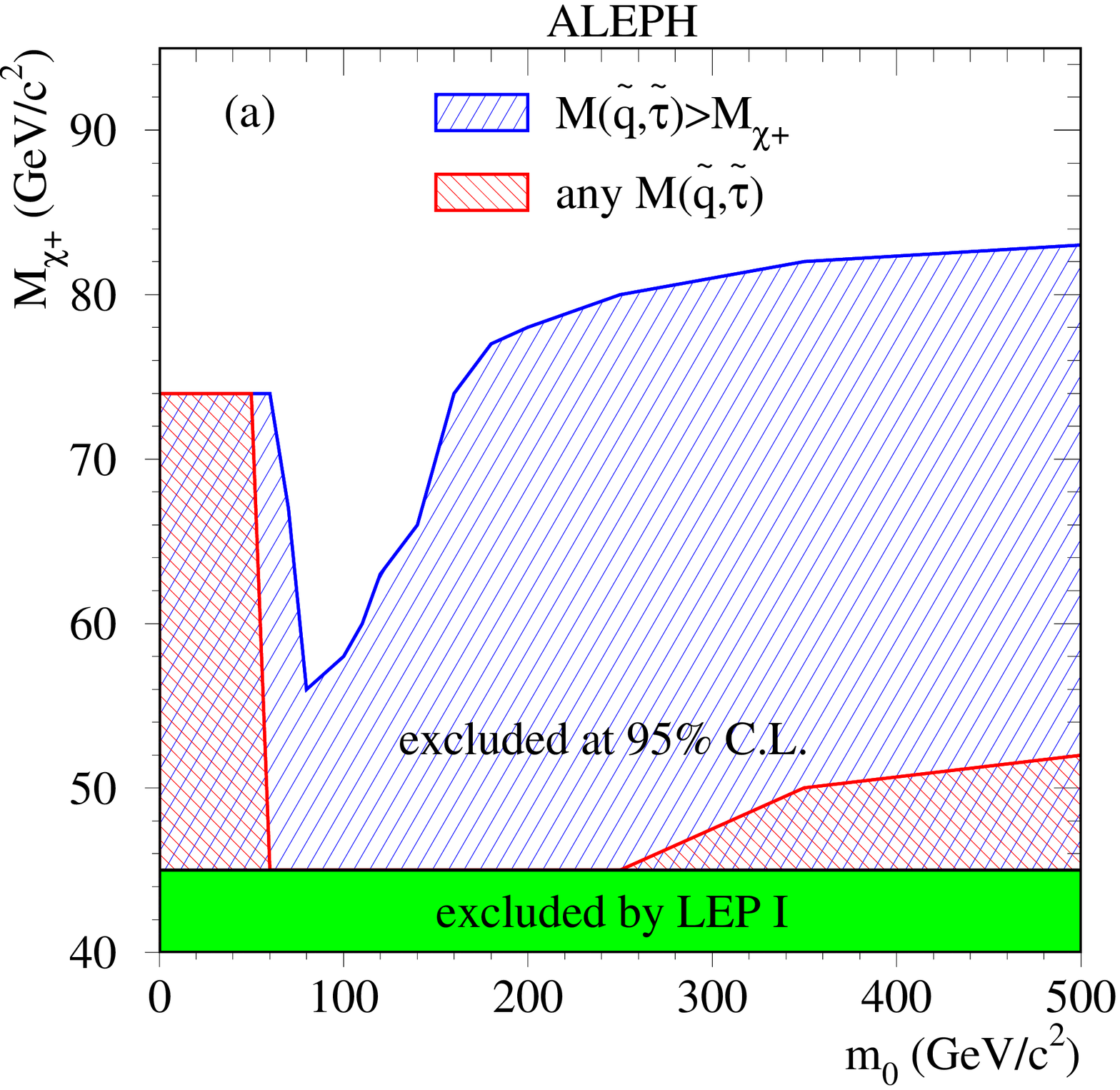,width=0.5\textwidth}\hfill
\epsfig{figure=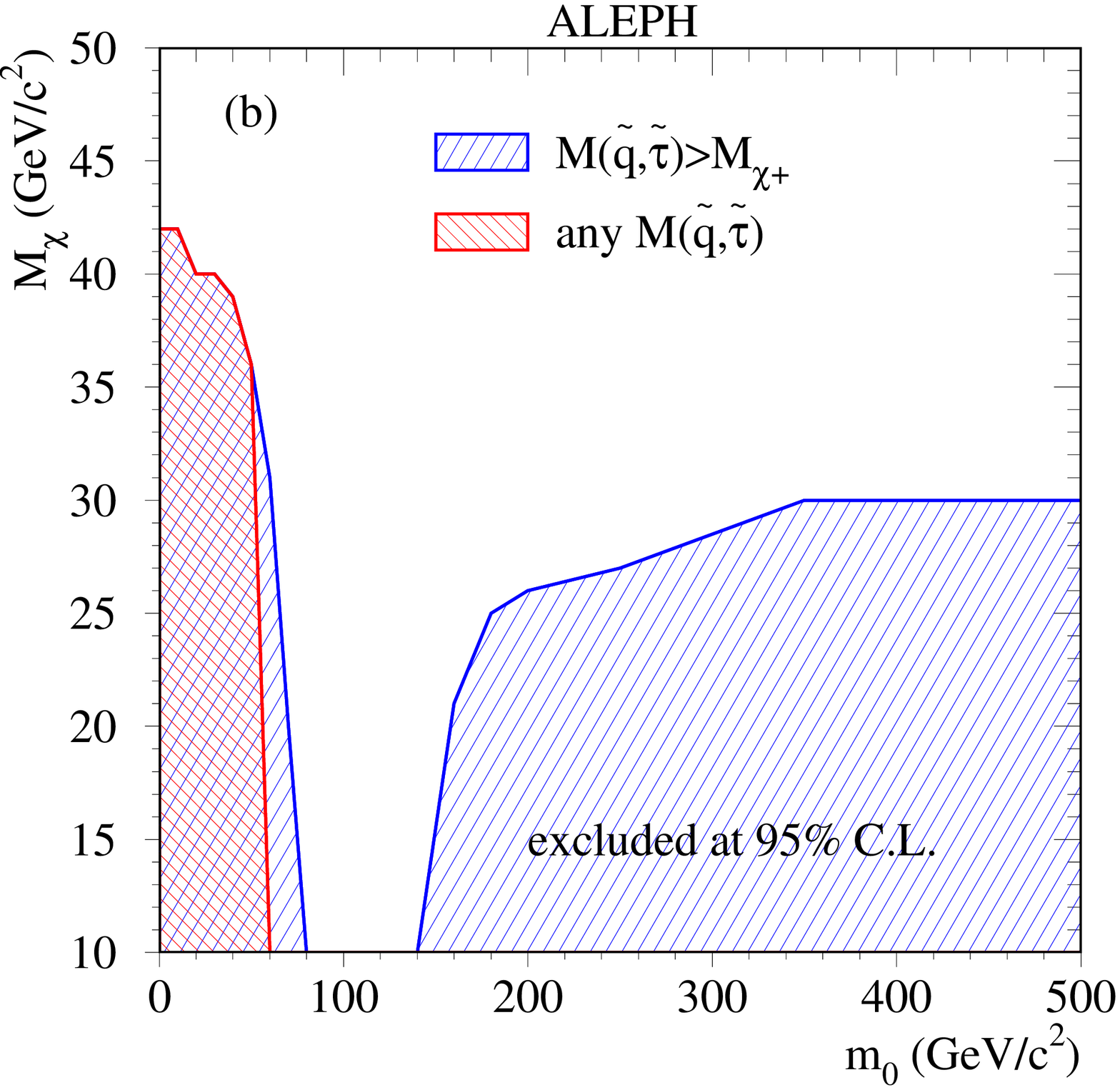,width=0.5\textwidth}}
\caption[.]{\em\label{mcha.vs.m0}{The limits on a) the chargino mass and
b) the neutralino mass as functions of $m_0$ obtained by scanning the 
($\mu$,$M_2$) plane at $\tan\beta=\sqrt{2}$, assuming a BR of $100\%$ for the
{\em indirect} decay modes.}}
\end{center}
\end{figure}

This effect is more apparent when 
limits on the masses of the lightest chargino and neutralino as a function 
of $m_0$ are obtained by scanning the ($\mu$,$M_2$) plane and fixing 
$\tan\beta=\sqrt{2}$. For this purpose the limits derived in 
Section~\ref{sec:slep} on {\em indirect} decays of sleptons and sneutrinos
are also used when $M_{\sfrm}<M_{\chi^+}$. The {\em indirect} decays of squarks
are not used because no limit improving on LEP~1 exists for general mixing angles.
The resulting limits are shown in Fig.~\ref{mcha.vs.m0} for the case
where the squarks and staus are constrained to be heavier than the lightest chargino
and for any squark and stau masses. The limits in the latter case are significantly
worse than those obtained in the former case. At low values of
$m_0$ the slepton and sneutrino limits contribute to constrain
the chargino and neutralino masses.

\subsubsection{Dominance of direct decays\label{dom.dir}}

In this section it is assumed that charginos and neutralinos decay {\em directly} to
6-fermion final states with a 100\% branching ratio. 
Charginos  typically decay {\em directly} if the sfermions are lighter than the
lightest neutralino, independent of the size of the coupling
 $\lambda'$. If the masses of the sfermions 
 lie between the mass of the chargino and the lightest neutralino, the {\em direct} decays
 of charginos can   dominate for large values of the R-parity violating
 coupling, and  if the neutralino couples higgsino-like. In small regions of
 parameter space {\em direct} chargino decays can dominate even when the sfermions are
 heavier than the chargino. Note that the lightest neutralino always decays
 {\em directly}.

\begin{figure}[bht]
\begin{center}
\makebox[\textwidth]{
\epsfig{figure=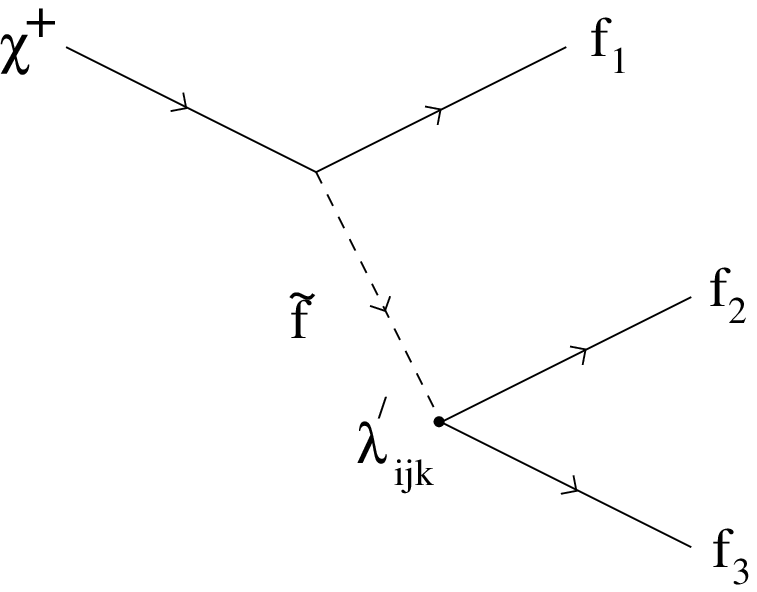,width=0.4\textwidth}}
\caption{\em{Generic diagram for a {\em direct} chargino decay. The exchanged
    sfermion $\Sf$ may either be on-shell (2-body decay), or off-shell
    (3-body decay).}}
\label{charg.decays}
\end{center}
\end{figure}

The experimental  signatures of the {\em direct} decays strongly depend on the mass
of the exchanged sfermion. Consider the generic diagram of a {\em direct} chargino
decay (Fig.~\ref{charg.decays}). 
If the sfermion is heavier than the chargino (or
neutralino) -- i.e. if the exchanged sfermion is virtual -- the momentum
distribution of the final state resembles  3-body kinematics, 
and shares the energy between the fermions
$f_1,f_2,f_3$ in roughly equal proportions. When
the exchanged sfermion is lighter than the chargino, the mass difference $\Delta
M = M_{\chi^+, \chi} - M_{\Sf}$ influences the decay distribution. 
 The fermion $f_1$ becomes softer as $\Delta M \ra 0$, and the signature of the
{\em direct} chargino/neutralino decays look more like a two-fermion rather than a
three-fermion final state.

Thus the overall signature from the pair-production of charginos and
neutralinos is best described as a 6-fermion final state at 
 large $\Delta M$ when the exchanged sfermions are virtual (3-body decays), 
or as a 4-fermion final state when $\Delta M$ is small. 

\clearpage

The possible topologies from the pair production of charginos are separated into 
eight different cases (see Table \ref{chargino.1}, \ref{chargino.2}).
The first four cases correspond to a dominant coupling of
type $\lambda'_{i3k}$, where the chargino can decay to an on-shell sneutrino or a
stop (Cases 2-4), or the chargino decay proceeds through a 3-body decay (Case
1). For example, in case 2 the sneutrino is lighter than the chargino, and the
 topology of the chargino signal may resemble  a four-jet final state for
 small $\Delta M = M_{\chi^+} - M_{\tilde \nu}$, or  a four-jet plus two
 lepton final state when $\Delta M$ is large. 
Note that the chargino decays to $\chi^+ \ra {\sbq} t$ are not kinematically
 allowed, and hence not listed. Similarly the  chargino cannot decay 
{\em directly} via sleptons for a non-zero $\lambda'_{i3k}$ coupling, since the
decay ${\slep}_i \ra {{\mathrm{t}}} {\bar {\mathrm{d}}}_k$ is kinematically inaccessible.

The Cases 5-8 correspond to the dominant couplings 
$\lambda'_{{i1k}}$ or $\lambda'_{{i2k}}$,
where the chargino can decay to an on-shell sneutrino or a
slepton (Cases 6-8), or the chargino decay proceeds through a 3-body decay (Case
5). 

Similarly,
the neutralino can decay to an on-shell slepton, sneutrino or sbottom, or the
neutralino decays may proceed via a 3-body decay. Depending on the mass of the 
exchanged sfermion a total of eight cases can be identified, and the topologies
corresponding to these cases are listed in Tables~\ref{neutralino.1} 
and \ref{neutralino.2}.

The analyses described in Section~\ref{searches} were developed and optimised for these 
topologies, and cover all the possible cases. 
 A set of chargino and neutralino  efficiencies for various gaugino and sfermion
masses  are shown in  Table~\ref{effics}. 
Given the relatively large expected background in each topology, it would be
impractical to simply take an OR of all the selections applicable to a given
case. Instead the optimal combination of selections is evaluated for each 
chargino/neutralino mass,  for each $\Delta M$ point and for a given 
branching ratio into the different topologies using the \nbar\ method. 

For each of the eight chargino cases and eight neutralino cases 
signal MC samples were generated for the two centre-of-mass energies 
$\sqrt{s}=161,172$, for different
chargino/neutralino masses, for various values of $\Delta M$, and different
generation indices of the coupling $\lambda'_{ijk}$. The topologies with taus in
the final state -- which correspond to a dominant coupling $\lambda'_{3jk}$ --
are the most difficult ones, and are used to set the most conservative limit.

If more than one topology is possible for a given $\Delta M$ point, as for
example in Case 1 of Table~\ref{chargino.1}, then the branching ratios of the allowed 
topologies are  varied
freely to determine the worst case exclusion. This approach again ensures that
the derived limit is conservative. 

\begin{table}[ht]
\centering
\begin{tabular}{c||c|c|c|c}
&  Case 1 & Case 2 & Case 3 & Case 4 \\
\hline
$M_{\tilde \nu}$ & heavy & light & heavy & light \\
$M_{\tilde t}$ & heavy & heavy & light & light \\
\hline
small $\Delta M$  & - & 4j & 2l+2j & 4j , 3j+l, 2j+2l \\
$\Da$                  & 4j+2l, 4j+2$\nu$, 4j+l+$\nu$&                          $\Da$ & $\Da$
                  & $\Da$ \\
large $\Delta M$ & - &  4j+2l &  4j+2l & 4j+2l
\end{tabular}
\caption[.]{\em Classification of the different chargino topologies for a
                  dominant coupling $\lambda'_{i3k}$, and the
                  transition from small $\Delta M$ to large $\Delta M$.        
 The attribute ``heavy''  denotes that the
                  exchanged sfermion is
                  heavier than the chargino, while  ``light'' 
                   denotes that the sfermion is lighter than the
                  chargino.}
\label{chargino.1}
\vspace{1cm}
\centering
\begin{tabular}{c||c|c|c|c}
&  Case 5 & Case 6 & Case 7 & Case 8 \\
\hline
$M_{\tilde l}$ & heavy & light & heavy & light \\
$M_{\tilde \nu}$ & heavy & heavy & light & light \\
\hline
small $\Delta M$  &  - & 4j & 4j & 4j  \\
$\Da$                  & 4j+2l, 4j+2$\nu$, 4j+l+$\nu$ &                          $\Da$ & $\Da$ & $\Da$ \\
large $\Delta M$ &  - &  4j+2$\nu$ &  4j+2l & 4j+2l, 4j+l+$\nu$, 4j+2$\nu$
\end{tabular}
\caption[.]{\em Classification of the different chargino topologies for a
                  dominant  $\lambda'_{i1k}$ or $\lambda'_{i2k}$ coupling.}
\label{chargino.2}
\vspace{1cm}
\centering
\begin{tabular}{c||c|c|c|c}
&  Case 9 & Case 10 & Case 11 & Case 12 \\
\hline
$M_{\tilde l}$ &   heavy & light & heavy & heavy \\
$M_{\tilde \nu}$ & heavy & heavy & light & heavy \\
$M_{\tilde b}$ &   heavy & heavy & heavy & light \\
\hline
small $\Delta M$  &  - & 4j & 4j & 2j+2l,2j+l+$\nu$,2j+2$\nu$  \\
$\Da$                  & 4j+2l, 4j+2$\nu$, 4j+l+$\nu$ &                          $\Da$ & $\Da$ & $\Da$ \\
large $\Delta M$ &  - &  4j+2l &  4j+2$\nu$ & 4j+2l, 4j+l+$\nu$, 4j+2$\nu$\\
\end{tabular}
\caption[.]{\em Classification of the different neutralino topologies.}
\label{neutralino.1}
\vspace{1cm}
\centering
\begin{tabular}{c||c|c|c|c}
&  Case 13 & Case 14 & Case 15 & Case 16 \\
\hline
$M_{\tilde l}$ &   light & light & heavy & light \\
$M_{\tilde \nu}$ & light & heavy & light & light \\
$M_{\tilde b}$ &   heavy & light & light & light \\
\hline
small $\Delta M$  &  4j & 4j,2j+2l,2j+l+$\nu$,2j+2$\nu$  & &    \\
$\Da$                  & $\Da$ &  $\Da$ & as Case 14 & as Case 14 \\
large $\Delta M$ &  4j+2l, 4j+l+$\nu$, 4j+2$\nu$  & 4j+2l, 4j+l+$\nu$, 4j+2$\nu$  & & 
\end{tabular}
\caption[.]{\em Classification of the different neutralino topologies.}
\label{neutralino.2}
\end{table}

\begin{figure}
\begin{center}
\makebox[\textwidth]{
\epsfig{figure=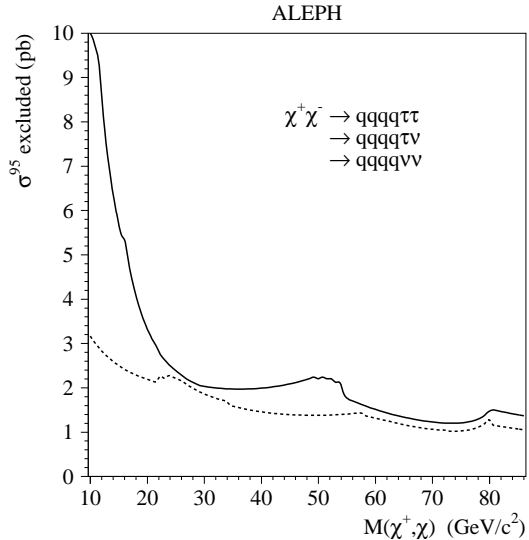,width=.47\textwidth}}
\caption{\em{The 95$\%$C.L. limit on the  chargino/neutralino cross section (at
$\sqrt{s}=172\gev$) for {\em direct} three body decays of
charginos/neutralinos and a non-zero coupling $\lambda'_{3jk}$. 
The dashed line corresponds to the most favourable chargino branching ratio,
 while the solid line
corresponds to the worst case branching ratio. The latter is used to derive a
conservative limit on the chargino/neutralino cross section.}} 
\label{case1_2}
\end{center}
\end{figure}

\begin{figure}
\begin{center}
\makebox[\textwidth]{
\epsfig{figure=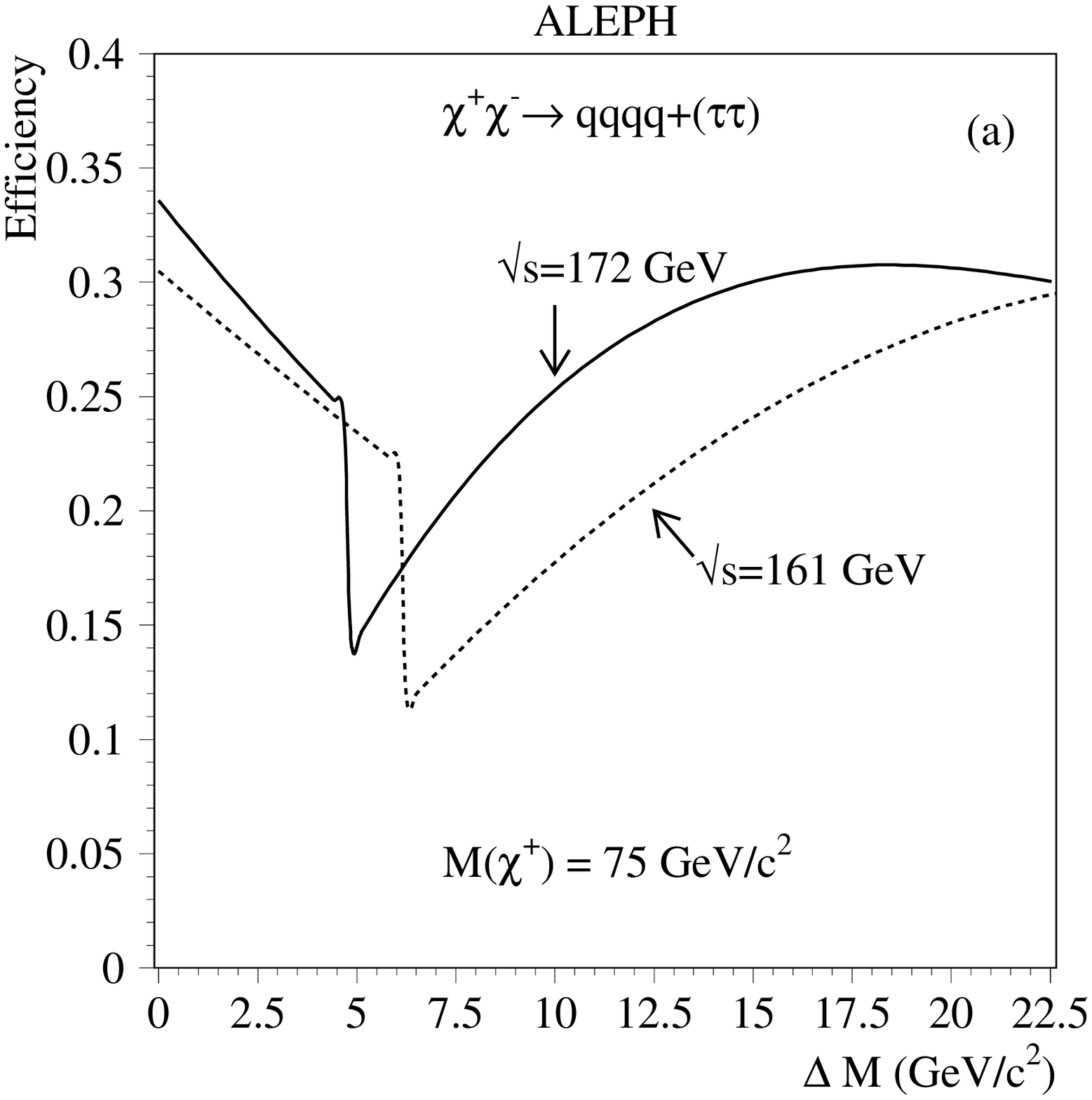,width=.47\textwidth}\hfill 
\epsfig{figure=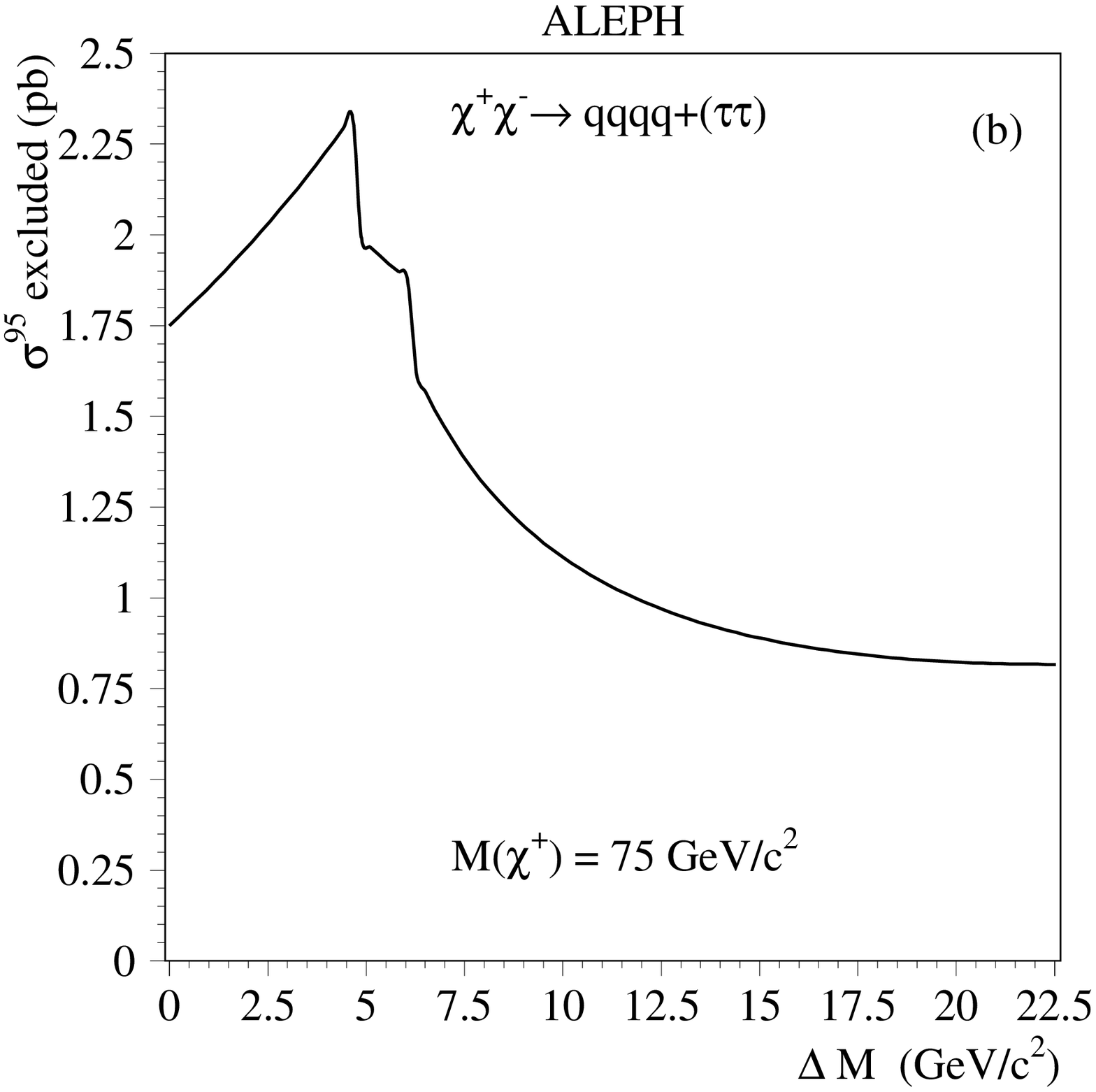,width=.47\textwidth}
}
\caption{\em {a) The selection efficiencies for the {\em direct}
chargino/neutralino decays  \mbox{$\chi^+ \chi^- \ra 4q
+ 2\tau$} via two-body decays to sleptons/sneutrinos. Here 
$\Delta M=M_{\chi^+} - M_{\tilde \nu}$. 
For $\Delta M \lsim\ 5\gevcc$ the Four Jet selection is used, while for 
$\Delta M \gsim\ 5\gevcc$ the 4J(2$\tau$) selection is used. 
 b) The corresponding 95$\%$C.L. excluded cross section (at 
$\sqrt{s}=172\gev$). }}
\label{case2_2}
\end{center}
\end{figure}

The following two examples illustrate the manner in which the limits were
derived:
\begin{itemize} 
\item {\bf Cases 1,5,9} \newline
The {\it three-body} chargino (or neutralino) decays lead to the 4j+2l,
4j+l+$\nu$, 4j+2$\nu$ topologies. At chargino/neutralino masses above $\approx
50\gevcc$ these topologies are selected by the 4J(2L) selections, the
4J(L$\nu$) selection and the 4J(2$\nu$) selection for final states with electrons
and muons (the $\lambda'_{(i=1,2)jk}$ couplings). 
For tau final states ($\lambda'_{3jk}$) a
combination of the 4J(2$\tau$) and the  4J(2$\nu$) selection is used. 
For masses below  $\approx 50\gevcc$ a combination of 
the 4J2L-low, 4JL$\nu$-low and the  4J2$\nu$-low selections
(or for tau final states the 4J2$\tau$-low, 4J2$\nu$-low selections)
  are used.  

\clearpage

The \nbar\ method is used to decide the optimal combination of selections 
for a given  chargino/neutralino mass and a fixed branching ratio into 4j+2l, 4j+l+$\nu$,
 4j+2$\nu$ topologies. Finally
 the branching ratios are varied freely to find  the most conservative exclusion
 limit. This procedure is illustrated in Fig.~\ref{case1_2}, which 
  shows the excluded cross section  as a function of
 $M_{{\chi^+},\chi}$ for the worst case coupling $\lambda'_{3jk}$. The dashed line
corresponds to the limit for the most favourable branching ratios, while the
solid line corresponds to the worst case branching ratio. The discontinuities in
the excluded cross section correspond to the points at which the combination of
selections changes.

\item {\bf Cases 2,7,10} \newline
These cases correspond to the {\it two-body} chargino (or neutralino) decays to
sneutrinos (sleptons), where the difference in mass $\Delta M=M_{\chi^+} -
M_{\tilde \nu}$ influences the event distributions of the $4j+2l$ topology. At
low $\Delta M$ the signal topologies are indistinguishable 
from four jet final states and the Four Jet selection is used.
At higher  $\Delta M$ the 4J(2L) (or 4J(2$\tau)$) selection is used. The 
 optimal point in $\Delta M$ at which the two selections switch is determined
using the \nbar\ method (c.f. Fig.~\ref{case2_2}).
\end{itemize} 

Limits on the chargino and neutralino masses are derived within the
MSSM assuming universal sfermion masses, but varying the third generation
sfermion mixing parameters $A_{\tau}, A_{\rm b}, A_{\rm t}$ between  -1$\tevcc$ and +1$\tevcc$.
For each point in $\mu-M_2-\tan{\beta}-m_0$ (and $A_{\tau}, A_{\rm b}, A_{\rm t}$) parameter
space the chargino, neutralino and sfermion masses are calculated, and hence
each point corresponds to  one of the neutralino 
topologies or cases (Tables~\ref{neutralino.1}, \ref{neutralino.2}), or two of the chargino
topologies/cases (Tables~\ref{chargino.1}-\ref{chargino.2}). 
 Again, for charginos the topology which
corresponds to the worst case exclusion is chosen to set a conservative
limit. Finally, the \nbar\ method is used to decide if one or more of
the following production processes  are combined to set the overall limit:
\begin{center} 
$e^+ e^- \ra \chi^+ \chi^-$, $\chi \chi$, $\chi \chi'$, $\chi' \chi'$.
\end{center} 

\begin{figure}[t]
\begin{center}
\makebox[\textwidth][l]{
\epsfig{figure=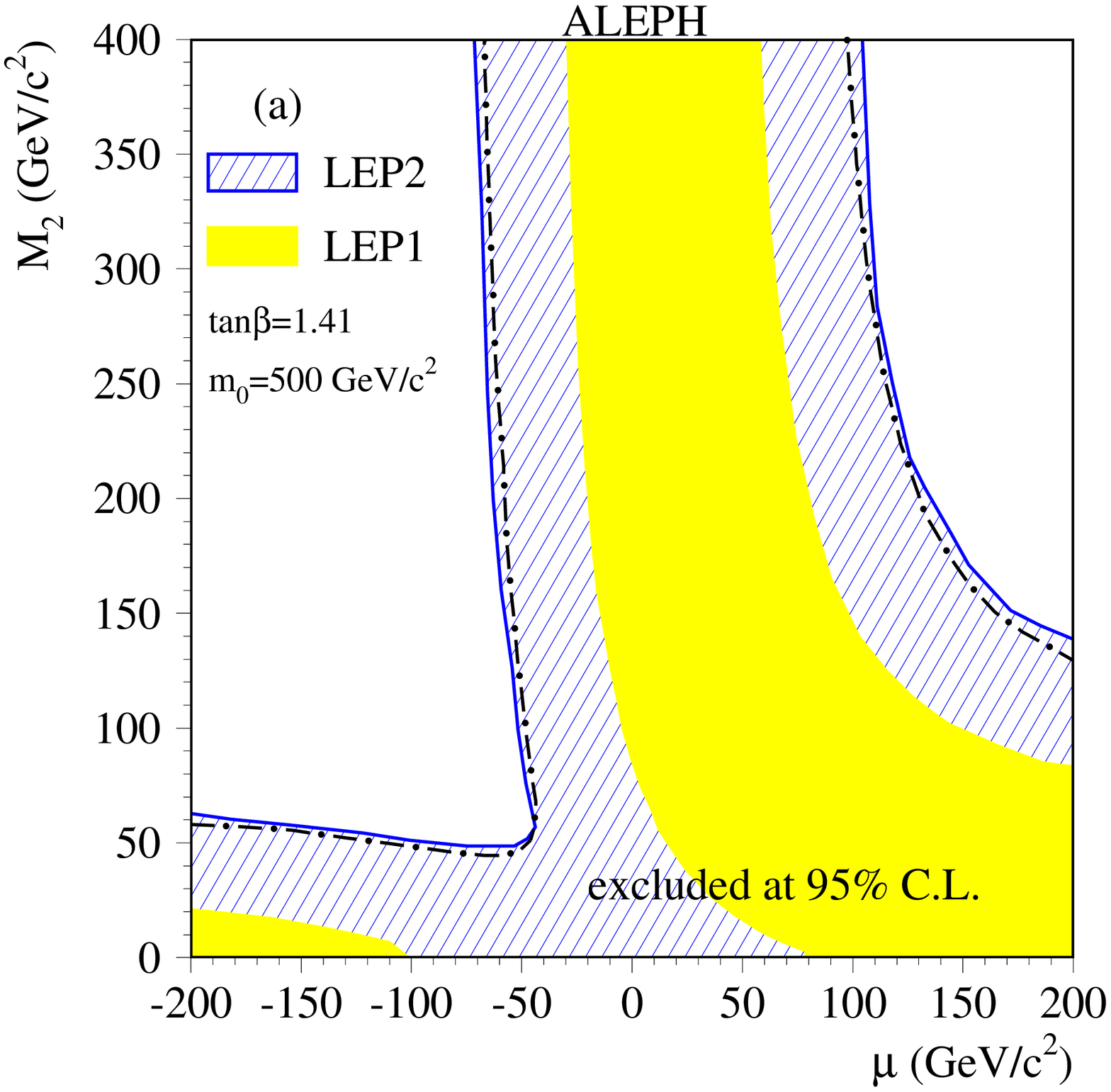,width=0.5\textwidth}\hfill
\epsfig{figure=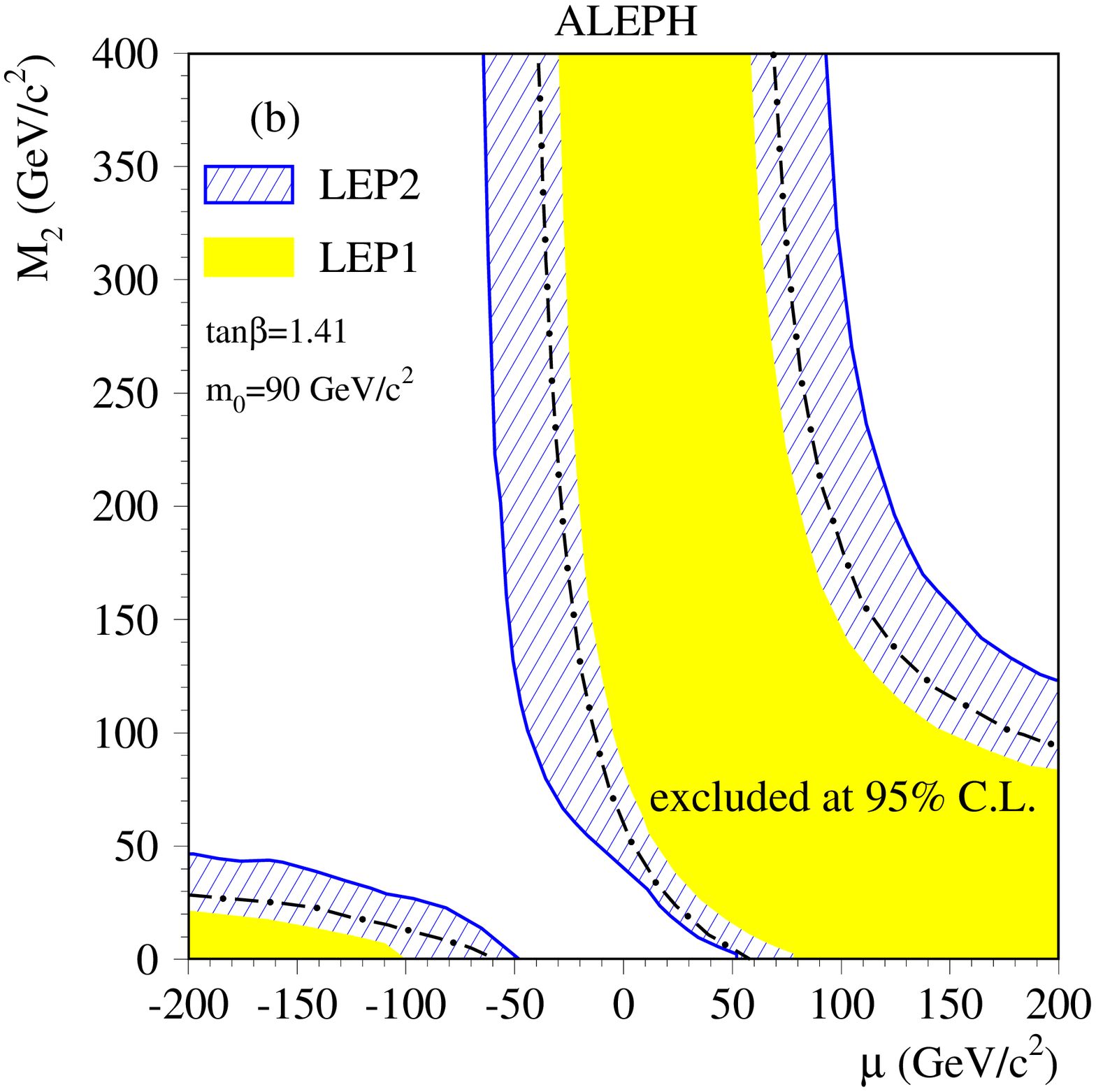,width=0.5\textwidth}}
\caption[.]{\em{
 Regions in the $(\mu,M_2)$ plane excluded
 at $95\%$ C.L. at $\tan\beta=\sqrt{2}$ assuming that charginos/neutralinos
decay {\em directly} with a BR of 100\%. a) for $m_0=500$\gevcc{}, where the inner dash-dotted
lines show the overall chargino mass limit of $M_{\chi^+}=81 \gevcc$. b) for
$m_0=90$\gevcc{},  where the inner dash-dotted
lines show the overall chargino mass limit of $M_{\chi^+}=52\gevcc$.}\label{directmum2}}
\end{center}
\end{figure}

\begin{figure}[t]
\begin{center}
\makebox[\textwidth][l]{
\epsfig{figure=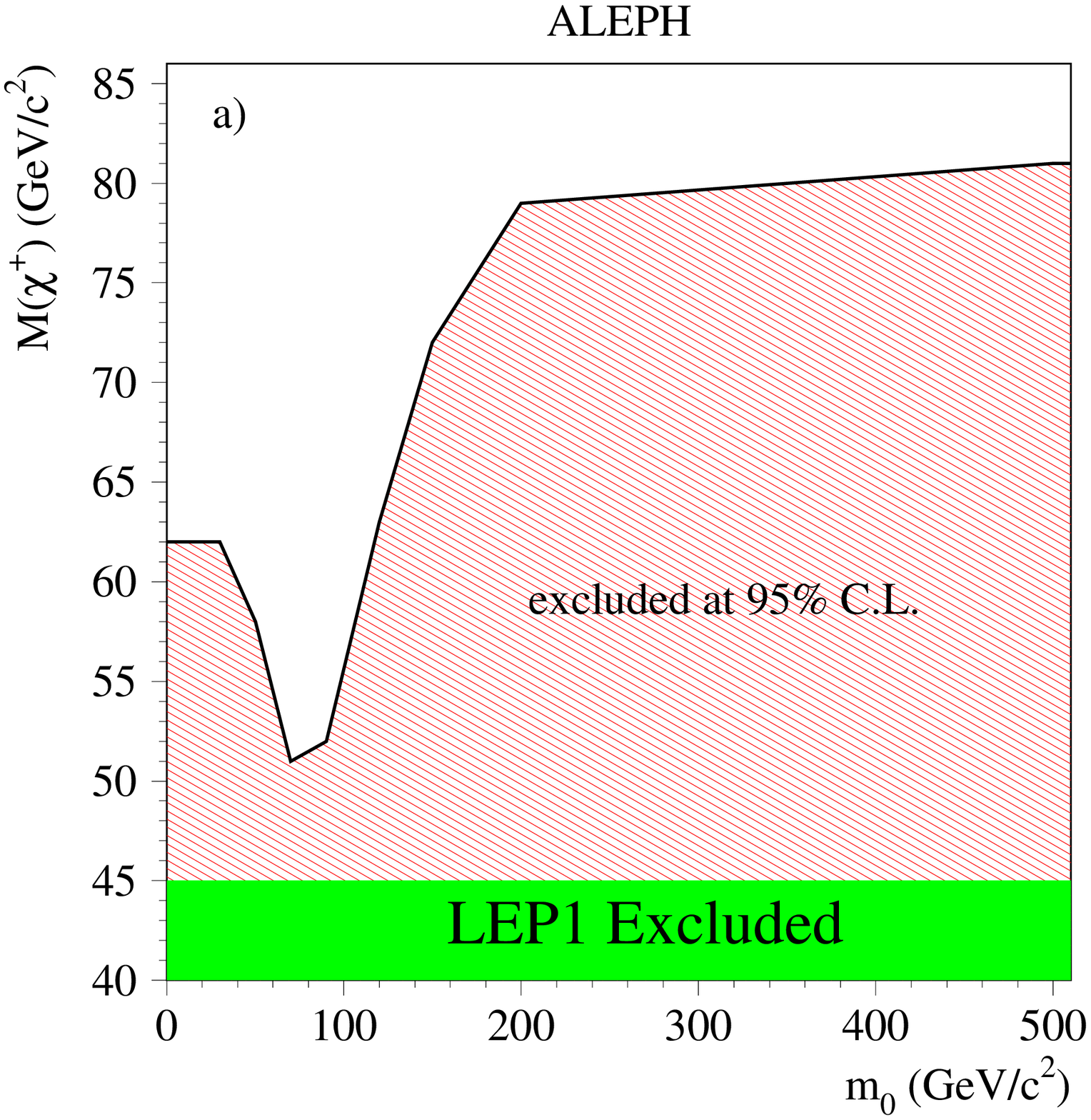,width=0.5\textwidth}\hfill
\epsfig{figure=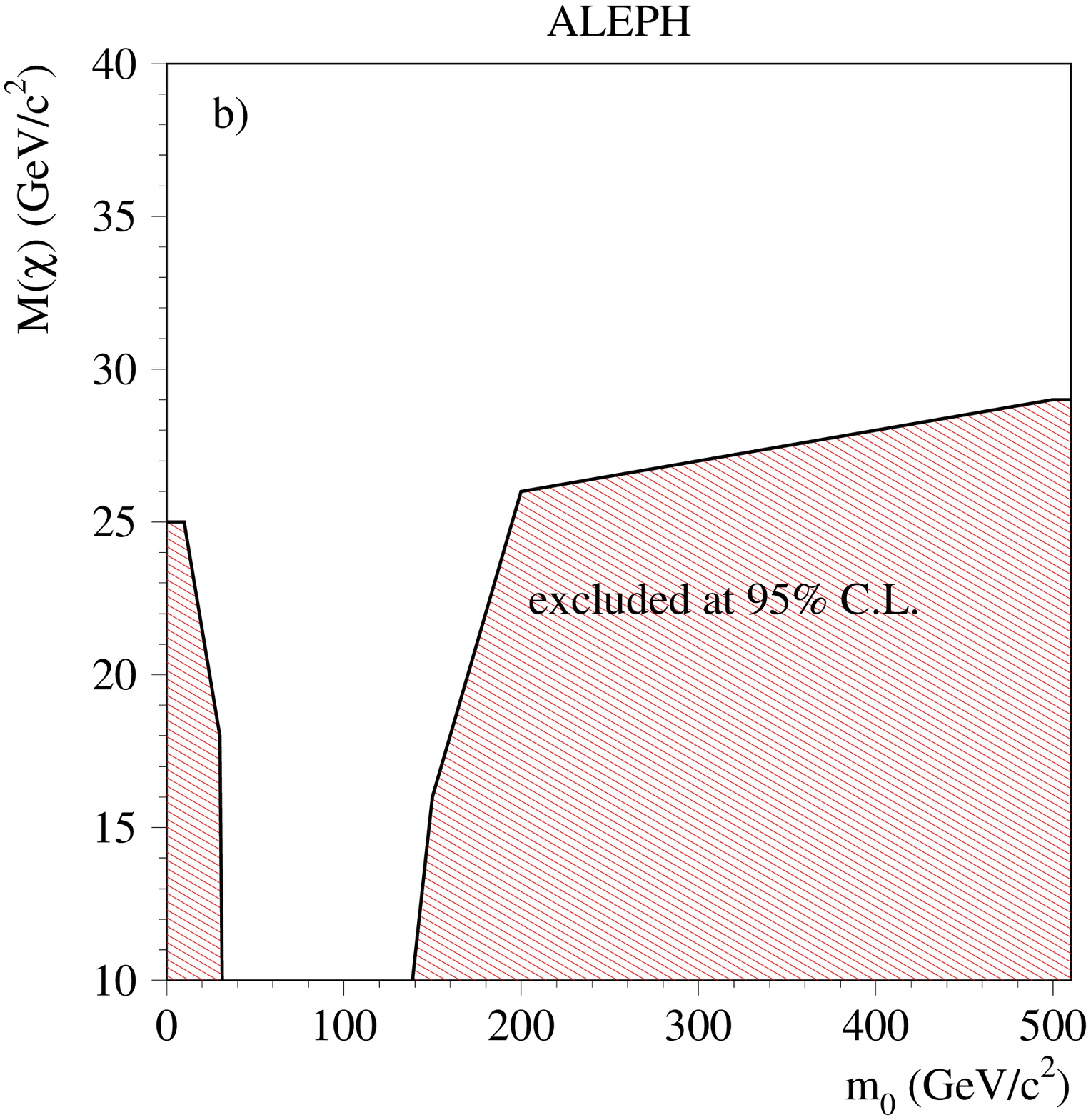,width=0.5\textwidth}}
\caption[.]{\em{The limits on a) the chargino mass and
b) the neutralino mass as functions of $m_0$ obtained by scanning the 
($\mu$,$M_2$) plane at $\tan\beta=\sqrt{2}$, assuming a BR of $100\%$ for the
{\em direct} decay modes.}}
\label{direct_gaugino_limit}
\end{center}
\end{figure}

The exclusion limit in the ($\mu, M_2$) plane for $\tan{\beta}=\sqrt{2}$
is shown in Fig.~\ref{directmum2}  for the two values of  $m_0=500$ , $90 \gevcc$. At $m_0=500
\gevcc$ the corresponding mass limits on the chargino and the neutralino are
$M_{\chi^+}>81\gevcc, M_{\chi}>29\gevcc$. The mass limits are weaker for low $m_{0}$
 since negative t-channel interference reduces the chargino
cross section in the gaugino region. This trend may be seen in
Fig.~\ref{direct_gaugino_limit}, which shows the mass limits as a function of
$m_0$ for fixed $\tan{\beta}=\sqrt{2}$. At $m_0\lsim\ 50\gevcc$ the gaugino mass limits
 increase again due to a combination of positive t-channel interference for the 
$\chi \chi$ production cross section in the gaugino region, and the LEP~1
slepton/sneutrino limits derived from the Z-width.

\subsection{Sleptons and Sneutrinos}\label{sec:slep}
A slepton can decay either {\em directly} to a pair of quarks or {\em indirectly} to 
a lepton and a  neutralino, which subsequently decays to two quarks and lepton or 
neutrino. Decays to charginos or heavier neutralinos are not considered. 
Right-handed sleptons may only decay {\em indirectly}.

\begin{figure}
\begin{center}
\makebox[\textwidth]{
\epsfig{figure=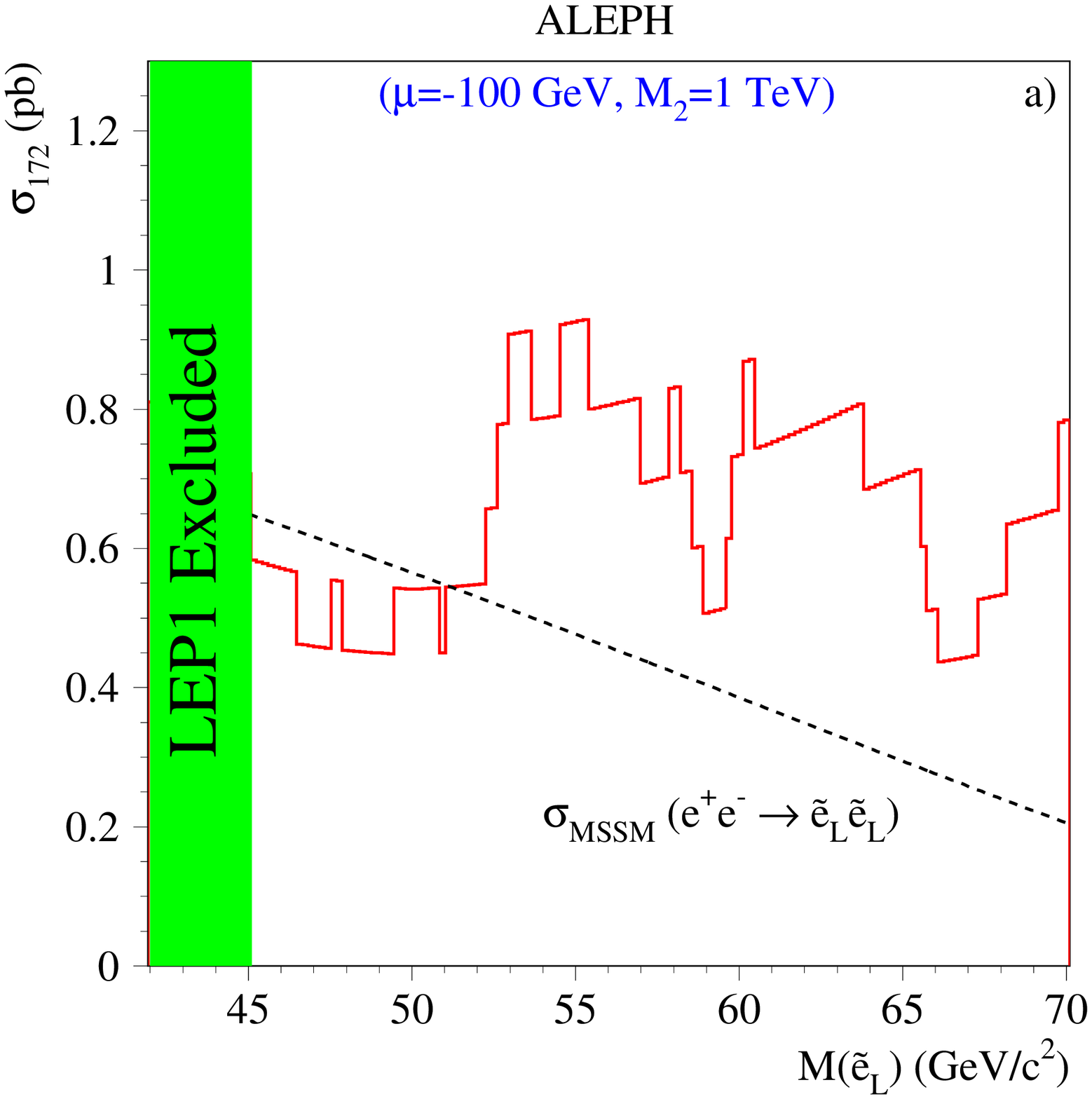,width=0.5\textwidth}\hfill
\epsfig{figure=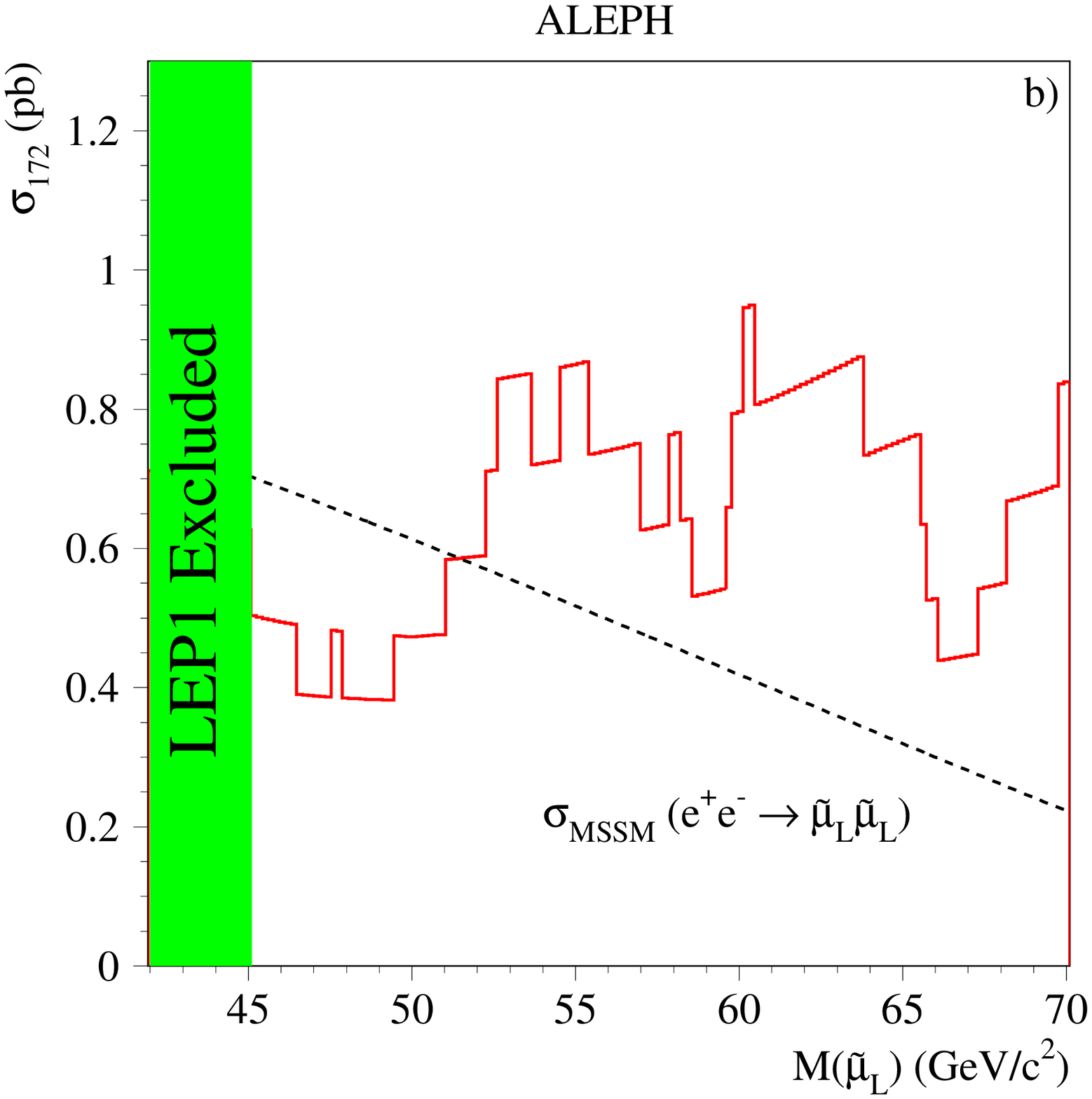,width=0.5\textwidth}}
\makebox[0.5\textwidth]{
\epsfig{figure=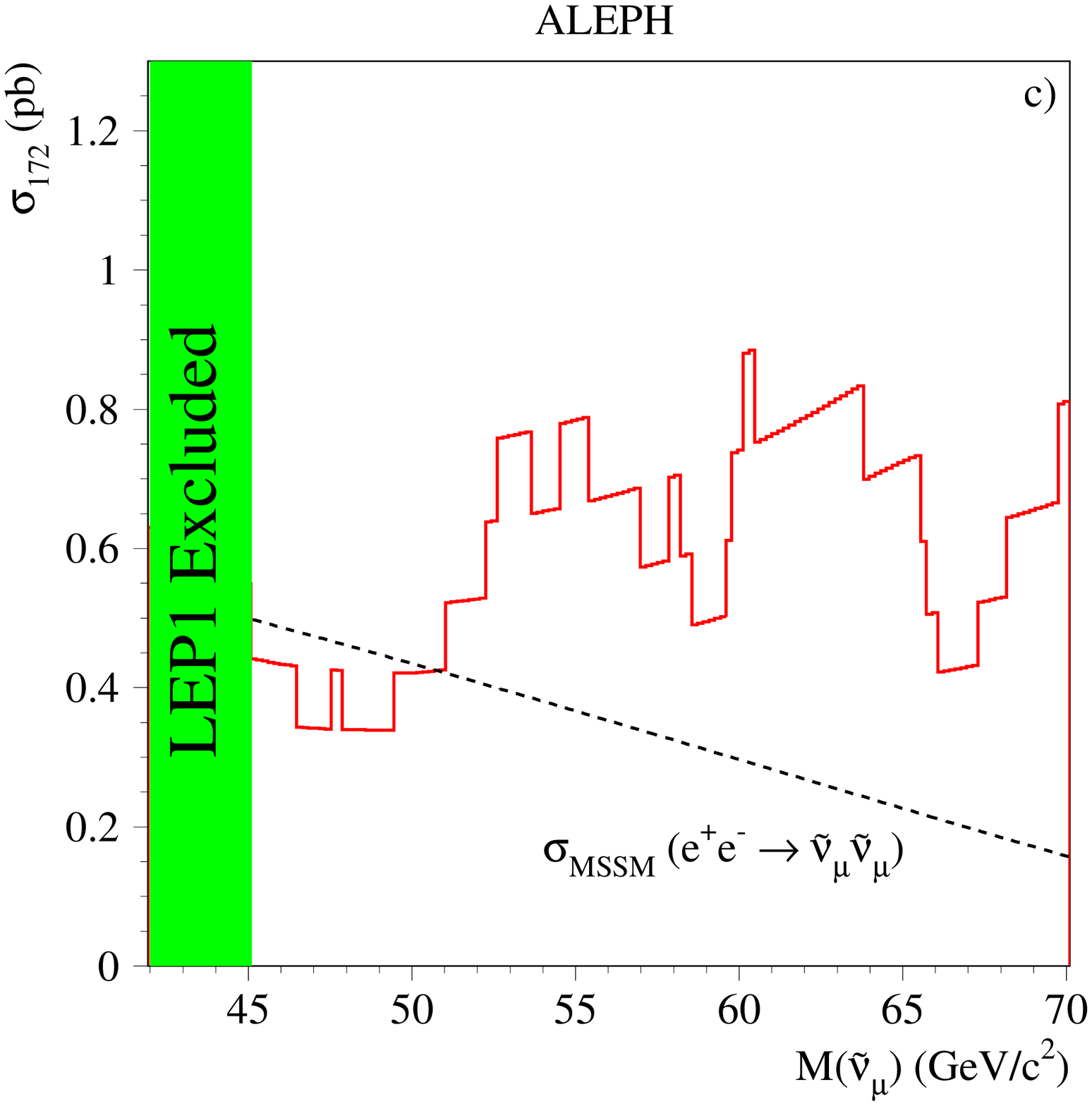,width=0.5\textwidth}\hfill}
\caption[.]{\label{slep.4jet}\em{The $95\%$ C.L. slepton exclusion
  cross sections scaled to $\sqrt{s}=172\gev$ for the {\em direct} decays. For the purpose of these plots
  a $\beta^3/s$ cross section dependence, valid for 
 scalar pair-production in the {\em s}-channel,  was assumed.}}
\end{center}
\end{figure}

The {\em direct} topology is defined as that when both sleptons decay {\em directly} leading 
to a four jet final state. To select this topology the Four Jet selection was 
employed. Events are counted if the di-jet mass $M_{\rm 5C}$ 
obtained from the 5C fit is within  $3\gevcc$ of the slepton or sneutrino mass.
The efficiencies for the signal to fall inside this window are determined at the three
centre-of-mass energies $\sqrt{s}=133, 161, 172\gev$ (see also
Table~\ref{effics}). 
The efficiencies are relatively
constant as functions of $M_{\slep}$, $M_{\snu}$. 
Limits on slepton and sneutrino production are set by
sliding a mass window across the di-jet mass distribution, counting the number
of events seen and subtracting the expected background according to the prescription 
given in \cite{PDG}. For this purpose the expected background has been assigned
a conservative error of $20\%$ and has been reduced by this amount.
The results are shown in Fig.~\ref{slep.4jet}.

\begin{figure}
\begin{center}
\makebox[\textwidth]{
\epsfig{figure=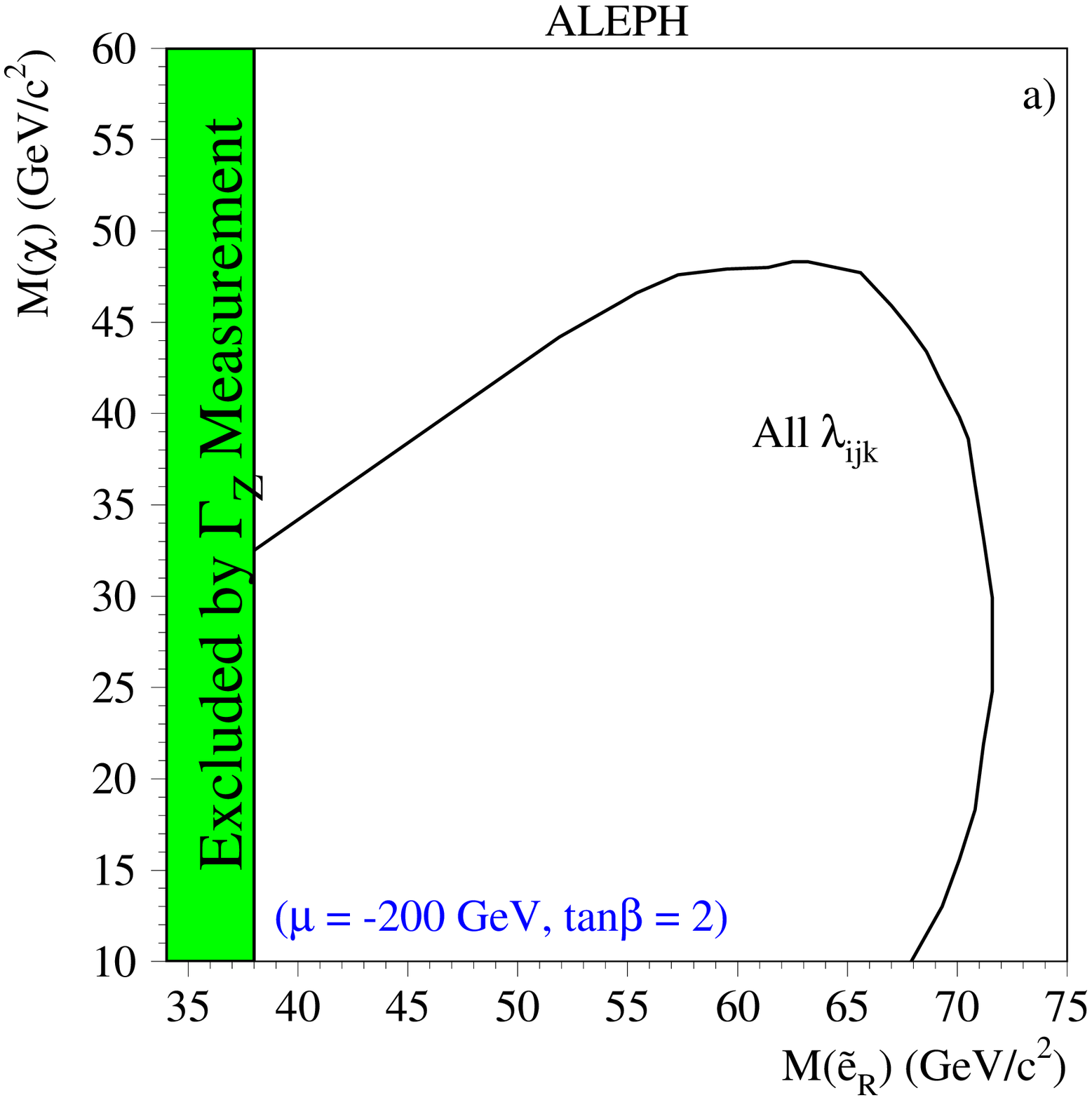,width=0.5\textwidth}\hfill
\epsfig{figure=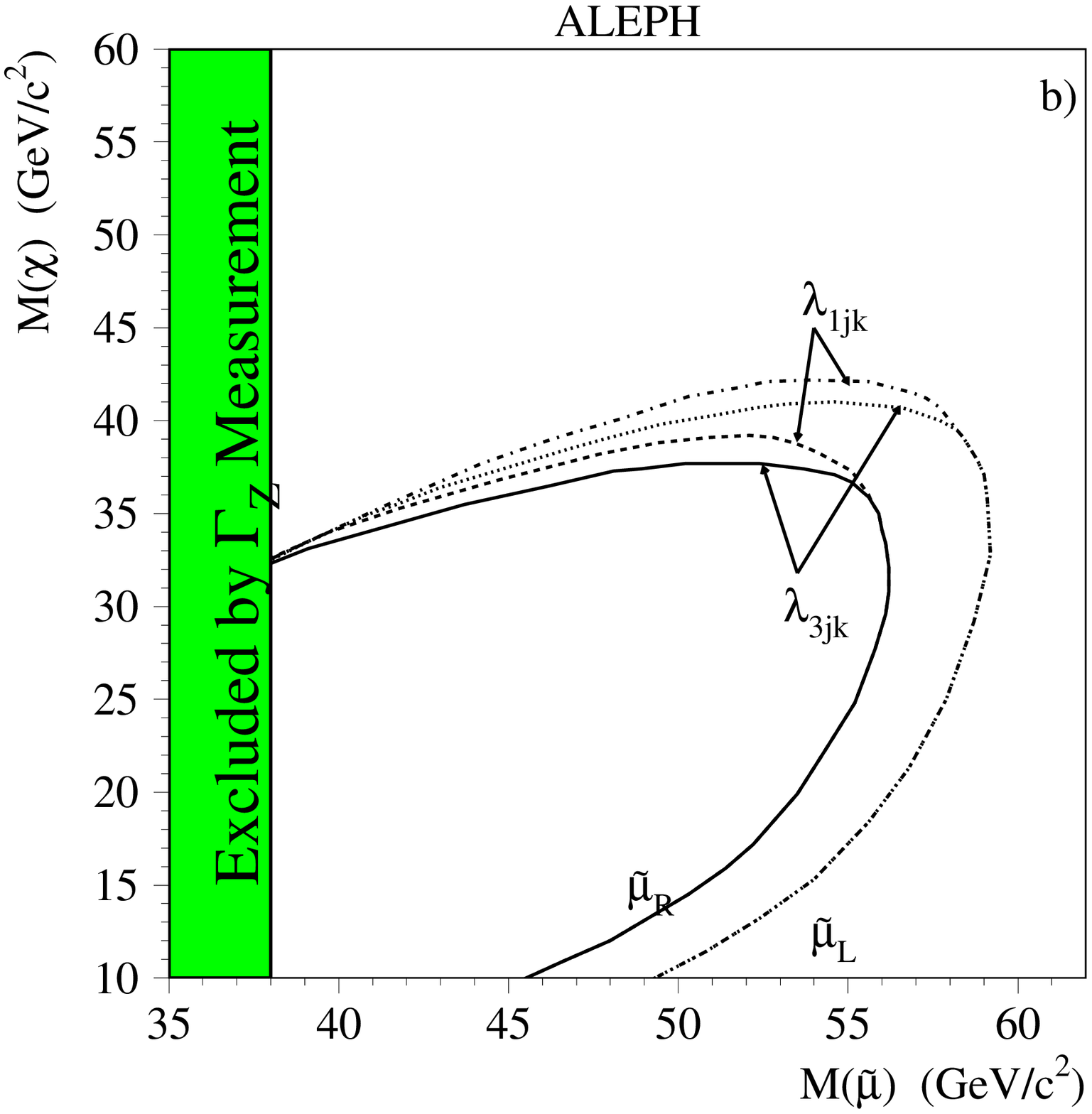,width=0.5\textwidth}}
\makebox[\textwidth]{
\epsfig{figure=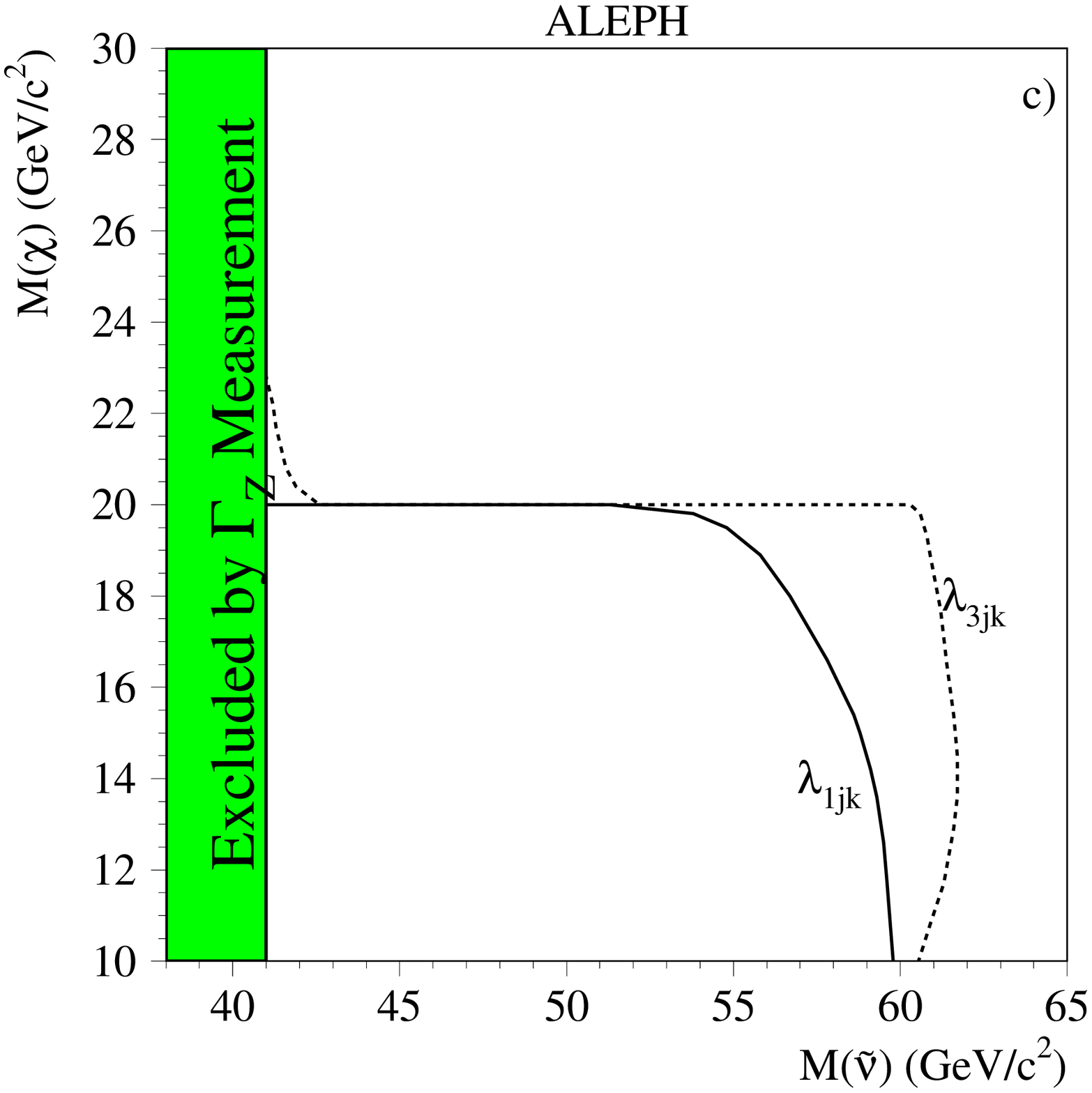,width=0.5\textwidth}\hfill
\epsfig{figure=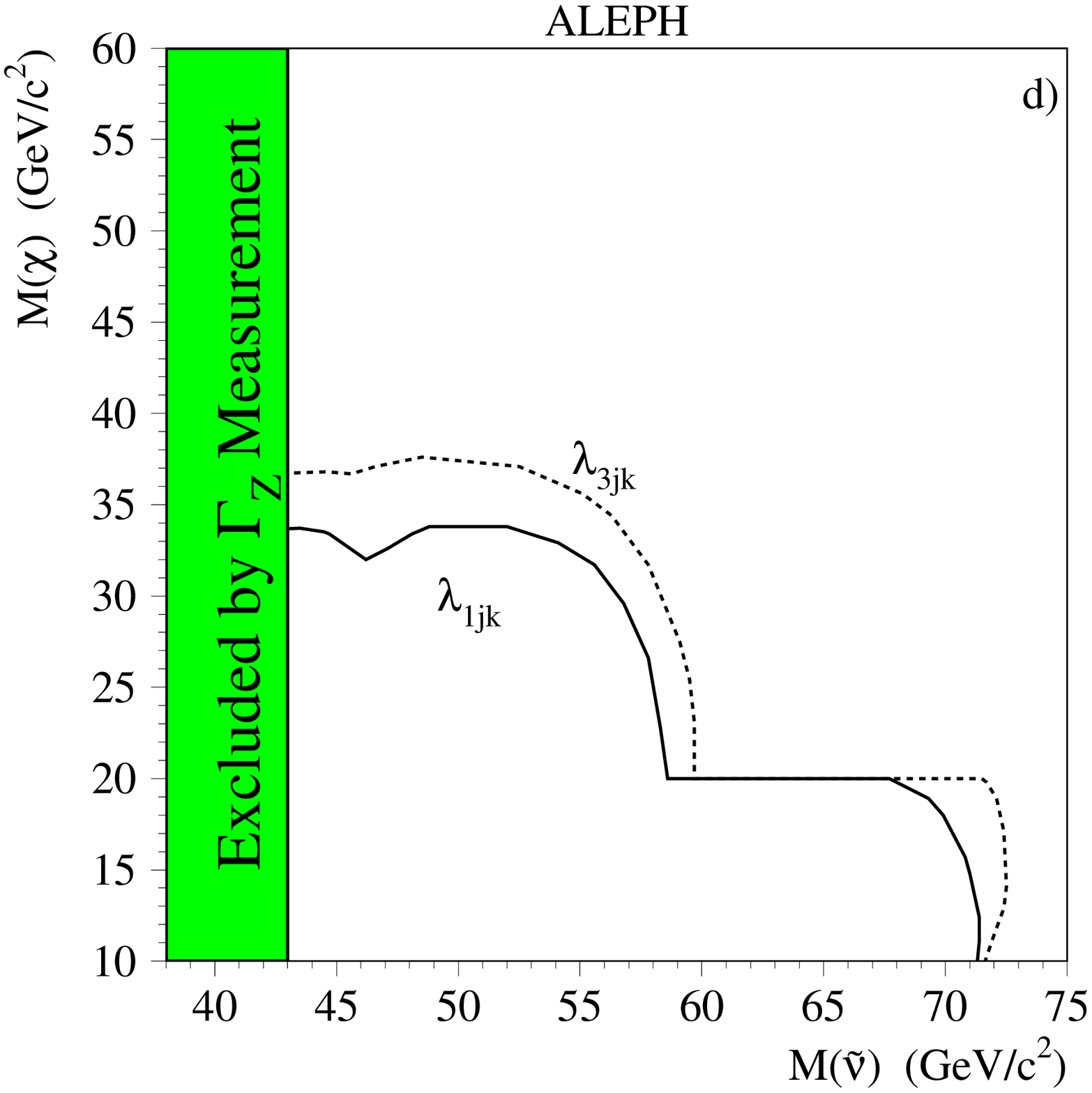,width=0.5\textwidth}}
\caption[.]{\em {The $95\%$ C.L. limits on the 
selectron, smuon and sneutrino in the $(M_{\chi},M_{\slep})$ 
or $(M_{\chi},M_{\tilde {\nu}})$ plane for the {\em indirect} decay modes.
d) shows the limit for three degenerate sneutrinos. 
The selectron cross section is evaluated at $\mu=-200\gev$ and
$\tan\beta=2$.}}
\label{slep.ind}
\end{center}
\end{figure}

The selections employed to analyse the {\em indirect} decays of sleptons and sneutrinos
were chosen to optimise \nbar\. The selectron and smuon signals share the property
of an easily identified, energetic lepton and are efficiently selected by the 
4J(2L) selection over much of the parameter space (c.f. Table~\ref{effics}). 
When the mass difference is 
small the leptons are less energetic and the signal efficiency is reduced. In this region
the inclusive combination of the 4J-VH and 4J(2$\tau$) selection is used. 
The excluded regions in the plane $(M_{\chi},M_{\slep})$
are shown in Fig.~\ref{slep.ind}a and \ref{slep.ind}b. The selectron cross section is evaluated at a
typical point in the gaugino region ($\mu=-200\gev$, $\tan\beta=2$) to show the 
effect of the constructive t-channel interference.

Stau events were selected with the reoptimised 4J(2$\tau$) selection 
across most of the parameter space. For small neutralino masses ($M_{\chi}<20\gev$)
the signal is similar to two energetic taus and two jets. In this region the 2J+2$\tau$
selection is used. The efficiencies are too low and the number of candidates
too large for any mass limits to be set that improve upon the Z width measurement.

The sneutrino signal 
is similar to pair production of the lightest neutralino, but with additional missing 
energy. The signal is therefore  similar to some R-parity conserving signals.
The inclusive combination of the Multi-jets plus Leptons selection with 4J-H and 4J-L
is employed for neutralino masses greater than $20\gevcc$. For neutralino masses less than
$20 \gevcc$ the signal is acoplanar jets and the AJ-H selection is used. The excluded regions 
in the plane $(M_{\chi},M_{\snu})$ are shown in Fig.~\ref{slep.ind}c. Assuming that
all three sneutrinos are degenerate in mass the improved limit shown in 
Fig.~\ref{slep.ind}d is obtained.

\clearpage

\subsection{Squarks}\label{sec:sq}
A squark can decay either {\em directly} to a quark and a lepton/neutrino or {\em indirectly} to 
a quark and a neutralino, which subsequently decays to two quarks and a lepton or 
neutrino. Decays to charginos or heavier neutralinos are not considered. 

\begin{figure}
\centering
\makebox[\textwidth]{
\epsfig{figure=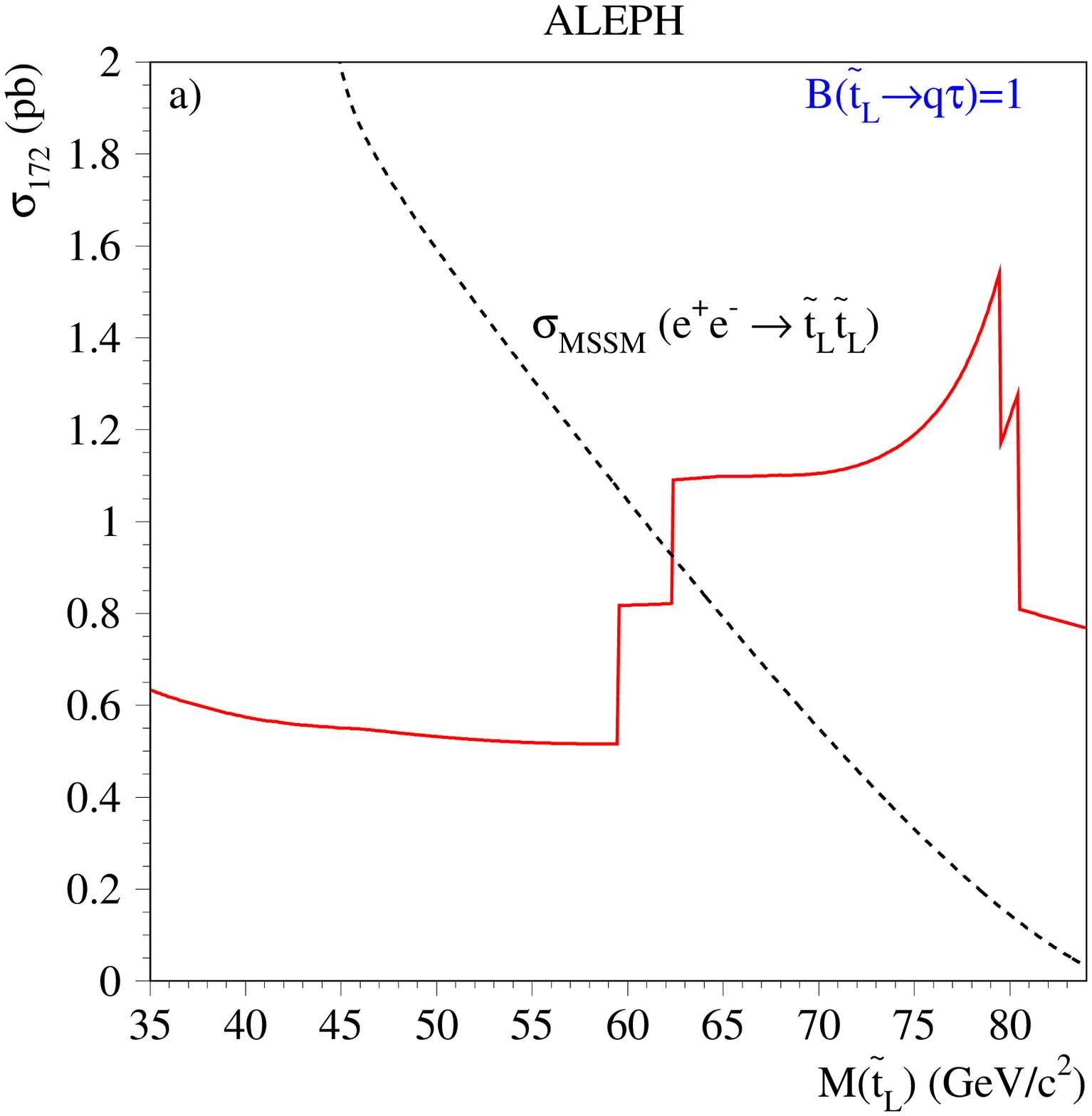,width=0.5\textwidth}
\epsfig{figure=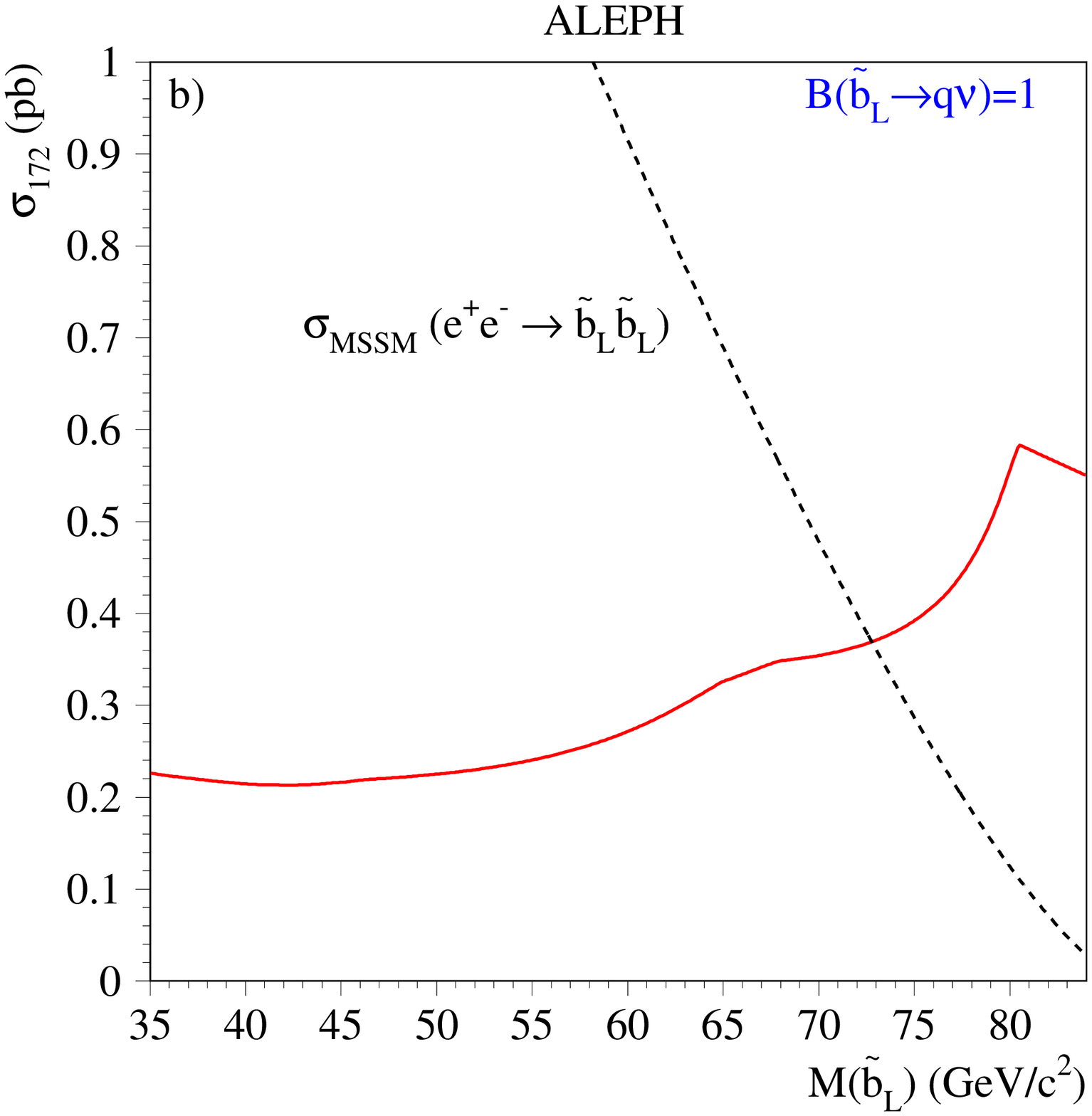,width=0.5\textwidth}\hfill}
\makebox[\textwidth]{
\epsfig{figure=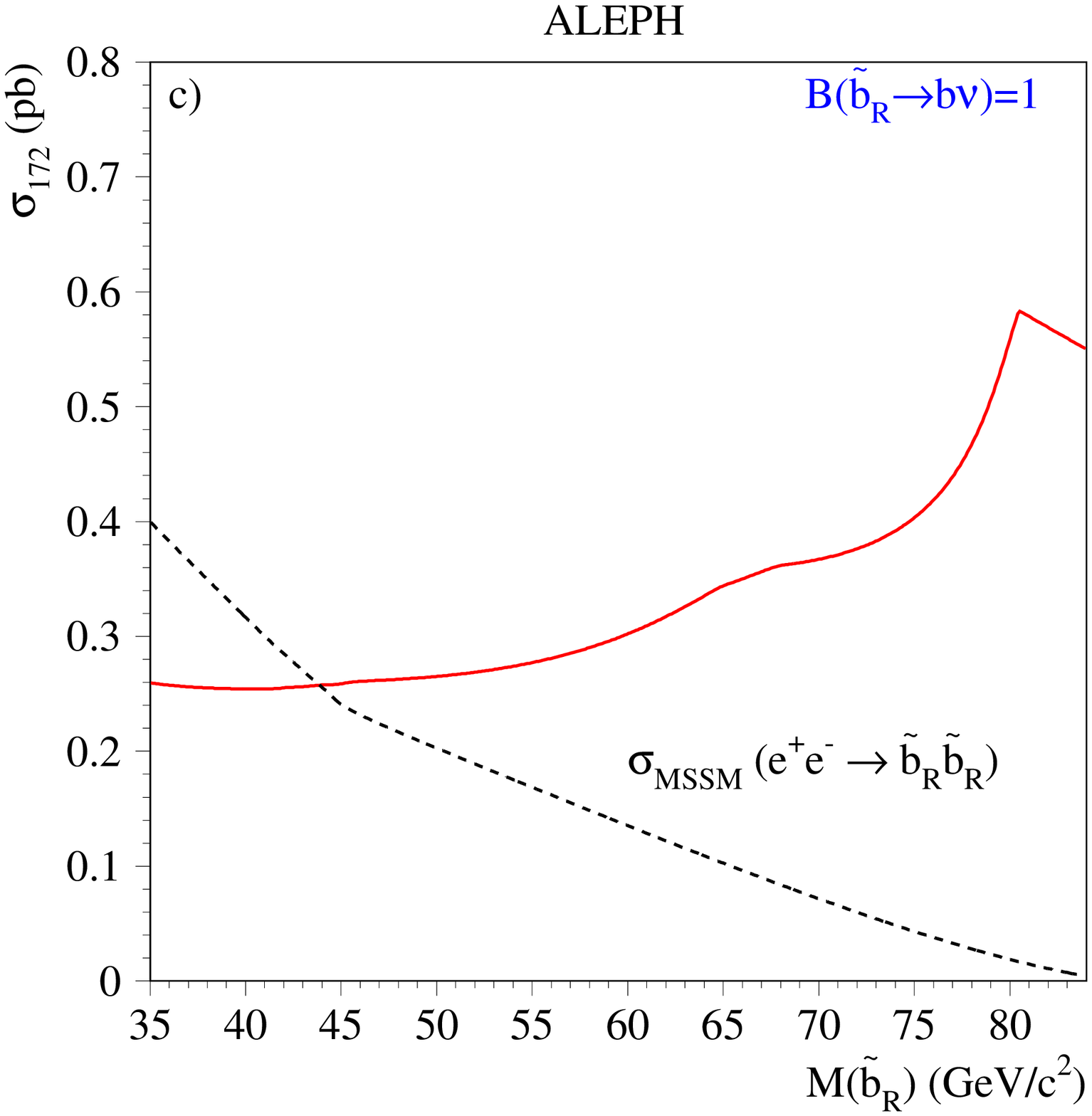,width=0.5\textwidth}
\epsfig{figure=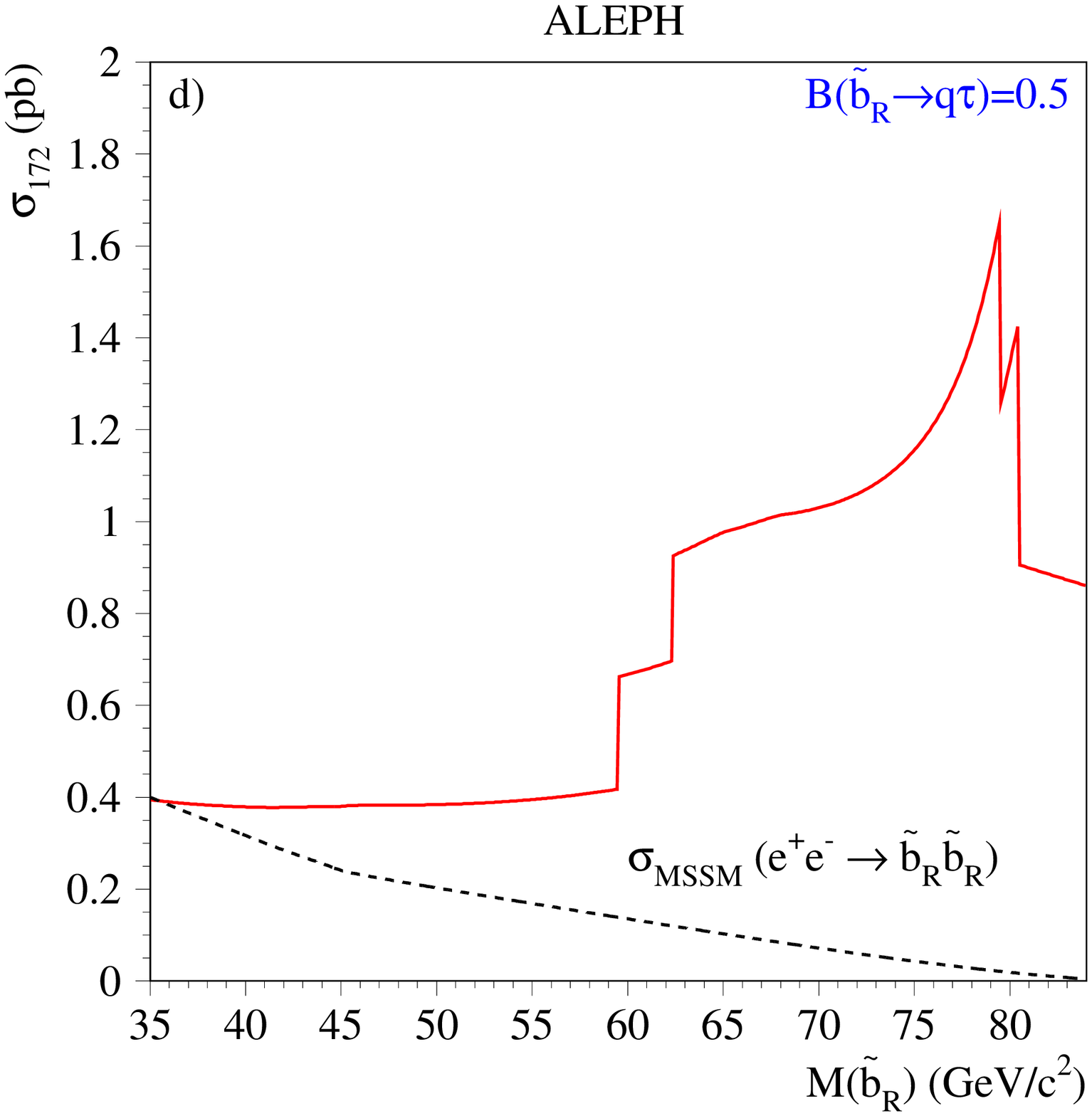,width=0.5\textwidth}\hfill}
\caption[.]{\label{sq.direct}\em{The $95\%$ C.L. excluded cross sections
for the {\em direct} decays of 
a) $\stq_L\stq_L \to \tau \q\tau \q$, b)
$\sbq_L\sbq_L \to \q\nu \q\nu$, c) $\sbq_R\sbq_R \to \bottom\nu \bottom\nu$ and d) 
$\sbq_R$ production with a $50\%$ branching ratio 
into $\tau \q$ and $\nu \q$. The MSSM cross sections
are superposed as dashed lines.}}
\end{figure}

The {\em direct} topology is defined as that when both squarks decay {\em
directly}  
leading to a topology of acoplanar jets and up to two leptons. Couplings leading
to electrons or muons in the final state are neglected as existing limits from
the Tevatron \cite{TeVLQ} exclude the possibility of seeing this signal at LEP.
To select $\sq \ra \q \tau$ and $\sq \ra \q\nu$,  the Two Jets (plus Leptons)
selections were used, and typical signal efficiencies  are shown in 
Table~\ref{effics}.  For the 
2J+2$\tau$ selection the limit is set by sliding a mass window of width $20\gevcc$
centred on the squark mass over the mass spectrum. The resulting limits are
shown in Fig.~\ref{sq.direct}.

To select the {\em indirect} topology the reoptimised subselection I from the 
Multi-jets plus Leptons was employed.
The efficiencies for the stop and sbottom signals (c.f. Table~\ref{effics}) are
determined as functions of
the squark and neutralino masses and the decay mode of the $\chi$ at each of 
the three energies. In the region where the neutralino mass is close to the squark
mass the efficiency is reduced because one of the jets is very soft. In this region
the expected exclusion limit, as determined from \nbar\, is improved by switching
to the inclusive combination of 4J-VH and the 4J(2$\tau$) selection.

\begin{figure}
\begin{center}
\makebox[\textwidth]{
\epsfig{figure=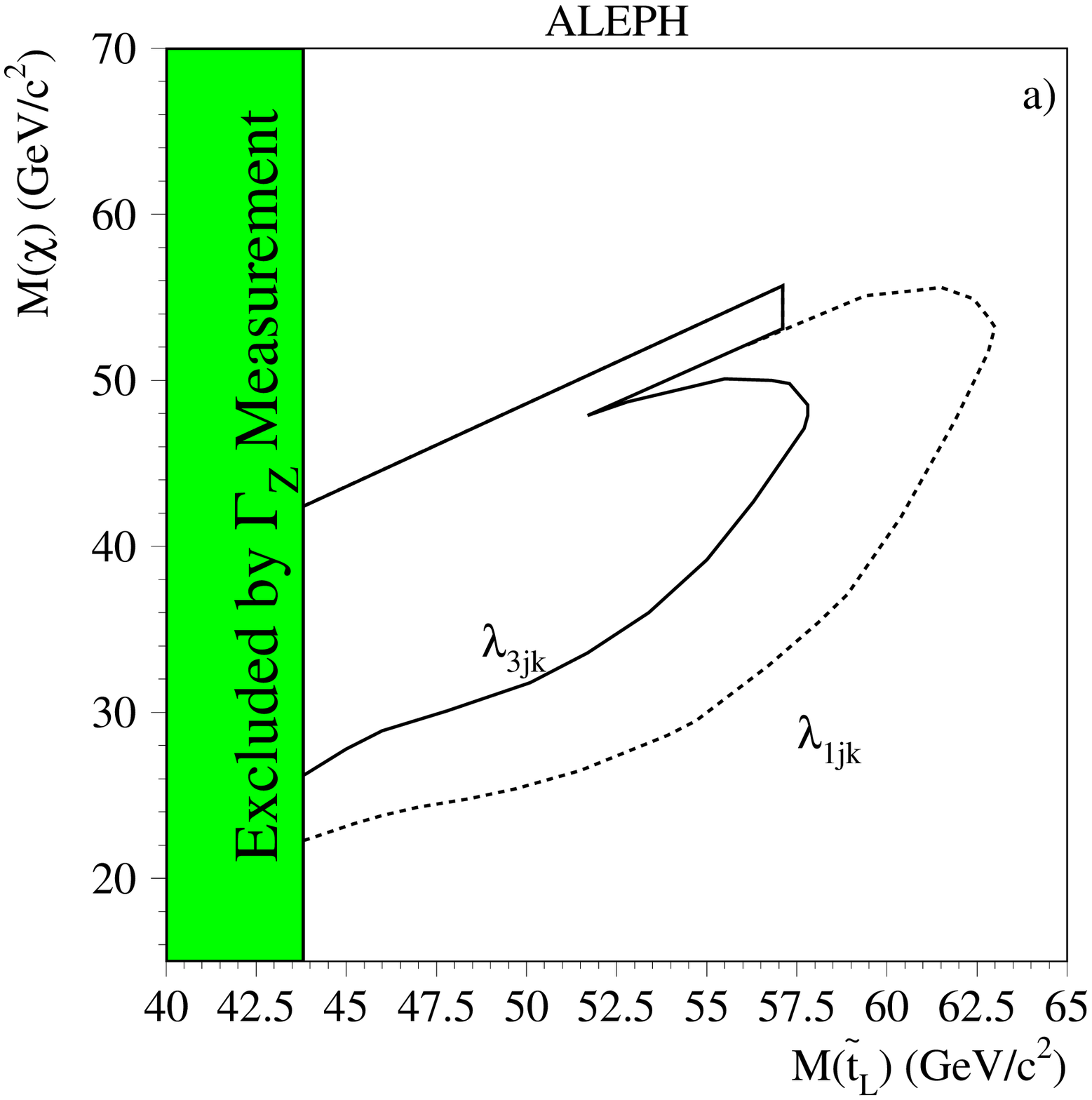,width=0.5\textwidth}
\epsfig{figure=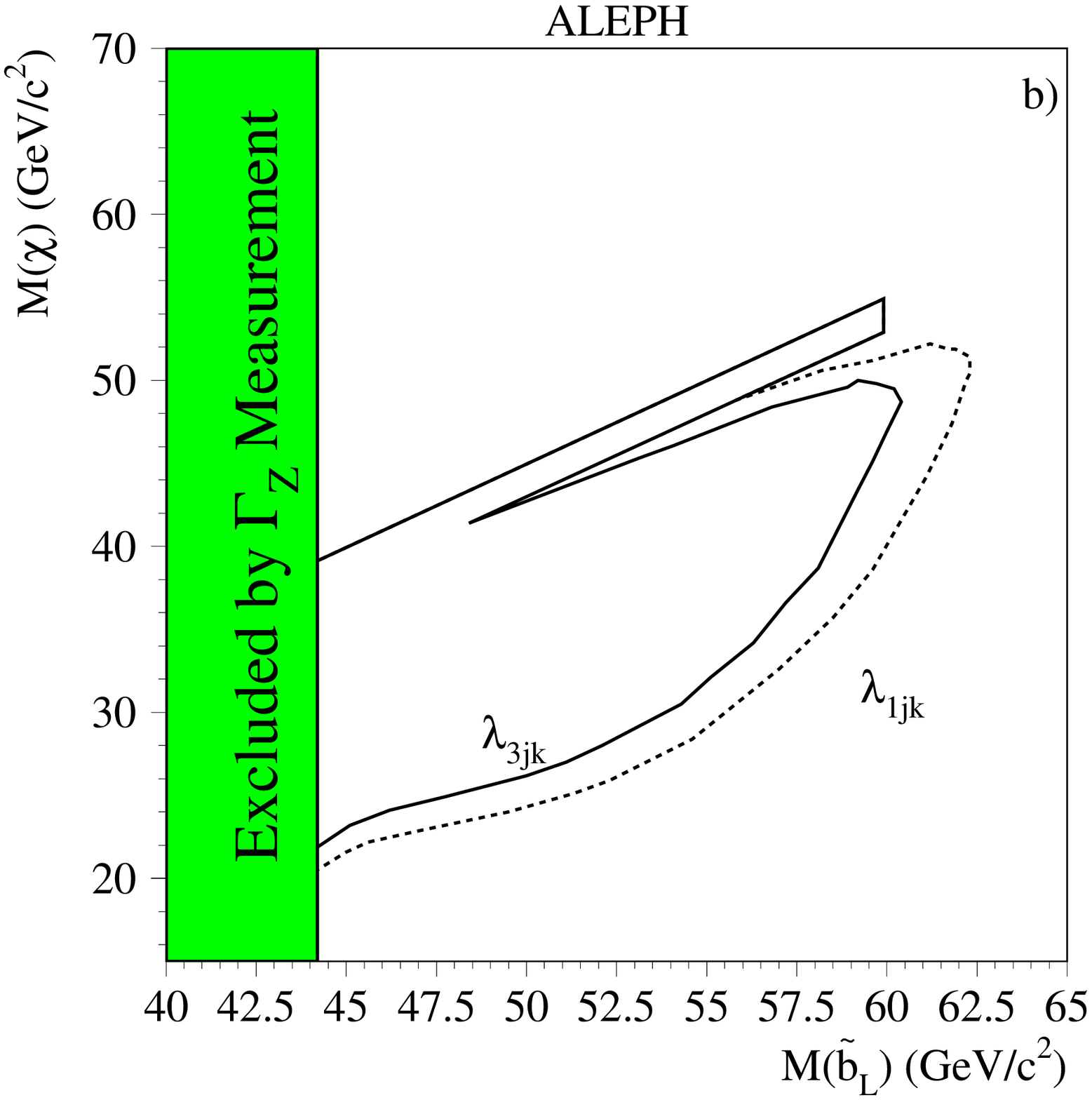,width=0.5\textwidth}\hfill}
\caption[.]{\label{sq.ind}\em{The $95\%$ C.L. limits on the 
stop and sbottom in the $(M_{\chi},M_{\sq})$ plane for the {\em indirect} decay modes. The limits 
are shown for the optimistic case of left-handed squarks. }}
\end{center}
\end{figure}

The limits in the $(M_{\chi},M_{\sq})$ plane obtained within the MSSM are shown
in Fig. \ref{sq.ind}. No limit is obtained for the general mixing angles of the
squarks.

\section{Conclusions}\label{conclusions}
A number of search analyses have been developed to select R-parity violating
SUSY topologies from the pair-production of sparticles.
It was assumed that the 
LSP has a negligible lifetime, and that only the  $LQ{\bar D}$ couplings are
non-zero. Limits were derived under the assumption that only one coupling
$\lambda'_{ijk}$ is non-zero.
 The search analyses for the various topologies find no evidence for
R-parity violating Supersymmetry in the data collected at
$\sqrt{s}=$130--172$\gev$, and limits have been set within the framework of the
MSSM. 

For the {\em indirect} decay modes charginos are excluded at the $95\%$ C.L. for
\mbox{$\charginom > 82 \gevcc$} at $m_0 = 500 \gevcc$ and $\tan{\beta}=\sqrt{2}$, and \mbox{$\charginom > 56 \gevcc$}
for $m_0=80\gevcc$ (the worst case), assuming that \mbox{$M_{\sq},M_{\tilde
\tau}>\charginom$}. For the {\em direct} decay modes \mbox{$\charginom > 82 \gevcc$} at
$m_0 = 500 \gevcc$, and $\charginom > 51 \gevcc$ for $m_0=70\gevcc$. Neutralinos
are excluded up to $30(29)\gevcc$ at $m_0=500\gevcc$ for the {\em indirect} ({\em direct})
decay modes, and up to $42(25)\gevcc$ at $m_0=0\gevcc$. For the worst case
$m_0\sim 100\gevcc$ no limit can be set on the neutralino mass. The above limits hold for any
 generation structure of the $LQ{\bar D}$
coupling.

The mass limits for the sfermions are highly dependent on the choice of
the indices $i,j,k$ and the nature of the LSP, mainly owing to the much smaller
production cross section of scalars compared to the fermionic cross sections. 
For the {\em indirect} decay modes  and the most conservative choice of coupling,
the slepton mass limits for $M_{\slep} - M_{\chi}>10\gevcc$ are:
\begin{itemize}
\item{$M_{\rm{\tilde  e}_R}>57\gevcc$ (gaugino region),}
\item{$M_{{\tilde \mu}_R}>45\gevcc$,}
\item{$M_{{\tilde \tau}_R}>45\gevcc$.}
\end{itemize}
For the {\em indirect} decays of  squarks and $M_{\chi}>30\gevcc$ the mass limits are:
\begin{itemize}
\item{$M_{\rm{\tilde  b}_L}>54\gevcc$,}
\item{$M_{\rm{\tilde  t}_L}>48\gevcc$.}
\end{itemize}
These mass limits  improve considerably upon existing limits.

\section{Acknowledgements}
It is a pleasure to congratulate our colleagues from the accelerator divisions
for the successful operation of LEP at high energy. 
We would like to express our gratitude to the engineers and 
support people at our home institutes without whose dedicated help
this work would not have been possible. 
Those of us from non-member states wish to thank CERN for its hospitality
and support.

\end{document}